\documentclass{aa}  
\usepackage{float}
\usepackage{graphicx}
\usepackage{txfonts}
\usepackage{hyperref}
\hypersetup{
     colorlinks   = true,
     citecolor    = blue
}

\usepackage{placeins}
\usepackage{siunitx}
\usepackage{gensymb}
\usepackage{caption,subcaption}
\usepackage{subcaption}
\usepackage{pgffor}
\usepackage{dblfloatfix}    % To enable figures at the bottom of page
\usepackage{lscape} % for 'landscape' environment
\usepackage{multirow}
\usepackage[export]{adjustbox}
\usepackage{numprint}
\npthousandsep{\,}

\def \Ha{H$\alpha$}
\def \agc{AGC 226178}
\def \msun{$\,M_{\odot}\,$}
\def \mstar{$\,M_{\star}\,$}
\def \chisq{$\chi^2$}
\def \magperarcsec{mag arcsec$^{-2}$}
\def \Hi{\ion{H}{i}}

\def \trps{$t_{rps}$}
\def \Vc{$V_{C}$}
\def \spin{$\lambda$}

\def \agc{AGC 226178}

\def \kms{km s$^{-1}$}
\def \Re{$R_e$}

\usepackage{pgffor}
\usepackage{readarray}
\def\data{% the data
67  186  227  261  321  421  466  590  604  646  796  892  935  964
 1008 1017 1160 1164 1346 1352 1397 1405 1424 1476 1479 1529 1593 1633
 1687 1719 1846 1968 1993 2001 2046 2079 2269 2343 2351 2365 2458 2531
 2572 2621 2690 2731 2887 2999 3032 3088 3112 3116 3128 3146 3190 3225
 3233 3265 3272 3356 3365 3379 3548 3633}
\readarray\data\ID[1,64] %read the data to \dataA

\def\datahidetected{% the data
186  261  1405  1424  1968}
\readarray\datahidetected\IDhidetected[1,5] %read the data to \dataA

\def\datahalphaextended{% the data
186  261  1405   1968}
\readarray\datahalphaextended\IDhalphaextended[1,4] %read the data to \dataA

\def\dataflags{% the data
Diffuse Diffuse UDG Diffuse Diffuse UDG Diffuse Diffuse
 Diffuse Diffuse Diffuse Diffuse UDG Diffuse Diffuse UDG
 UDG Diffuse Diffuse UDG UDG Diffuse Diffuse Diffuse UDG
 UDG UDG UDG Diffuse UDG Diffuse Diffuse UDG Diffuse
 UDG UDG UDG Diffuse Diffuse UDG UDG UDG Diffuse
 Diffuse Diffuse UDG Diffuse UDG UDG Diffuse Diffuse
 Diffuse Diffuse UDG Diffuse UDG Diffuse Diffuse UDG
 Diffuse Diffuse Diffuse UDG Diffuse}
\readarray\dataflags\flag[1,64] %read the data to \dataA

\def\dataflagscolor{% the data
blue blue red blue blue red blue blue blue blue
 blue blue red blue blue red red blue blue red red
 blue blue blue red red red red blue red blue blue
 red blue red red red blue blue red red red blue
 blue blue red blue red red blue blue blue blue red
 blue red blue blue red blue blue blue red blue
}
\readarray\dataflagscolor\flagcolor[1,64] %read the data to \dataA

\def\ncount{% the data
3  5  7  9 11 13 15 17 19 21 23 25 27 29 31 33 35 37 39 41 43 45 47 49
 51 53 55 57 59 61 63
}
\readarray\ncount\ncounts[1,31] %read the data to \dataA

\begin{document} 

\title{A Virgo Environmental Survey Tracing Ionised Gas Emission (VESTIGE)\\ XIII. The role of ram-pressure stripping in transforming the diffuse and ultra-diffuse galaxies in the Virgo cluster%\thanks{Based on observations obtained with MegaPrime/MegaCam, a joint project of CFHT and CEA/DAPNIA, at the Canada-French-Hawaii Telescope (CFHT) which is operated by the National Research Council (NRC) of Canada, the Institut National des Sciences de l'Univers of the Centre National de la Recherche Scientifique (CNRS) of France and the University of Hawaii.}
}

    \titlerunning{The role of the environment in the evolution of UDGs in the Virgo cluster}

   \author{
            Junais\inst{1,2}, 
            S. Boissier\inst{1},
            A. Boselli\inst{1}\thanks{Scientific associate at INAF-Osservatorio di Brera, Milano, Italy},
            L. Ferrarese\inst{3},
            P. C{\^o}t{\'e}\inst{3},
            S. Gwyn\inst{3},
            J. Roediger\inst{3},
            S. Lim\inst{4},
            E.W. Peng\inst{5,6},
            J.-C. Cuillandre\inst{7}, 
            A. Longobardi\inst{8},
            M. Fossati\inst{8,9},
            G. Hensler\inst{10},
            J. Koda\inst{11},
            J. Bautista\inst{11},
            M. Boquien\inst{12},            
            K. Ma\l{}ek\inst{2},
            P. Amram\inst{1},
            Y. Roehlly\inst{1}
            }
    \authorrunning{Junais et al.}
  \institute{Aix Marseille Univ, CNRS, CNES, LAM, Marseille, France\\   %1
             \email{samuel.boissier@lam.fr, alessandro.boselli@lam.fr}
             \and
             National Centre for Nuclear Research, Pasteura 7, PL-02-093 Warsaw, Poland\\ %2
             \email{junais@ncbj.gov.pl}
            \and
            National Research Council of Canada, Herzberg Astronomy and Astrophysics, 5071 West Saanich Road, Victoria, BC, V9E 2E7, Canada %3
            \and
            Department of Astronomy, Yonsei University, 50 Yonsei-ro, Seodaemun-gu, Seoul 03722, Republic of Korea %4
            \and
            Department of Astronomy, Peking University, Beijing 100871, PR China %5 
            \and
            Kavli Institute of Astronomy and Astrophysics, Peking University, Beijing 100871, PR China  %6
            \and
            AIM, CEA, CNRS, Université Paris-Saclay, Université Paris Diderot, Sorbonne Paris Cité, Observatoire de Paris, PSL University, F-91191 Gif-sur-Yvette Cedex, France %7
            \and
            Dipartimento di Fisica G. Occhialini, Universit\`a degli Studi di Milano-Bicocca, Piazza della Scienza 3, 20126 Milano, Italy %8
            \and
            INAF-Osservatorio Astronomico di Brera, via Brera 28, I-20121 Milano, Italy %9
            \and
            Department of Astrophysics, University of Vienna, T\"urkenschanzstrasse 17, 1180 Vienna, Austria %10
            \and
            Department of Physics and Astronomy, Stony Brook University, Stony Brook, NY 11794-3800, USA %11
            \and
            Centro de Astronom\'a (CITEVA), Universidad de Antofagasta, Avenida Angamos 601, Antofagasta, Chile %12            
             }

   \date{Received 10 June 2022 / Accepted 29 July 2022}

% \abstract{}{}{}{}{} 
% 5 {} token are mandatory
 
  \abstract
%   {} leave it empty if necessary  
  {Low-surface-brightness galaxies (LSBs) contribute to a significant fraction of all the galaxies in the Universe. Ultra-diffuse galaxies (UDGs) form a subclass of LSBs that has attracted a lot of attention in recent years (although its definition may vary between studies). Although UDGs are found in large numbers in galaxy clusters, groups, and in the field, their formation and evolution are still very much debated.}
  {Using a comprehensive set of multiwavelength data from the NGVS (optical), VESTIGE (\Ha{} narrowband), and GUViCS (UV) surveys, we studied a sample of 64 diffuse galaxies and UDGs  
  in the Virgo cluster to investigate their formation history.}
  {We analyzed the photometric colors and surface-brightness profiles of these galaxies and then compared them to models of galaxy evolution, including ram-pressure stripping (RPS) events to infer any possible strong interactions with the hot cluster gas in the past.}
  {While our sample consists mainly of red LSBs, which is typical in cluster environments, we found evidence of a color variation with the cluster-centric distance. Blue, \Hi{}-bearing, star-forming diffuse galaxies are found at larger distances from the cluster center than the rest of the sample.
  The comparison of our models with multifrequency observations suggests that most of the galaxies of the sample might have undergone a strong RPS event in their lifetime, on average 1.6 Gyr ago (with a large dispersion, and RPS still ongoing for some of them). This process resulted in the transformation of initially gas-rich diffuse blue galaxies into gas-poor and red ones that form the dominant population now, the more extreme UDGs having undergone the process in a more distant past on average.}
  {The RPS in dense environments could be one of the major mechanisms for the formation of the large number of quiescent UDGs we observe in galaxy clusters.}
   \keywords{Galaxies: clusters: general; Galaxies: clusters: individual: Virgo; Galaxies: evolution; Galaxies: interactions; Galaxies : star formation}
   \maketitle
%
%-------------------------------------------------------------------

\section{Introduction}

Most extragalactic surveys are focused toward high-surface-brightness galaxies (HSBs) that are not affected by the brightness of our night sky, and are likely to be detected by standard techniques. This leads to a partial understanding of the nature of the galactic population as a whole. Low-surface-brightness galaxies (LSBs), of which we know little, may be an important population.

Low-surface-brightness galaxies are diffuse galaxies that are fainter than the typical night sky surface-brightness level of $\sim$23 \magperarcsec{} in the B band \citep{bothun1997}. Astronomers have only known about the existence of the LSB population for about four decades \citep{sandage1984}, and, until the beginning of the 21st century, only a handful of them had been identified \citep{bothun1987,impey1988,bothun1990,delcanton1997}. The extreme faintness of LSBs hindered in-depth observations for a long time. Now it is estimated that LSBs may represent about 50\% (or more) of all the galaxies in the Universe \citep{oneil2000,galaz2011,martin2019}. In recent years, with advancements in technology, it has become possible to obtain deeper observations, allowing astronomers to study this very significant population with a new perspective (e.g., owing to instruments such as CFHT Megacam, Subaru Suprime-Cam, the Dragonfly Telescope Array, and VLT-MUSE). Ultra-diffuse galaxies (UDGs), a subclass of LSBs, have attracted a lot of attention in the past few years \citep{vandokkum2015a,koda2015,leisman2017,grishin2021}. While UDGs are usually broadly defined as galaxies with a lower surface brightness and a larger extent than other galaxies or dwarfs, various working definitions have been used (depending especially on available data). 

For instance, in the Coma cluster, \cite{vandokkum2015a} and \cite{koda2015} selected galaxies with a central surface brightness ($\mu_{0,g}$) > 24 \magperarcsec{} and an effective radius ($R_{e,g}$) > 1.5 kpc, this size limit being dictated by the limited angular resolution of the Dragonfly Telescope Array. 
In the Virgo cluster, where a complete set of high-quality deep imaging data is available thanks to the Next Generation Virgo Cluster Survey \citep[NGVS;][]{ferrarese2012}, \citet{lim2020} defined UDGs as galaxies at least $2.5\sigma$ away from the Virgo scaling relationships (see Sect. \ref{sect:lsb_udg_sample_selection}).

Ultra-diffuse galaxies  are found in abundance in a variety of environments, including galaxy clusters, groups, and the field \citep{koda2015,prole2019,tanoglidis2021,zaritsky2022}. Cluster UDGs tend to be quiescent \citep{koda2015,van_der_burg2016}, whereas UDGs in low density environments are gas-rich and blue \citep{leisman2017,prole2019}. Despite their abundance, the extreme nature of UDGs poses serious questions as to their formation and evolution, which are still debated.

Several UDG formation scenarios have been proposed. For instance, \citet{vandokkum2015a} suggested that UDGs could be failed Milky Way-like galaxies residing in large halos that have experienced a truncation of their star formation history in the past. Another scenario considered UDGs as "puffed-up dwarf" galaxies that are an extension of the dwarf galaxy population whose stellar and gas components were puffed up by internal processes, such as supernova feedback \citep{chan2018,dicintio2019}, or external processes, such as tidal interaction, mergers, or ram-pressure stripping \citep[RPS;][]{yozin_bekki2015,zaritsky2017,conselice2018,bennet2018,baushev2018, carleton2019,lim2020}. Recent work by \citet{grishin2021} suggests that UDGs might be the outcome of an early RPS event followed by a passive expansion in size over a long time, which removed any signatures of the RPS from their morphology. \citet{amorisco_loeb2016} put forward yet another scenario where UDGs could be formed in dwarf-sized halos with an intrinsically large initial angular momentum, irrespective of their environment. Considering all the above UDG formation mechanisms, it is likely that UDGs are a mix of galaxies with multiple evolutionary paths, as also suggested by the various amounts of dark matter (low or high dark matter content) determined for a few of them in the literature \citep{vandokkum2015a,vandokkum2018,toloba2018}.

Most of the discovered and identified UDGs are in several nearby galaxy clusters and groups \citep{koda2015,van_der_burg2016,yagi2016,lim2020}. The Virgo cluster, being one of the richest clusters of galaxies in the nearby Universe and with abundant deep multiwavelength data \citep{boselli2011,ferrarese2012,boselli2018_vestige1,haynes2018}, is therefore a perfect laboratory for studying UDGs and other LSBs in general. The purpose of this work is to study a sample of LSBs in the Virgo cluster using a multiwavelength set of photometric data in the optical, UV, and \Ha{} narrowband (NB) with the aim of understanding the role of the cluster environment in shaping their evolution.

In Sect. \ref{DATA_AND_MEASUREMENTS} we present the data used in this work, and we describe our sample selection in Sect. \ref{sect:lsb_udg_sample_selection}. Section \ref{sect:lsb_udg_measurements} discusses the preparation of the data and the photometric measurements performed on the sample. In Sect. \ref{sect:analyis_of_the_measurements} we analyze the observed photometric properties of the sample. Sections \ref{sect:alfalfa_crossmatch_subsample} and \ref{sect:halpha_detection} present a subsample of sources with \Hi{} and \Ha{} detections. Section \ref{sect:modelling_rps} is dedicated to an extensive comparison of the observed properties of the sample with a suite of chemo-spectrophotometric galaxy evolutionary models.
After a discussion in Sect. 7, conclusions are given in Sect. \ref{conclusion}. We note that results obtained from the analysis of other galaxies selected to have high angular momentum (but which do not form a complete sample) are given in Appendix \ref{appendix:high_spin_sample}.

Throughout this paper we assume that the Virgo cluster is centered on M87, has a virial radius ($R_{vir}$) of 1.55 Mpc \citep{ferrarese2012}, and is located at a distance of 16.5 Mpc \citep{gavazzi1999,mei2007}. At this distance, 1 arcsec corresponds to 80 pc.

\section{Data}\label{DATA_AND_MEASUREMENTS}
This work makes use of multiwavelength photometric data from various surveys of the Virgo cluster as discussed below.

\subsection{NGVS}\label{sect:ngvs_survey}
The NGVS (\citealt{ferrarese2012}) is a deep broadband imaging survey of the Virgo cluster in the \textit{u}, \textit{g}, \textit{i}, and \textit{z} bands, carried out with the MegaCam instrument on the 3.6m Canada-France-Hawaii Telescope (CFHT). The survey spans an area of 104 deg$^2$, covering the whole Virgo cluster region from its core to one virial radius ($R_{vir}=1.55$ Mpc; \citealt{ferrarese2012}). The NGVS images were processed with the Elixir-LSB pipeline optimized for the recovery of low-surface-brightness features, reaching a surface-brightness limit of $\mu_{g}\sim29$ AB \magperarcsec{} ($2\sigma$ above the mean sky level; \citealt{ferrarese2012}). The survey has a typical full-width-half-maximum (FWHM) resolution of 0.54\arcsec{} in the \textit{i} band and $\sim$0.8\arcsec{} in the other bands (see Table \ref{table:survey_fwhm_wavelength}). Full details on the survey, including observations and data processing, are discussed in \citet{ferrarese2012}.

%-------------------------------------------------------------
% SURVEY BANDS AND RESOLUTION TABLE
%-------------------------------------------------------------
%
\begin{table}
\centering                          % used for centering table
\caption{Filters and resolution of the data used in this work.}             % title of Table
\begin{tabular}{c c c c c}        % centered columns (4 columns)
\hline\hline                 % inserts double horizontal lines
Survey & Filter & Central wavelength (\AA) & FWHM\\    % table heading 
\hline                        % inserts single horizontal line
   GUViCS & FUV & 1524 & $\sim$5$\arcsec$ \\ 
   GUViCS & NUV & 2309 & $\sim$5$\arcsec$ \\ 
   NGVS & \textit{u} & 3811 & 0.88$\arcsec$ \\
   NGVS & \textit{g} & 4862 & 0.80$\arcsec$ \\
   VESTIGE & \textit{r} & 6258 & 0.76$\arcsec$ \\
   VESTIGE & \Ha{} & 6591 & 0.76$\arcsec$ \\
   NGVS & \textit{i} & 7552 & 0.54$\arcsec$ \\
   NGVS & \textit{z} & 8871 & 0.75$\arcsec$ \\
\hline                                   %inserts single line
\end{tabular}
\tablebib{NGVS \citep{ferrarese2012}; GUViCS \citep{boselli2011}; VESTIGE \citep{boselli2018_vestige1}}
\label{table:survey_fwhm_wavelength}      % is used to refer this table in the text
\end{table}
%
%-------------------------------------------------------------

Using the deep data covering the entire cluster, the NGVS catalog is the most up-to-date catalog of the Virgo cluster \citep{ferrarese2020}. For every source in the NGVS catalog, the cluster membership probabilities are designated into three categories as a {\tt certain}, {\tt likely} or {\tt possible} member. This was done using a rigorous algorithm involving multiple distance indicators to compute the probability that a given galaxy is a member of the cluster. This process utilizes several scaling relationships (of magnitudes, colors, and structural parameters), photometric redshift estimates, and visual inspections to confidently identify potential Virgo cluster members. The final NGVS catalog consists of a total of 3689 galaxies, out of which only 1483 are in the Virgo Cluster Catalogue (VCC) of \citet{binggeli1985}.

The NGVS catalog is used as the basis for the LSB sample selection and analysis discussed in this work. Sources with only the {\tt certain} and {\tt likely} cluster membership flag from NGVS were used, to avoid possible contaminations (see Sect. \ref{sect:lsb_udg_sample_selection} for details on the sample selection). For reference, the mean membership probabilities for the class of {\tt certain} and {\tt likely} members are $84\pm23\%$ and $77\pm21\%$, respectively \citep{lim2020}. Among the 3689 NGVS galaxies, 1651 galaxies are {\tt certain} members (1280 of them in the VCC) and 842 galaxies are {\tt likely} members (166 of them in the VCC). We adopt the cluster memberships of the NGVS catalog and assume that all the cluster members are at the same distance (16.5 Mpc).

For all the NGVS galaxies, stellar masses were computed (Roediger et al. in prep), using spectral energy distribution modeling based on the {\tt PROSPECTOR} code \citep{conroy2009,johnson2021}, comparing %The SEDs were modeled on 
the $ugriz$ integrated photometric fluxes from the NGVS catalog, with %a simple 
stellar population spectra based on the MIST isochrones \citep{choi2016}, MILES stellar library \citep{sanchez2006} and a Chabrier initial mass function (IMF; \citealt{chabrier2003}). The median stellar mass values obtained via this procedure are used in our sample selection discussed in Sect. \ref{sect:lsb_udg_sample_selection}.

\subsection{GUViCS}

The GALEX Ultraviolet Virgo Cluster Survey (GUViCS; \citealt{boselli2011}) is a blind survey of the Virgo cluster carried out with GALEX in the far-UV (FUV) and near-UV (NUV) bands. GUViCS combines data from the GALEX All-Sky Imaging Survey, typically with an exposure of 100s, Medium Imaging Survey with deeper exposure times of at least 1500s \citep{morrissey2005}, which corresponds to an NUV surface-brightness limit of $\sim$28.5 AB \magperarcsec{}, and dedicated observations of the Virgo cluster \citep{boselli2011}. With the GALEX field of view of $\sim$1.2\degr{} and a resolution of $\sim$5\arcsec{}, GUViCS covers almost the entire Virgo cluster region with multiple overlapping exposures in the NUV band, while only $\sim$ 40\%\ in the FUV band (see Fig. 1 of \citealt{boselli2014}).

\subsection{VESTIGE}\label{sect:vestige_survey}

The Virgo Environmental Survey Tracing Ionised Gas Emission (VESTIGE; \citealt{boselli2018_vestige1}) is a blind \Ha{} NB, and broadband \textit{r} imaging survey of the Virgo cluster carried out with MegaCam at the CFHT. It is designed to cover an area of 104 deg$^2$ in the Virgo cluster (the same area as that of NGVS). 
The \Ha{} NB filter\footnote{VESTIGE \Ha{} NB filter contains the \Ha{} line and the two nearby [\ion{N}{ii}] emission lines at $\lambda6548$ and 6583 \AA. Hereafter we refer to the \Ha{}+[\ion{N}{ii}] contribution simply as \Ha{}, unless otherwise stated} of VESTIGE covers a wavelength range of $6538 < \lambda < 6644$ \AA\, with a central wavelength of 6591 \AA\, and filter width of 106 \AA.
Currently, the survey covers $\sim$75\% of the designed area at full depth (exposure of 7200 s in \Ha{}) with observations of high imaging quality (resolution of $\sim$0.76\arcsec{}; see Table \ref{table:survey_fwhm_wavelength}). The depth and extremely high image quality of the survey make it perfectly suitable for studying the effects of the environment on the star formation process in galaxies down to scales of $\sim$100 pc, since \Ha{} is a perfect tracer of star formation on a short timescale of $\sim$10 Myr \citep{kennicutt1998,boissier2013}. Moreover, the VESTIGE \Ha{} filter is optimal to detect the line emission of galaxies at the distance of the Virgo cluster with a typical recessional velocity of $-500\leq cz \leq 3000$ km s$^{-1}$. Therefore, in the case of detection, VESTIGE also provides further confirmation of the Virgo membership of NGVS galaxies.

The line sensitivity limit of VESTIGE is $f(\text{\Ha{}})\sim 4\times10^{-17}$ erg s$^{-1}$ cm$^{-2}$ (5$\sigma$ detection limit) for point sources and \textit{$\Sigma$(\Ha{})} $\sim$2$\times10^{-18}$ erg s$^{-1}$ cm$^{-2}$ arcsec$^{-2}$ ($1\sigma$ detection limit at 3$\arcsec$ resolution) for extended sources \citep{boselli2018_vestige1}. The contribution of the stellar continuum emission in the NB \Ha{} filter is determined and removed using a combination of the VESTIGE $r$-band and NGVS $g$-band images \citep{fossati2018,boselli2019_vestige5}.

\section{Sample selection}\label{sect:lsb_udg_sample_selection}

\begin{figure*}
    \centering
    \includegraphics[width=\hsize]{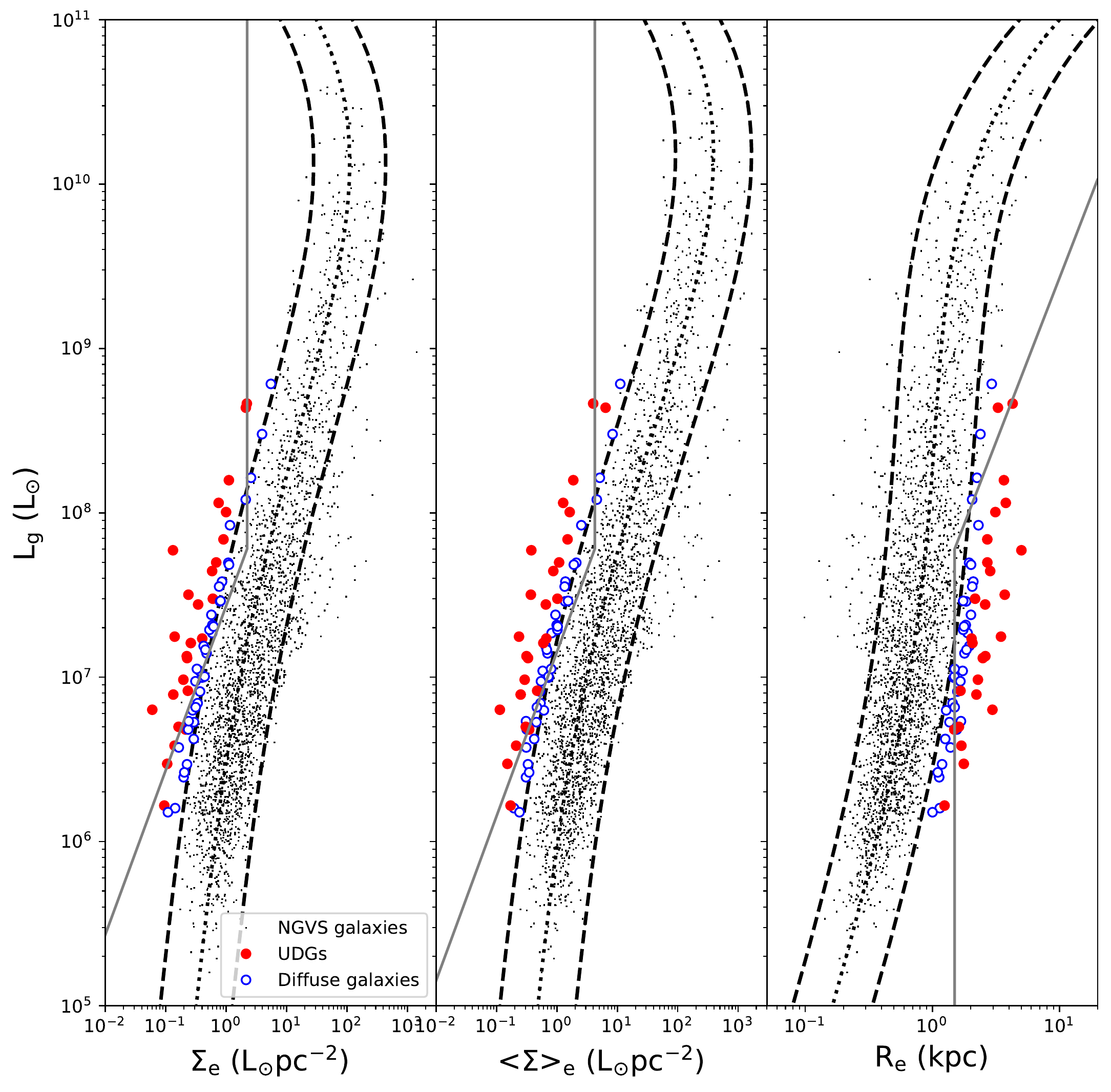}
    \caption{Scaling relations for the Virgo cluster galaxies (in the \textit{g} band). In the three panels from left to right, the luminosity ($L_g$) is plotted with respect to the surface brightness at the effective radius ($\Sigma_{e}$), the mean surface brightness within the effective radius (<$\Sigma$>$_{e}$), and the effective radius ($R_{e}$), respectively. The black dots are all the galaxies in the NGVS catalog. The filled red and open blue circles show the selected samples of UDGs and diffuse galaxies, respectively. The dotted and dashed black curves show the mean scaling relations and their $2\sigma$ confidence limits, respectively, based on \citet{lim2020}. The solid gray line marks the UDG selection cut of \citet{vandokkum2015a} with $\mu_{0,g}>24$ \magperarcsec{} and $R_{e,g}>1.5$ kpc for comparison.}
    \label{fig:lsb_sample_selection_plot}
\end{figure*}

The target of this work are large and diffuse galaxies. \citet{lim2020} identified 26 UDGs in the Virgo cluster as the most extreme extended and diffuse objects outliers by $>2.5\sigma$ from multiple galaxy scaling relationships (see Fig. 1 of \citealt{lim2020}). In this work, we aim to study the effects of the cluster environment on the evolution of LSBs, and possibly understand the evolutionary path that gave birth to the UDG population. 
We are inspired by the fact that several models and simulations of galaxy evolution in a cluster environment suggest that gravitational or hydrodynamic interactions might strongly affect the baryonic matter distribution on relatively short timescales ($\sim$ 1 Gyr) and thus significantly modifying the stellar surface brightness of the perturbed galaxies  \citep[e.g.,][]{Mastropietro2005, Boselli2008b}.

We thus decided to relax the \cite{lim2020} selection criteria to include a larger number of LSB objects, still avoiding any possible progenitor bias. We selected all objects located simultaneously at $>$ 2$\sigma$ from the same three scaling relations (consistent with the "primary" sample selection of \citealt{lim2020}). We also introduced a further criterion on the galaxy stellar mass \mstar{}$ < 10^9$ \msun{} \citep[similar to the luminosity cut used by][]{lim2020} to avoid any possible contamination by massive HSBs (mostly spirals).

Figure \ref{fig:lsb_sample_selection_plot} shows the scaling relations we used, where the \textit{g}-band luminosity of all the 3689 galaxies in the NGVS catalog \citep{ferrarese2020} is plotted as a function of the following three quanties in the \textit{g} band: 1) surface brightness measured at the effective radius ($\Sigma_{e}$); 2) mean surface brightness measured within the effective radius (<$\Sigma$>$_{e}$); 3) effective radius ($R_{e}$). The mean distribution of each scaling relation (dotted lines in Fig. \ref{fig:lsb_sample_selection_plot}) is a fourth-order polynomial, as given in Côté et al. (in prep.), which was obtained by a maximum likelihood fitting of the observed scaling relations.

Our selected sample %of LSBs 
consists of 64 galaxies, including the 26 primary UDGs from \citet{lim2020} and 38 additional galaxies. From now on, we address these subsamples as the UDGs and the diffuse galaxies, respectively. For simplicity when we consider all of them together in this work, we refer to them as LSBs. Table \ref{Table:lsb_sample_ngvs_params} provides the basic parameters for the selected sample.

\section{Measurements}\label{sect:lsb_udg_measurements}

We present the preparation of the available data and the measurements performed on the sample defined in Sect.  \ref{sect:lsb_udg_sample_selection}. 
We followed the same general steps as in \citet{junais2021}. For  completeness, we provide a brief description of the main steps and a few specific details that are different.

\subsection{Creation of stamps}

We collected the photometric data from the GUViCS, NGVS, and VESTIGE surveys (FUV, NUV, \textit{u}, \textit{g}, \textit{r}, \textit{i}, \textit{z,} and \Ha{} NB) to create separate cut-out stamps for all the sources in each band. This procedure was done using the {\tt Montage} tool in Python \citep{jacob2010}, which co-adds the multiple exposures of the same source to a single stamp. For comparison purposes, all the stamps, including the UV images (with GALEX pixels of 1.5\arcsec{}), were projected onto a pixel scale of the optical images (NGVS and VESTIGE) with pixels of 0.187\arcsec{}. From here onward we call the optical stamps (excluding the GALEX ones) the "high resolution" (HR) stamps, keeping their original resolution. We also convolved them with Gaussian kernels to match the GALEX spatial resolution (see Table \ref{table:survey_fwhm_wavelength} for the resolution of the data sets) to obtain a set of "low resolution" (LR) stamps all matching the GALEX resolution.

\begin{figure*}%[h]
    \centering
    %\ContinuedFloat
    \foreach \i in {1,...,35} {%
        \begin{subfigure}{0.18\textwidth}
            \includegraphics[width=\linewidth]{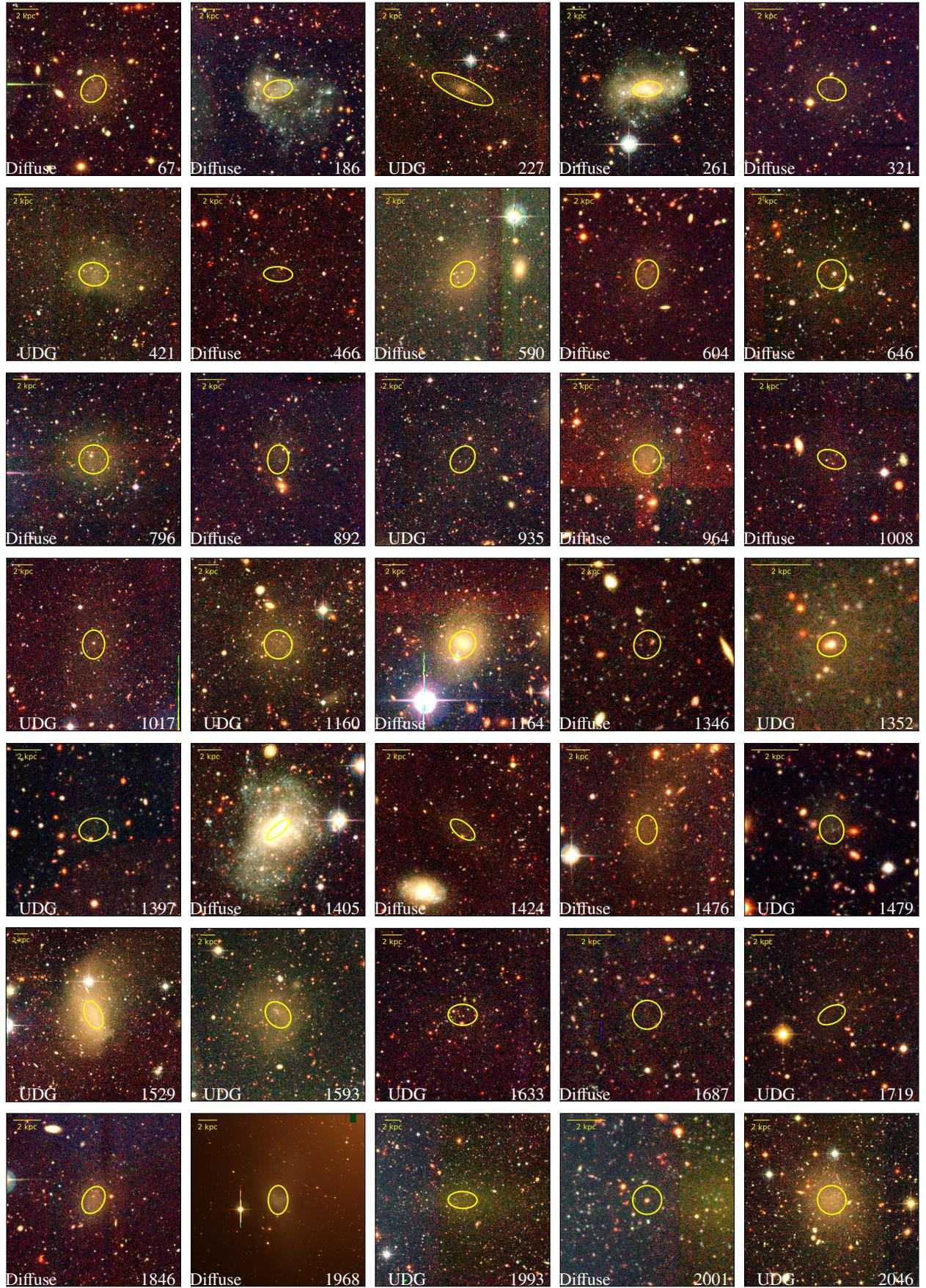}%
            \raisebox{3.5pt}{\makebox[-3pt][r]{\footnotesize \color{white} \ID[1,\i]
            }}
            % \raisebox{3.5pt}{\makebox[-63pt][r]{\footnotesize \color{\flagcolor[1,\i]} \flag[1,\i]}}
            \raisebox{3.5pt}{\makebox[-63pt][r]{\footnotesize \color{white} \flag[1,\i]
            }}
        \end{subfigure}\hspace{0em}}
    \caption{NGVS \textit{u, g, i} color composite images of the selected sample. The images are smoothed with a Gaussian kernel with $\sigma = 3$ pixels (0.56\arcsec{}) to enhance the low-surface-brightness features. The yellow ellipse in each image shows the effective radius and position angle of the galaxy from the NGVS catalog. The ID and the type of each galaxy are marked at the bottom of the images. The size of the stamps is 6\Re{} of each galaxy.}
    \label{fig:selected_sample_colour_images}
\end{figure*}

\begin{figure*}%[h]
    \centering
    \ContinuedFloat
    \foreach \i in {36,...,64} {%
        \begin{subfigure}{0.18\textwidth}
            \includegraphics[width=\linewidth]{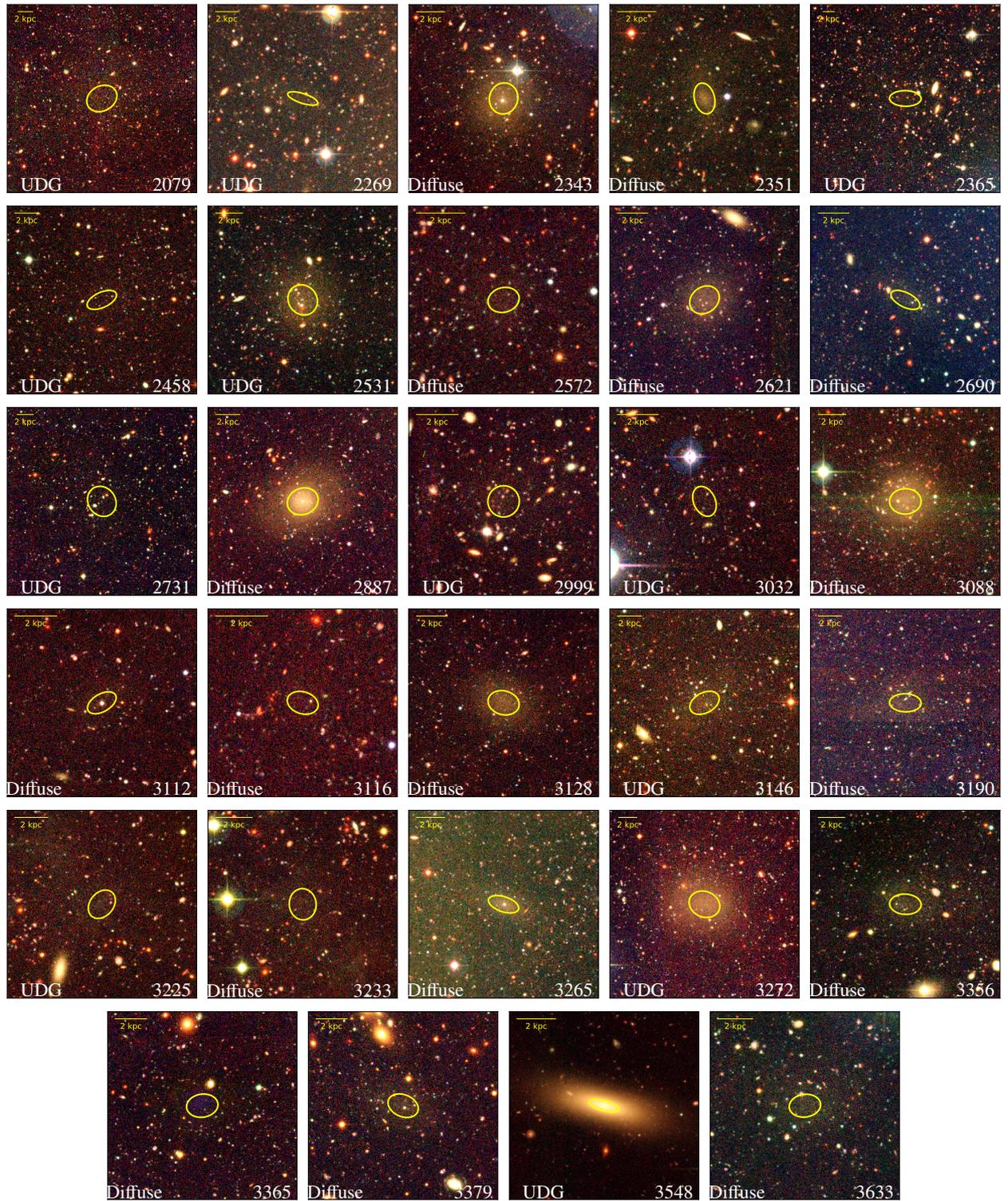}%
            \raisebox{3.5pt}{\makebox[-3pt][r]{\footnotesize \color{white} \ID[1,\i]
            }}
            % \raisebox{3.5pt}{\makebox[-63pt][r]{\footnotesize \color{\flagcolor[1,\i]} \flag[1,\i]}}
            \raisebox{3.5pt}{\makebox[-63pt][r]{\footnotesize \color{white} \flag[1,\i]
            }}
        \end{subfigure}\hspace{0em}}
\caption{ continued.}
\end{figure*}

Figure \ref{fig:selected_sample_colour_images} shows the \textit{u, g, i} color composite images of the HR stamps of all the galaxies in the sample.

\subsection{Preparation of masks and further processing of the stamps}\label{sect:preparation_of_masks}

We used the masks provided by the NGVS team \citep{ferrarese2020} to clean the stamps from foreground or background sources. NGVS masked artifacts, foreground stars, stellar halos, background galaxies, as well as globular clusters in the field. Whenever necessary, we modified the NGVS masks after inspection of the images to remove any residual artifacts and faint stars.
For the UV images, we first created a separate mask after smoothing the NGVS masks to the GALEX point source resolution of 5\arcsec{}. The UV masks were also manually edited to remove any background source not masked in the NGVS masks, and unmask areas that were not affected by artifacts in UV. The stamps were then cleaned using these two sets of masks (optical and UV) by the IRAF {\tt fixpix} procedure that linearly interpolates over the masked regions.

\subsection{Background sky measurements}\label{sect:sky_measurements}

We measured the local and global sky variation around each source following the procedure from \citet{gildepaz_madore2005}. This was done by placing 24 equidistant and equal-sized "skyboxes" vertically and horizontally around each source. The boxes were arbitrarily placed at a distance of 3\Re{} from the source to be far away from any of its light (for a sanity check, we repeated the exercise by placing them at 4\Re{} without significant changes in our results). We visually inspected all these boxes and moved or removed them if any problem or artifact could affect the measurements. 
Following Eq. 4 of \citet{gildepaz_madore2005}, we combined the mean and standard deviation of each of the skyboxes to get an estimate of the global sky level and uncertainty at any point within the field of a galaxy.

\subsection{Extraction of surface-brightness profiles}\label{sect:sb_profiles_measurements}

The surface-brightness profile measurements were separately performed for the HR and LR stamps using the {\tt Ellipse} task in the {\tt Photutils} python package \citep{photutils}. This procedure measures the average flux along concentric elliptical isophotes on the stamp of each galaxy. We fixed the geometrical parameters of the galaxy (central coordinates, position angle, and axis ratio) as given in Table \ref{Table:lsb_sample_ngvs_params} based on the NGVS catalog. The choice of concentric elliptical isophotes (rather than trying to fit the geometrical parameters in each band) was made to be consistent with previous studies \citep{munosmateos2011, boissier2016}, which allowed us to compare various band measurements made in the same physical area, which is important for color analysis and comparison to model predictions (Sect.  \ref{sect:model_fitting_with_sample_profiles}).
We applied a foreground Galactic extinction correction to the measured profiles using the E($B-V$) values from \citet{schlegel1998} given in Table \ref{Table:lsb_sample_ngvs_params}, and adopting a \citet{cardelli1989} extinction curve. We assumed that there is no internal extinction in these objects, as it is generally found in LSBs \citep{hinz2007,rahman2007}. The profiles were also corrected for the galaxy inclination using their corresponding axis ratios from Table \ref{Table:lsb_sample_ngvs_params}. The measured LR profiles of all the sources are given in Appendix \ref{appendix:profiles_and_bestfits}.
We verified that our surface-brightness profiles in the $g$ band are consistent with the profiles obtained with the NGVS parameters given in Table \ref{Table:lsb_sample_ngvs_params}.

\subsection{Surface-brightness profile decomposition}\label{sect:sb_decomposition}

We performed a simple two-component decomposition of all the profiles into a de Vaucouleurs central component (S\'ersic with index $n=4$) and an exponential disk, using the {\tt Profit} python routine developed by \citet{barbosa2015}. The fitting algorithm performs a weighted \chisq{} minimization procedure, with a Gaussian point spread function (corresponding to the FWHM of the data given in Table \ref{table:survey_fwhm_wavelength}) convolved with the model light profiles (both HR and LR). The initial guesses for the S\'ersic and disk components in the fitting were provided based on the values from the NGVS catalog.

\begin{figure}[!ht]
    \centering
    \includegraphics[width=\hsize]{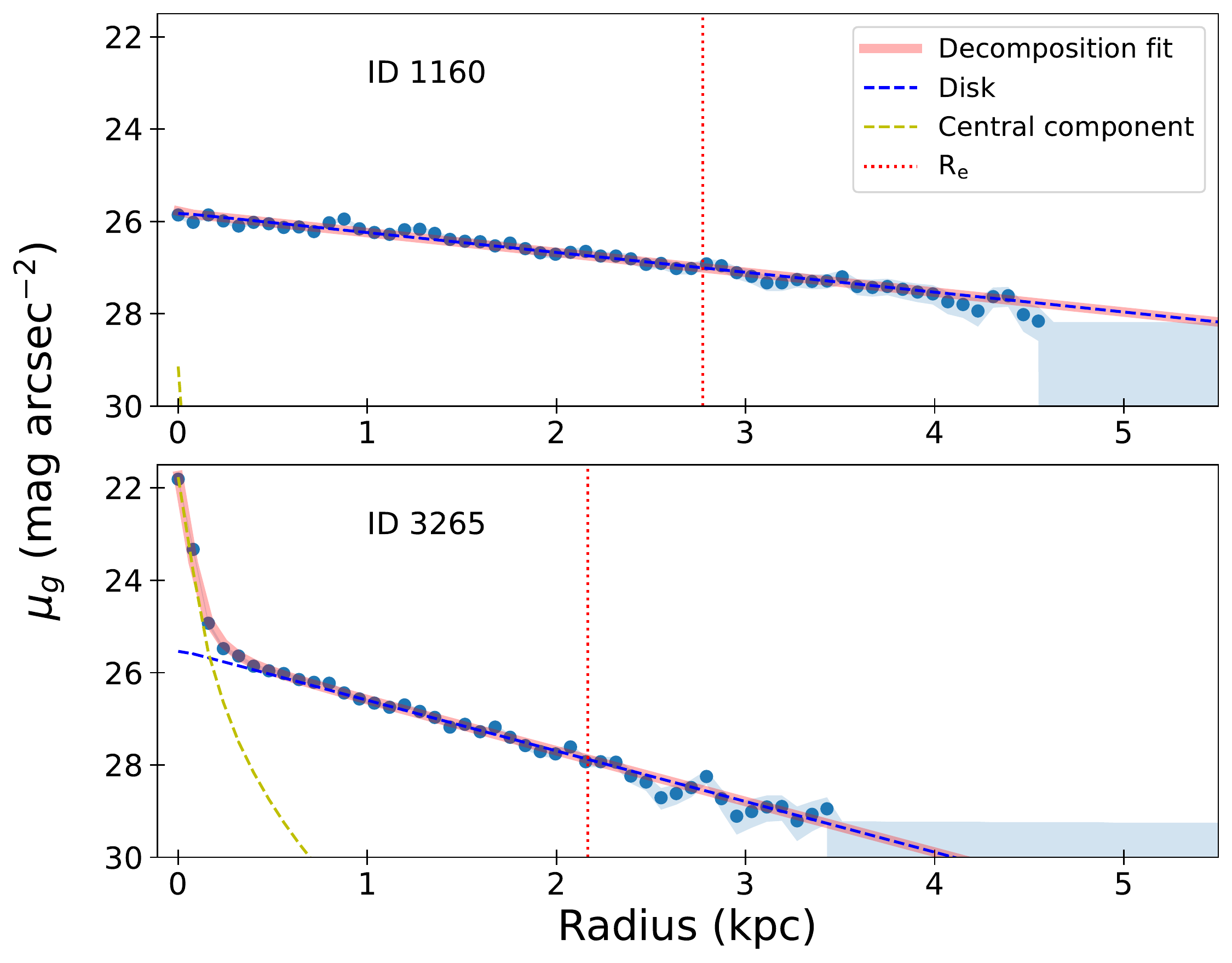}
    \caption{Example of the \textit{g}-band surface-brightness profile decomposition of two sources from the sample, one with a pure exponential disk, more typical of our sample (top panel), and the other with a significant central component (bottom panel). The blue circles and the shaded areas are the observed data points and the $3\sigma$ upper limits, respectively. The dashed yellow, dashed blue, and solid red lines are, respectively, the S\'ersic central component, the exponential disk, and the total best fit from the decomposition. The vertical dotted red line marks the effective radius of the galaxies.}
    \label{fig:decomposition_bulge_disk}
\end{figure}

For the majority of the sources in the sample, this procedure provided a good decomposition (about 93\% of the sample has a reduced $\chi^2_{\nu}<3$ in the \textit{g} band).
Figure \ref{fig:decomposition_bulge_disk} shows two examples of  decomposition, one for a galaxy with a bright central core, and another one for a galaxy more typical of our sample with a low central-to-total-light ratio.
Our sample predominantly consists galaxies with a central-to-total-light ratio lower than 0.1 (for $\sim$92\% of the sample), similar to what is generally observed for other LSBs and UDGs in the literature \citep{rong2017,pahwa2018}. For simplicity, we call them "disks" (since fitted by a disk component) although intrinsic shapes of dwarf galaxies may not exactly be thin disks, but rather oblate spheroids as shown by \citet{sanchez-janssen2016}, and seldom have perfectly exponential surface-brightness profiles \citep{ferrarese2020}.
The rotation measured in a few UDGs from the literature also indicate that at least some of them may not be rotating disks even if rotation velocities up to 60 \kms{} can be found in some cases \citep{ruiz-lara2018,toloba2018,vanDokkum2019,mancera-pina2020}.

The mean disk component central surface brightness ($\mu_{0}$) and scale length ($r_{s}$) of the sample in the \textit{g} band are of order 26 \magperarcsec{} and 1.9 kpc, respectively. The \textit{g}-band decomposition results of the HR profiles obtained here were used for identifying the radial range dominated by the disk component (in the cases where a few sources have a central component brighter than the disk, as shown in Fig. \ref{fig:decomposition_bulge_disk}, it could be a core or, in some case, a background object). For all the remaining analyses in this paper, such as the magnitude measurements (Sect. \ref{sect:magnitude_measurements}) and the model fitting (Sect. \ref{sect:model_fitting_with_sample_profiles}), we only use the LR profiles to have a matching resolution from UV to optical.

\subsection{Integrated magnitudes}\label{sect:magnitude_measurements}

In each photometric band, we integrated the surface-brightness profiles until the last measured radius in the \textit{g} band above 3$\sigma$ of the sky level, to have a uniform aperture for all bands (similar to the approach by \citealt{roediger2017}). This allowed us to compare colors in the same aperture, and to include all the light detectable in each band (the \textit{g} band being the deepest one). We call this magnitude the "integrated" magnitude. Our integrated magnitudes agree with those reported in the NGVS catalog, except for the few faintest sources ($g > 19$ mag) where our integrated magnitudes are systematically fainter than the NGVS ones. This can be attributed to the difference in the measurement procedures (e.g., integrated light profiles within a fixed aperture in our case and extrapolated S\'ersic fits in the case of NGVS). 

Table \ref{table:lsb_sample_magnitudes} (see page \pageref{page:photometry_table}) gives our measured integrated magnitudes in the optical and UV bands\footnote{The \Ha{} flux measurements for our sources will be published in Boselli et al. (in prep), as part of a unified catalog of the VESTIGE survey.}. Most of the sources are well detected in the optical bands (\textit{u}, \textit{g}, \textit{r}, \textit{i}, \textit{z}), whereas in \Ha{} and UV bands, there are primarily upper limits, which are still quite important in constraining the nature of the objects studied in this work.

\section{Analysis of the properties of the sample}\label{sect:analyis_of_the_measurements}

\subsection{Spatial distribution}\label{sect:spatial_distribution}

\begin{figure*}
    \centering
    \includegraphics[width=0.49\hsize]{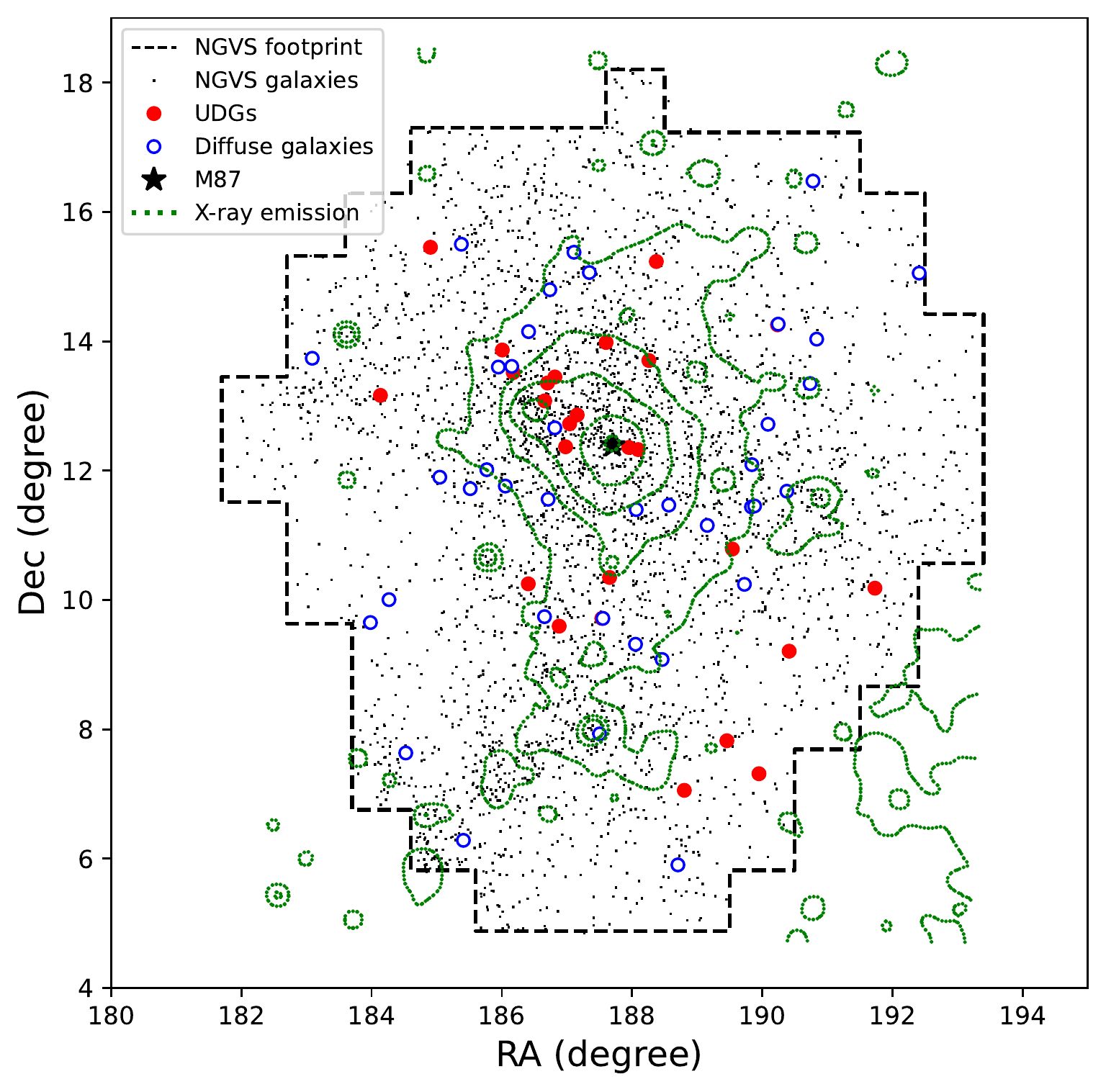}
    \includegraphics[width=0.49\hsize]{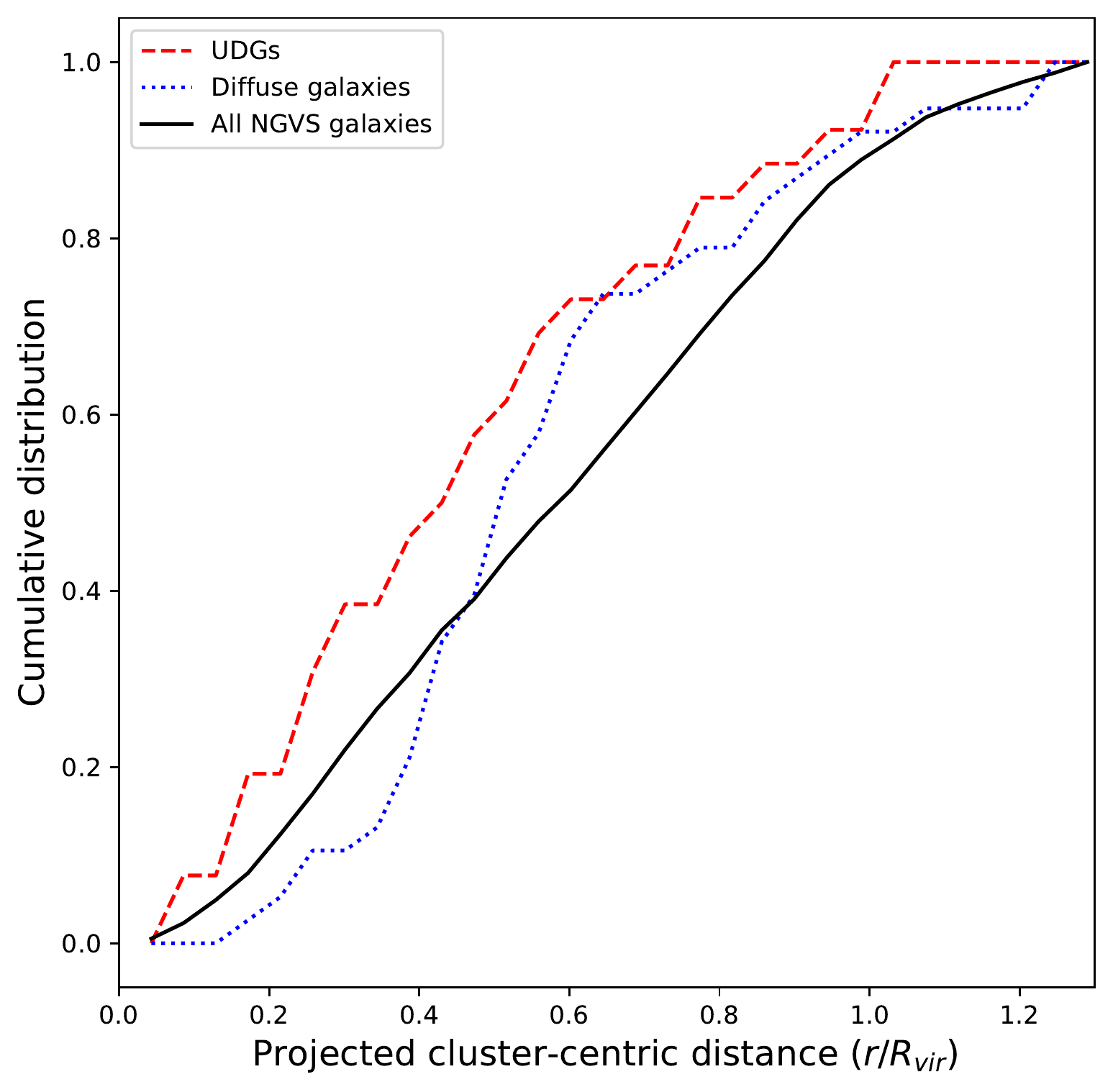}
    \caption{Distribution of the sample within the Virgo cluster. \textit{Left:} On-sky distribution of the selected sample of LSBs. The green contours mark the X-ray emission of the hot cluster gas obtained by ROSAT \citep{bohringer1994}. \textit{Right:} Cumulative distribution of the projected cluster-centric distance of the galaxies in units of the cluster virial radius ($R_{vir}=1.55$ Mpc; \citealt{ferrarese2012}). The dashed red, dotted blue, and solid black lines are the UDGs, diffuse galaxies, and all the NGVS galaxies, respectively.}
    \label{fig:radec_cdf}
\end{figure*}

The distribution of galaxies within a cluster could indicate their nature and evolutionary stage \citep[e.g.,][]{raichoor2012,Beyoro-Amado2021}. Figure \ref{fig:radec_cdf} shows the distribution of our selected sample of galaxies within the cluster. The left panel of Fig. \ref{fig:radec_cdf} shows the on-sky distribution where we can see that both the UDGs and the diffuse galaxies are found at all cluster-centric distances. However, an inspection of the cumulative distribution of the sample with respect to the cluster-centric distance\footnote{In this work, for simplicity, we assume that the Virgo cluster is centered on the galaxy M87.} (right panel of Fig. \ref{fig:radec_cdf}) reveals that UDGs are more centrally located in the cluster than the diffuse galaxies, which favor instead the cluster outskirts. 
This could be a first clue that the UDGs have fallen into the cluster at an earlier epoch, whereas the diffuse galaxies have entered the cluster only recently. However, one should be cautious that some galaxies that appear close to the cluster center could be due to projection effects, although it is unlikely to be the case for all the galaxies in our sample.

\subsection{Optical color distribution}
\label{sec_opticalcolor}
The optical colors of galaxies can be used to infer the nature of their underlying stellar population. Galaxies are generally separated into a red and blue sequence based on their optical colors \citep[e.g.,][]{strateva2001} or UV color \citep[e.g.,][]{boselli2014}. Such a bimodal distribution is also observed in LSBs \citep{greco2018,tanoglidis2021}. In a recent study, \citet{tanoglidis2021} classified red and blue LSBs based on their $g-i$ color ($g-i>0.6$ and $g-i<0.6$ mag for red and blue LSBs, respectively) from the Dark Energy Survey, covering a large area of the sky and thus sampling a wide range of environments (dense and less dense regions). 
They observed that the red LSBs are mainly located within denser 
regions whereas the blue ones are distributed uniformly across all environments. 

\begin{figure}
    \centering
    \includegraphics[width=\hsize]{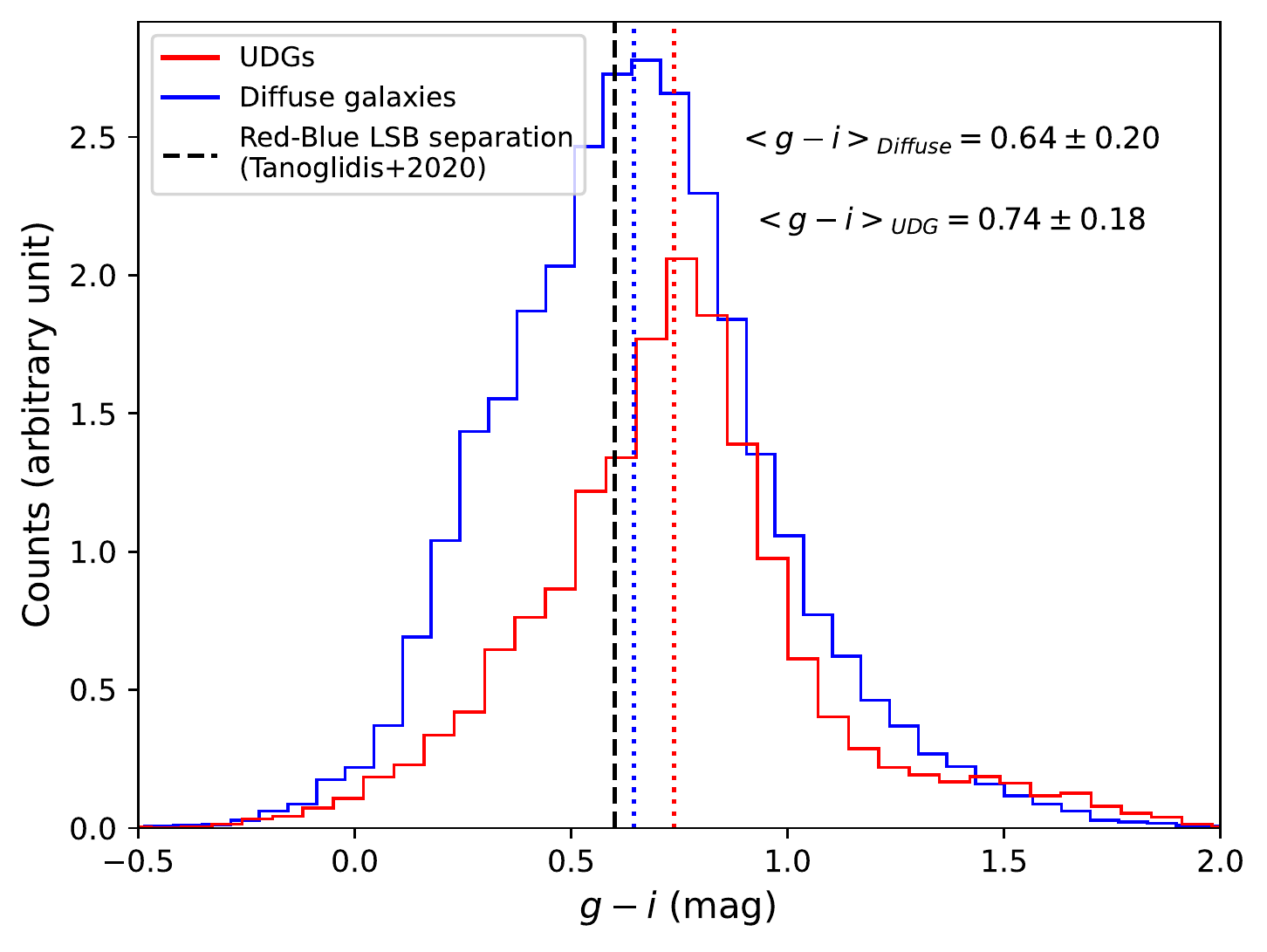}
    \caption{Observed $g-i$ color distribution of the sample after redistributing the galaxies within their uncertainties. The solid red and blue lines show the UDGs and diffuse galaxies, respectively, along with their median values (given at the top-right corner), marked as the dotted vertical lines. Sources with only upper limits on the color are removed from this distribution. The vertical dashed black line marks the separation of red and blue LSBs from \citet{tanoglidis2021}.
    }
    \label{fig:g-i_histogram}
\end{figure}

Figure \ref{fig:g-i_histogram} shows the distribution of the observed $g-i$ color of our sample of LSBs after redistributing them within their uncertainties using 1000 Monte Carlo chains. Such a redistribution was done to take into account the difference in the uncertainties of the measured color for each galaxy\footnote{All the histograms shown in this work are redistributed within the uncertainties of the quantities in a similar way}. The UDGs and the diffuse galaxies have a median $g-i$ value of 0.74 and 0.64 mag, respectively. This indicates a predominantly red LSB population consistent with what is generally found for cluster UDGs \citep{koda2015,van_der_burg2016,roman_trujillo2017a}. Moreover, the UDGs seems to be slightly redder than the diffuse galaxy population, although they agree within the color uncertainty. This again could indicate a difference (like for their spatial distribution discussed in Sect. \ref{sect:spatial_distribution}) between UDGs, being already red and central, while diffuse galaxies are bluer and less centrally concentrated in the cluster.

\subsection{Color variation with the cluster-centric distance}\label{sect:distance_color_gradient}

\begin{figure}
    \centering
    \includegraphics[width=\hsize]{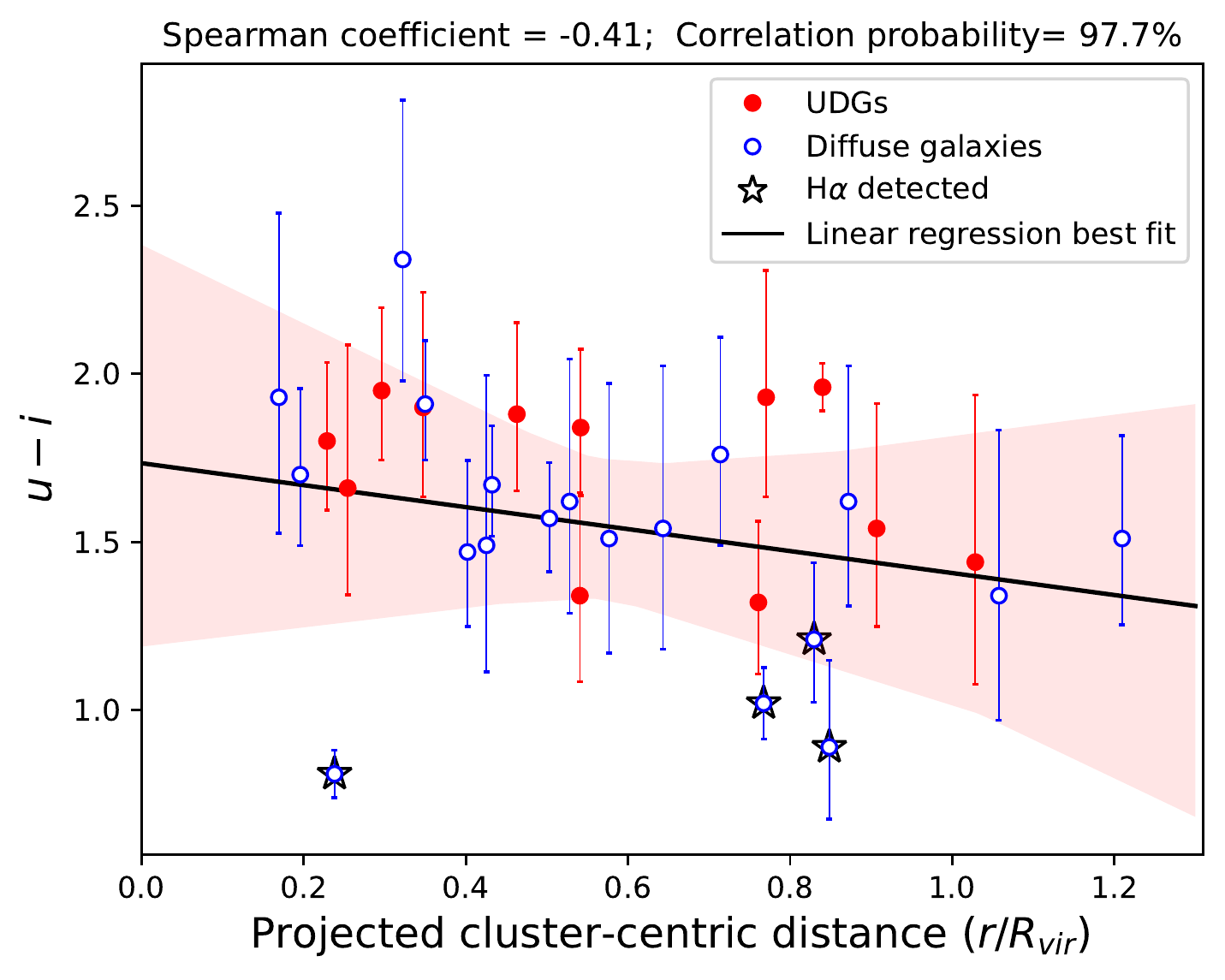}
    \caption{$u-i$ color of the sample as a function of the projected distance from the Virgo cluster center (distance from M87) in units of the cluster virial radius ($R_{vir}=1.55$ Mpc; \citealt{ferrarese2012}). The UDGs and the diffuse galaxy subsamples are marked with the filled red and open blue circles, respectively. The black star symbols are the sources detected in \Ha{}  (see Sect. \ref{sect:halpha_detection}). 
    The solid black line and the red-shaded region give the linear regression best fit and the $3\sigma$ scatter, respectively, obtained using the {\tt linmix} Python package \citep{kelly2007}. Sources with only an upper or lower limit in the color are excluded from the plot.}
    \label{fig:u-i_distance_gradient}
\end{figure}

Exploring the properties of cluster galaxies as a function of their distance from the cluster center is another useful tool in studying the role of the cluster environment in their evolution. Figure \ref{fig:u-i_distance_gradient} shows the variation of the $u-i$ color of the sample with respect to the projected cluster-centric distance. There is an indication of a color variation, with redder sources found more frequently toward the cluster center, whereas bluer sources more in the cluster outskirts (outside half the virial radius). This visual impression was further verified by a linear regression fitting of the entire sample using the {\tt linmix} Python package \citep{kelly2007}, as shown in Fig. \ref{fig:u-i_distance_gradient}.% and Eq. \ref{eqn:ug_gradient}: 

The linear fit has a slope of $-0.21\pm0.18$ and an intercept $1.73\pm0.18$ (see Table \ref{table:sample_gradientfits}). This indicates a negative color-distance correlation (although there is a large uncertainty in the slope). If we perform the same procedure on all the galaxies in the NGVS catalog within the same stellar mass range as our sample (\mstar{}$ < 10^9$ \msun{}), we obtain an almost flat linear fit with a slope of $-0.03\pm0.01$, indicating that there is no color-distance correlation among regular dwarfs in the NGVS catalog, but the trend we observe is more specific to our sample. To be more confident of our results, we computed the Spearman coefficient of our sample to obtain a value of $\rho_{\rm sample} = -0.41$, with a correlation probability of 97.7\%, reflecting a similar trend as obtained from the linear fit. 

Similar color variations were reported in the literature both from observations and simulations of group environments \citep{roman_trujillo2017b,jiang2019}, where LSBs with bluer colors tend to reside at farther distance from the group center.
This radial trend is discussed again in section \ref{sect:model_fitting_with_sample_profiles} in the context of models including an RPS effect.

\subsection{\Hi{} gas content}\label{sect:alfalfa_crossmatch_subsample}

\begin{figure*}
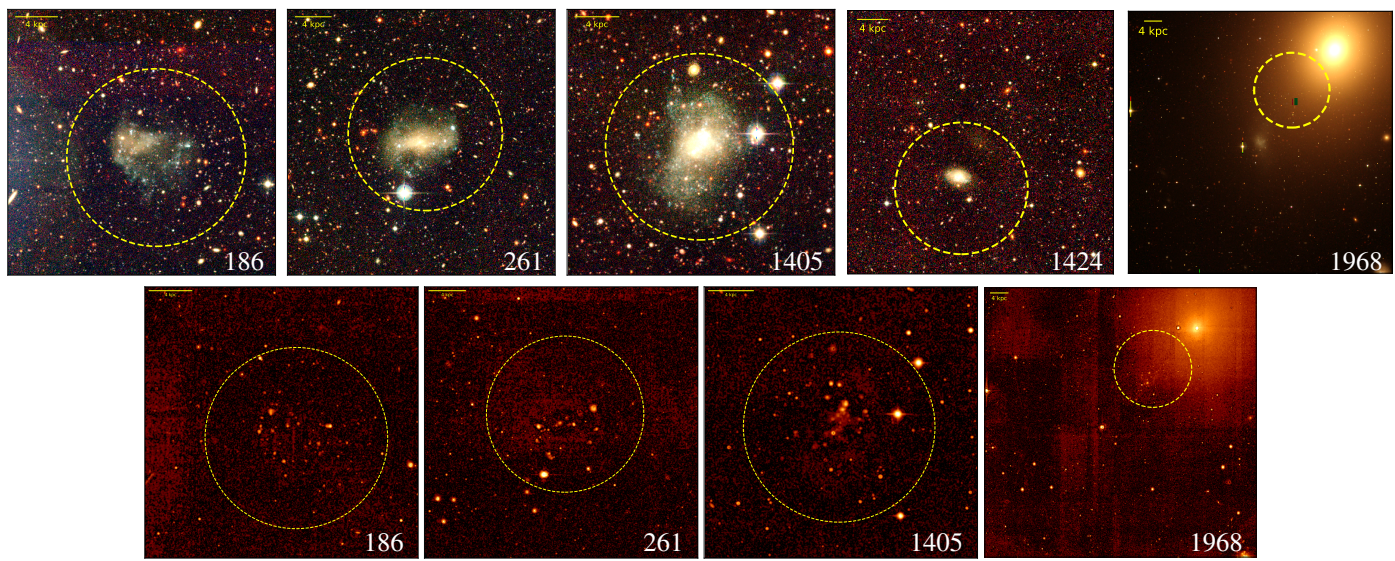
%[H]
    \centering
    %\ContinuedFloat
    \foreach \i in {1,...,5} {%
        \begin{subfigure}{0.2\textwidth}
            \includegraphics[width=\linewidth]{colour_images_alfalfa_detected/galaxy\IDhidetected[1,\i]_colour_alfalfa_detected.pdf}%
            \raisebox{4pt}{\makebox[-4pt][r]{\color{white} \IDhidetected[1,\i]
            }}
            %\caption{}
        \end{subfigure}\hspace{0em}}

    \foreach \i in {1,...,4} {%
        \begin{subfigure}{0.2\textwidth}
            \includegraphics[width=\linewidth]{images_halpha_extended/galaxy\IDhalphaextended[1,\i]_halpha_extended.pdf}%
            \raisebox{4pt}{\makebox[-4pt][r]{\color{white} \IDhalphaextended[1,\i]
            }}
            %\caption{}
        \end{subfigure}\hspace{0em}}
        
\caption{Diffuse galaxies with detection in ALFALFA \Hi{} (top row) and VESTIGE \Ha{} (bottom row). The color composite images are from the NGVS \textit{u}, \textit{g}, and \textit{i} bands. The dashed yellow circles show the ALFALFA beam (3.5\arcmin{} diameter). The \Ha{} images (bottom row) were smoothed with a Gaussian kernel of $\sigma=3$ pixels, and the image color contrast range is from an \Ha{} surface-brightness level of $10^{-18}$ to $10^{-14}$ erg s$^{-1}$ cm$^{-2}$ arcsec$^{-2}$.}
\label{fig:alfalfa_detected_source}
\end{figure*}

The Arecibo Legacy Fast ALFA (ALFALFA; \citealt{giovanelli2005,haynes2018}) survey is a blind extragalactic \Hi{} survey covering a wide area of the sky up to $\sim$7000 deg$^2$, including the Virgo cluster. At the Virgo distance, ALFALFA can detect galaxies with \Hi{} masses as low as $M_{\rm HI} \sim 10^{7}$ \msun{} \citep{giovanelli2005}. Therefore, to investigate the presence of \Hi{} gas in our LSBs, we cross-matched all the \Hi{} detections in the catalog of \citet{haynes2018} with an \Hi{} recessional velocity $cz_{\odot} < 3000$ \kms{} (Virgo members) with our catalog
and found that only about 8\% of our sample (5 galaxies) have an \Hi{} counterpart within the beam of Arecibo ($\sim$3.5\arcmin{} diameter). This means that the majority of the galaxies in our sample have a \Hi{} mass $<10^{7}$ \msun{}. Table \ref{table:alfalfa_crossmatch_subsample} and Fig. \ref{fig:alfalfa_detected_source} shows the 5 galaxies with ALFALFA \Hi{} detection.

%-------------------------------------------------------
%
\begin{table}[!ht]
\centering                          % used for centering table
\footnotesize
\caption{Subsample of LSBs at the distance of Virgo detected in \Hi{} by ALFALFA \citep{haynes2018}.}  
\begin{tabular}{c c c c c c}        % centered columns (4 columns)
\hline\hline                 % inserts double horizontal lines
ID & AGC ID & $cz_{\odot}$ & $W_{50}$  & $\log M_{HI}$ & Offset\\
& & (\kms{}) & (\kms{}) & (\msun{})& (arcmin)\\

(1) & (2) & (3) &  (4) & (5) & (6)\\
\hline                        % inserts single horizontal line
186 & 220258 & $2219$ & $24\pm3$ & $8.8\pm0.1$ & 0.4 \\
261 & 7307 & $1183$ & $52\pm3$ & $8.6\pm0.1$ & 0.2 \\
1405 & 7547 & $1100$ & $71\pm3$ & $9.1\pm0.1$ & 0.1 \\
1424 & 220597 & $1860$ & $48\pm2$ & $7.7\pm0.1$ & 1.3\tablefootmark{a} \\
1968 & 227874 & $473$ & $30\pm10$ & $7.9\pm0.2$ & 2.8\tablefootmark{b} \\
\hline                                   %inserts single line
\end{tabular}          % title of Table
\label{table:alfalfa_crossmatch_subsample}      % is used to refer this table in the text
\tablefoot{(1) Name of the source ; (2) Arecibo General Catalogue (AGC) ID; (3) Heliocentric velocity of the \Hi{} line profile midpoint; (4) Velocity width of the \Hi{} line profile at 50\% of the peak ($W_{50}$); (5) \Hi{} mass; (6) Offset of the ALFALFA beam centroid with respect to the NGVS coordinate of the source.
\tablefoottext{a}{The ALFALFA beam includes also the bright galaxy VCC 963 to which the \Hi{} detection is probably associated.}\\
\tablefoottext{b}{\Hi{} detection is between our diffuse galaxy and the elliptical galaxy M49 \citep{sancisi1987,patterson1992,henning1993,battaia2012}}
}
\end{table}
%
%-------------------------------------------------------

All of the \Hi{} detected galaxies belong to the diffuse galaxy subsample (none of the UDGs in our sample have an \Hi{} counterpart) and have \Hi{} masses in the range $7.7 < \log M_{HI} < 9.1$ \msun{}. 
For two galaxies among them (ID 1424 and 1968), we found the presence of a large nearby companion. The \Hi{} centroid close to the ID 1424 coincides with another Virgo galaxy VCC 963 (see Fig. \ref{fig:alfalfa_detected_source}), and therefore the \Hi{} is probably not associated with the diffuse galaxy. In the case of the ID 1968 (VCC 1249), the \Hi{} detection was found to be located midway between the diffuse galaxy and the large elliptical galaxy M49 (see Sect. \ref{sect:analysis_individual_galaxies}).

\subsection{Narrowband \Ha{} imaging}\label{sect:halpha_detection}

\Ha{} NB imaging data are critical for the following analysis since constraining the recent star formation activity on timescales of $\lesssim$10 Myr \citep{kennicutt1998,boselli2009_halpha_UV, boissier2013}. These timescales are much shorter than those inferred by any other star formation indicator and are thus crucial to accurately reconstruct the star formation history of our target galaxies. 

We cross-matched our sample with the \Ha{} VESTIGE catalog (Boselli et al. in prep.). Only four galaxies in our sample are confidently detected in \Ha{} (see Fig. \ref{fig:alfalfa_detected_source}) and all of them are also detected in \Hi{}. For a few other sources (ID 1352, 1529, 2343, and 3265) we see the presence of a tentative \Ha{} detection in their central region (see Fig. \ref{fig:all_profiles}). However, they are likely an artifact due to the continuum subtraction\footnote{VESTIGE continuum subtraction procedure uses optical images, and therefore any defaults in alignments or not perfectly identical PSF in images with strong gradients in the center might induce artifacts in the \Ha{} image (see \citealt{boselli2019_vestige5} for more details).} from the VESTIGE NB filter, since these galaxies have a strong central component in their optical images (see Fig. \ref{fig:selected_sample_colour_images}). Therefore, we consider all the galaxies in our sample, except for four, as \Ha{} non-detections. The lack of any \Ha{} emission for most of the galaxies indicates that they have either not undergone any recent star formation or have a very low activity 
($SFR < 2\times10^{-5}$ \msun{} yr$^{-1}$, the detection limit of VESTIGE).
This result is consistent with the dominant red colors, suggesting they did not undergo any star formation event in the last few hundred megayears. For all of these galaxies, the \Ha{} non-detection still provides confident upper limits ($3\sigma$) useful for our analysis.

\section{Modeling the evolution of LSBs in clusters}\label{sect:modelling_rps}

\subsection{Models without environmental effects}
\label{sect:Models_without_environmental_effect}

We compared the photometric properties of the 64 galaxies of our sample to the multi-zone chemo-spectrophotometric models of galaxy evolution by \citet{boissier2000}.
In the next section we consider the same models, but
modified to include the effects of an RPS event as previously done in \citealt{boselli2006,boselli2008a,Boselli2008b,boselli2014} and \citet{junais2021}. The main assumptions of these models are briefly reminded here.

Each model consists of independently evolving rings, accreting primordial infalling gas. These models are based on the model of the Milky Way \citep{boissier1999}, extended to nearby spiral galaxies \citep{boissier2000,munosmateos2011}, by considering galaxies with various circular velocities (\Vc{}) and spin parameters (\spin{}), and using scaling relationships such that the mass of the galaxies varies proportional to $V_{C}^{3}$, while the spin parameter controls the distribution of this mass within the disk. 
\citet{boissier2000} found that the B-band central surface brightness is tightly connected to the spin parameter. Models with large spins (larger than typically 0.1) were found to be well tuned to study the star formation history of LSBs \citep{boissier2003,boissier2016}, while usual HSBs have spins around the peak value of 0.05, expected in simple galaxy halo formation models \citep{mo1998}). The spin distribution in these models is log-normal with dispersion 0.05, leading to a significant fraction of large spin galaxies \citep[estimated to 25 \% in ][]{boissier2003dla}, although some phenomena like late gas accretion could also lead to large angular momentum  \citep{stewart2017}.   

The models follow only the evolution of the baryon content: the dark matter halo of the galaxy is assumed to scale with the baryons, and only affects the rotation curve adopted in the model.
We used the same basic assumptions as in \citet{munosmateos2011}: a universal \citet{kroupa2001} IMF and a universal star formation law, the star formation rate (SFR) being dependent on the gas density and the angular frequency. The rate of gas accretion as a function of surface density and circular velocity of the galaxy is variable, having been tuned to fit the properties of nearby galaxies.

\begin{figure}
    \centering
    \includegraphics[width=\hsize]{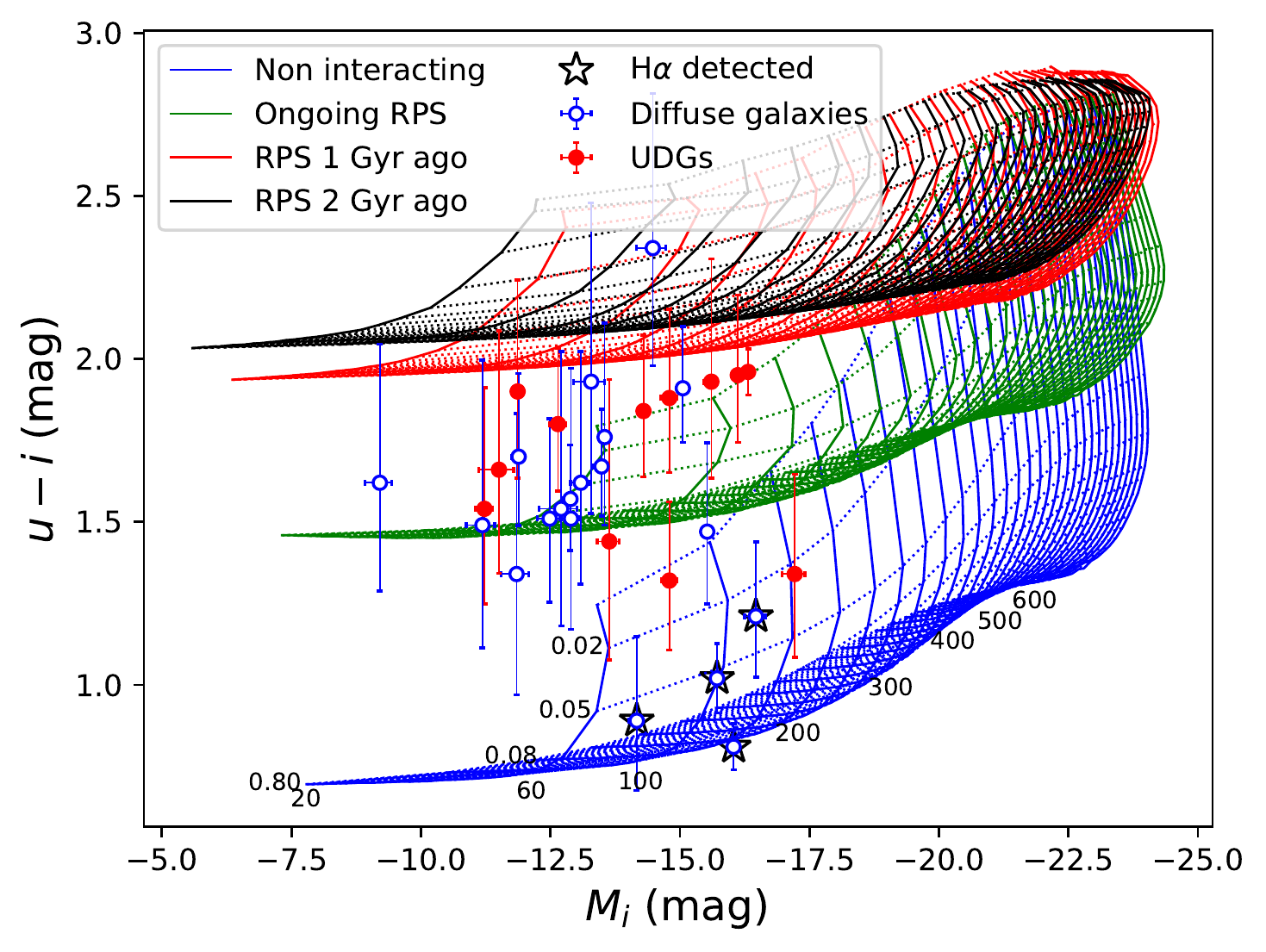}
    \caption{$u-i$ color versus the \textit{i}-band absolute magnitude of the sample in comparison with the grid of colors from the models discussed in Sections \ref{sect:Models_without_environmental_effect} and \ref{sect:Models_with_RPS}. The UDGs and the diffuse galaxy subsamples are marked with the filled red and open blue circles, respectively. The black star symbols are the sources detected in \Ha{}. Sources with only an upper or lower limit in the color are excluded from the plot. The blue grid corresponds to galaxies without any environmental interaction (non-RPS models; see Sect. \ref{sect:Models_without_environmental_effect}). The green, red, and black grids correspond to models with an RPS event at different epochs (ongoing, 1 Gyr ago, and 2 Gyr ago, respectively). The black labels marked along the blue are the different \spin{} (0.01 to 0.80) and \Vc{} (20 \kms{} to 600 \kms{}) values of the models, as given in Table. \ref{table:grid}. The dotted and solid lines in the grids give the variation for a fixed \spin{} and \Vc{}, respectively.}
    \label{fig:grid_color_magnitude_diagram}
\end{figure}

In Fig. \ref{fig:grid_color_magnitude_diagram} we compare the color-magnitude diagram for these models with the integrated colors of our sample. The full grid of models without any environmental interactions can reproduce the colors only for a few galaxies, while most of our sample is much redder than these models at the same magnitude. This already shows that these models not including any environmental effects are not realistic for most of our sample. In Sect. \ref{sect:Models_with_RPS} we introduce one of these effects.

\subsection{Models including the effect of ram-pressure stripping}
\label{sect:Models_with_RPS}

\citet{boselli2006} modified the reference models discussed in Sect. \ref{sect:Models_without_environmental_effect} to implement the effect of RPS due to the dense intra-cluster medium (ICM). With this addition, \citet{boselli2006,boselli2008a,Boselli2008b} successfully reproduced the properties of anemic and dwarf galaxies in the Virgo Cluster, while \citet{cortese2011} and \citet{boselli2014} those of larger samples including also massive objects in different environments.
The RPS scenario was inspired by the dynamical models of \citet{vollmer2001} following the ram-pressure exerted by the ICM during the crossing of the cluster in an elliptical orbit.
Our models assume that the gas in the galaxy is removed at a rate of $\epsilon \Sigma_{gas}/ \Sigma_{potential}$, which is proportional to the galaxy gas column density but is modulated by the gravitational potential well of the galaxy, measured by the total (baryonic) local density. To mimic the results of \citet{vollmer2001}, the efficiency of gas removal ($\epsilon$) follows a Gaussian with an FWHM of 150 Myr, and a maximum value $\epsilon_0$,
at the peak time, \trps{} (\trps{} is the time at which the RPS peaks, and this occurs at the pericenter of its orbit).
Consistent with other works, we adopted the efficiency $\epsilon_0 = 1.2\,M_{\odot}$ kpc$^{-2}$ yr$^{-1}$ derived by \citet{boselli2006} in the study of NGC4569. Under these assumption, this RPS model has only three free parameters, \Vc{}, \spin{} and \trps{}.

The fact that $\epsilon_0$ and the Gaussian FWHM are fixed is an over-simplification of the problem since these parameters should depend on the orbit of the galaxy within the cluster. However, it was chosen to explore a large grid of models for the other parameters, within reasonable computational time, and we do not have enough constraints for each galaxy to determine its precise trajectory in the cluster.
Moreover, \citet{junais2021} investigated the possible effect of a variation of $\epsilon_0$ and of the FWHM and found that it changed very little the results on the other parameters  (\trps{}, \Vc{} and \spin{} were modified by less than 0.1 Gyr, 2 \kms{} and 0.01, respectively for a typical UDG galaxy fitting).

In Fig. \ref{fig:grid_color_magnitude_diagram}, the effect of introducing an RPS event of various age on the color-magnitude diagram is shown for the \textit{u-i} color. We obtain much redder colors for the same magnitude than in the non-RPS case, sweeping over the observed colors of our sample when modifying the age of the event. From this crude comparison using just a single color, we can have a clue that our galaxies need a relatively old ram-pressure event to be fitted with our models, which will be tested in detail in Sect. \ref{sect:model_fitting_with_sample_profiles}.

\begin{table*}[h]
\centering
\caption{Grids of $V_C$, $\lambda$, and $t_{rps}$ values used for the modeling in this work.}
\scalebox{1}{
\begin{tabular}{c|cc|cc|cc} \hline \hline 
Grid type & \multicolumn{2}{c|}{\underline{\hspace{0.9cm}$V_C$ (\kms{})\hspace{0.9cm}}} & \multicolumn{2}{c|}{\underline{\hspace{1.5cm}$\lambda$\hspace{1.5cm}}} & \multicolumn{2}{c}{\underline{\hspace{1.5cm}$t_{rps}$ (Gyr)\hspace{1.5cm}}}\\

& Range & $\Delta V_C$ & Range & $\Delta\lambda$ & Range & $\Delta t_{rps}$\\

% (1) & (2) & (3) & (4) & (5) & (6) & (7) \\ 
\hline
Coarse & $20-600$ & 20 & $0.01-0.80$ & 0.1 & $8.0-13.5$ & 0.5 \\
% \hline
Fine & $20-220$ & 2 & $0.01-0.40$ & 0.01 & $8.0-13.6$ & 0.1 \\
% \hline
Hyper-fine & $20-220$ & 2 & $0.01-0.40$ & 0.01 & $13.40-13.60$ & 0.01 \\
\hline
\end{tabular}}
\tablefoot{For the three different types of grids, the range and corresponding spacing of each parameter are provided as separate columns. Apart from the models with RPS, all the grids also include models without RPS.}
\label{table:grid}
\end{table*}

\subsection{Grid of parameters used in this work}\label{sect:grid_of_parameters}

In \citet{junais2021}, we used two grids of models, a coarse grid and a fine grid. The coarse grid covers a very large range of spin and velocity to include both LSB, extreme LSBs such as Malin 1 \citep{boissier2016,junais2020}, and regular galaxies from dwarf to massive spirals, but with a 20 \kms{} resolution in velocity. Since we found that most galaxies in the current work fall within a smaller range of spin and velocities, we then constructed a "fine" grid with velocity steps of 2 \kms{}.
In these two grids, the \trps{} values  were chosen such that the models include the peaks of RPS events peaking at various epochs from very distant past (\trps{} = 8 Gyr) to the future onset (\trps{} = 13.6 Gyr, with the current time being assumed to be 13.5 Gyr), in steps of 0.1 Gyr. For the current work, we also computed a "hyper-fine" grid for the most recent or ongoing RPS events, with steps of 0.01 Gyr for the peak epoch. This is mostly needed for the four galaxies with \Ha{} detection (see Fig. \ref{fig:alfalfa_detected_source}) since the typical timescale for the emission of ionizing radiation is about 10 Myr. The range and steps of each parameter in the three grids are given in Table \ref{table:grid}. A total of \numprint{310550} different models were created from these grids, including models with and without RPS.

\subsection{Fitting of models}\label{sect:model_fitting_with_sample_profiles}

As in \citet{junais2021}, we made a \chisq{} fitting of the observed surface-brightness profiles with the surface-brightness profiles in the eight photometric bands from the models. Since the models were developed for disk profiles, we performed the fitting procedure only within the radial range where the disk is dominant and until the last observed radius ($>3\sigma$). This also excludes the \Ha{} central detections in a few sources that are artifacts, as discussed in Sect. \ref{sect:halpha_detection}. For the \Ha{} surface-brightness profiles, we also applied an additional correction for the [NII] line flux contamination using the standard [NII]/\Ha{}-stellar mass relation from \citet{boselli2009_halpha_UV}. We did not apply any dust attenuation correction in any of the eight bands because these LSB dwarf galaxies are known to be mainly dust free \citep{rahman2007,hinz2007}.
% (JUNAIS IS THAT TRUE? PUT A REFERENCE HERE).

We adopted a minimum error of 0.05 mag in the surface-brightness profiles to take into account systematic uncertainties associated with the models (e.g., IMF, stellar tracks, stellar libraries). Any model violating the $3\sigma$ upper limits of the photometry was rejected (this is particularly useful in the case of the \Ha{} data, which are mostly upper limits), but also allowing a tolerance of 0.1 mag above this level to avoid rejecting a good model that only marginally violates one upper limit. Modifying this tolerance within a range of a few tenths of dex changes the best-fit parameters within their error bars. The $3\sigma$ uncertainty associated with the best-fit model parameters are computed from this distribution following \citet{avni1976}. 

All the profiles and their fit are provided in the appendix (Fig. \ref{fig:all_profiles}). As can be seen in this figure, most of the profiles are well fitted.
The majority of the sample ($\sim75\%$) have a reduced $\chi^2_{\nu}<3$ with a median $\chi^2_{\nu}$ of 1.04, indicating a good fit (see Table \ref{table:model_params_full_sample} in page \pageref{page:models_table}). However, for a few sources with profiles close to the sky level with very few data points, the fit is poor. We verified that all the results presented in this work remain unchanged even if we exclude the poor fits.

We also made a comparison of our best-fit RPS models with that of the best-fit models without any RPS. The $\chi^2_{\nu}$ values of the non-RPS best fits are always larger than the best fits from the RPS models, except for one galaxy. For 92\% of the sample, the $\chi^2_{\nu}$ values of the non-RPS best fits are larger than the RPS best fits models by a factor of at least 3. 
However, in a few cases, such as ID 186 and ID 261, the \Ha{} and the FUV appear more consistent with a non-RPS model, whereas the NUV and the \textit{u} band are consistent with an RPS model (see Fig. \ref{fig:all_profiles}). Such an inconsistency where the models do not fit very well simultaneously the \Ha{}, FUV, NUV, and the \textit{u} band may indicate that the short timescale ($<10$ Myr) star formation history may be more complex in these two cases, which is hard to model. Since we perform a simultaneous multiwavelength fitting based on the least total \chisq{}, in the case of these two galaxies, a model with RPS is favored. Therefore, in general, we can say that the non-RPS models fail to reproduce all the observed properties of the galaxies in our sample, similar to what we saw from Fig. \ref{fig:grid_color_magnitude_diagram}. Table \ref{table:model_params_full_sample} gives the results of our model fitting.

\subsection{Distribution of \Vc{}, \spin,{} and \trps{}}
\label{sec_models_distribparam}

\begin{figure*}
    \centering
    \includegraphics[width=\hsize]{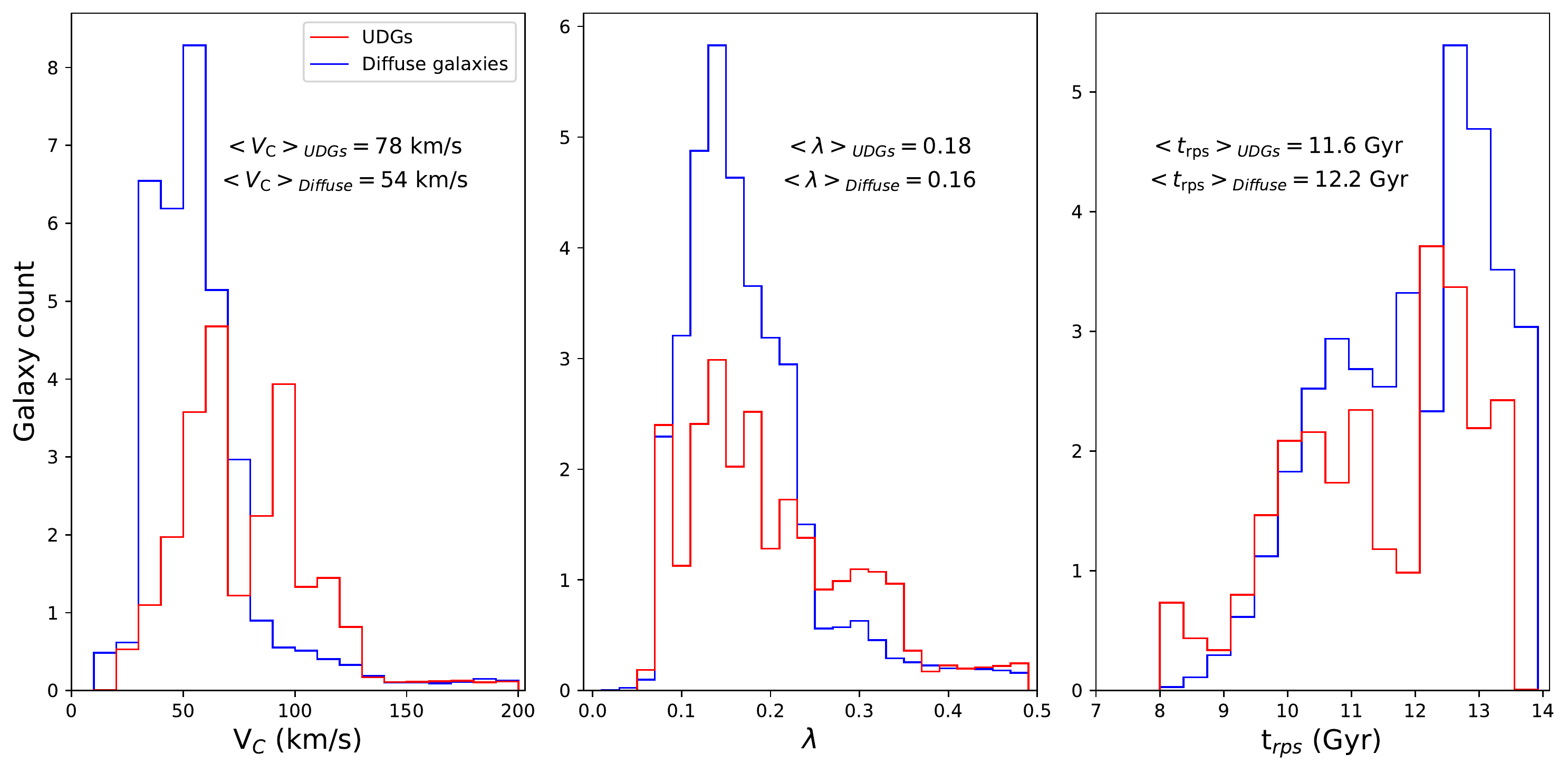}
    \caption{Distribution of the \Vc{}, \spin{}, and \trps{} parameters from the best-fit models. The red and blue lines correspond to the UDG and diffuse galaxy subsamples, respectively. The median values corresponding to each parameter are shown within each panel.}
    \label{fig:histogram_trps_vel_spin}
\end{figure*}

Figure \ref{fig:histogram_trps_vel_spin} shows the distribution of the model parameters \Vc{}, \spin{} and \trps{} that we obtained. The uncertainties associated with these parameters (shown in Table \ref{table:model_params_full_sample}) were also taken into account for this distribution. This was done using a Monte Carlo simulation of 1000 chains where we created pseudo sources corresponding to each source, with random parameter values generated based 
on its value and uncertainty.

The \Vc{} distribution from Fig. \ref{fig:histogram_trps_vel_spin} peaks along the range of low mass galaxies\footnote{Based on our models, an unperturbed galaxy with a \Vc{} value of 60 \kms{} approximately corresponds to a total baryonic mass of about $10^9$ \msun{} and a stellar mass of $10^8$ \msun{}.} (see Fig. \ref{fig:histogram_mgas_mstar_sfr}) with a median velocity value of $V_{C} = 78$ \kms{} and $V_{C} = 54$ \kms{}, for the UDGs and the diffuse galaxies, respectively. 
However, the UDGs present double peaked distributions, a few galaxies being relatively fast rotators 
\citep[around 100 \kms{}, the order of magnitude of the Large Magellanic Cloud,][]{olsen2011,vandermarel2016} while the rest of the UDGs have similar velocities as the diffuse galaxies. 

Regarding the spin distribution, we can see that the majority of the sources have large extended spins with a median \spin{} of 0.18 and 0.16 for the UDGs and diffuse galaxies, respectively. This is consistent with the typical spins observed in LSBs \citep{boissier2003, amorisco_loeb2016}, compared to a typical spin of 0.05 for a regular HSB. The very large spin tail ($\lambda > 0.4$) seen in this distribution results from the sources with poor fits and large uncertainties and should thus be considered with caution. 
The UDGs have a slightly higher spin than diffuse galaxies, which could contribute to their more extreme nature.

Almost all of the sources in the sample have undergone an RPS event in their lifetime (except for one source - ID 3365 from Table \ref{table:model_params_full_sample}, having the least reduced \chisq{} for a model without RPS, and only 2 free parameters, velocity and spin). The \trps{} distribution of the sample peaks at a median value of 11.6 Gyr\footnote{A \trps{} of 11.6 Gyr corresponds to 1.9 Gyr in the past (since our models assume the current age of the galaxy as 13.5 Gyr)} for the UDGs and 12.2 Gyr for the diffuse galaxies, with a large dispersion of several Gyr. This means that on average the galaxies in our sample have experienced a peak RPS $\sim$1.6 Gyr ago, with some of them having ongoing RPS too. However, we should note that the uncertainty on \trps{} depends a lot on the RPS event age. It is very small (on the order of 0.1 Gyr) for very recent events. For events peaking at 12 and 11 Gyr, the uncertainty can reach up to 1 and 2 Gyr, respectively.

\subsection{\trps{} gradient}
\label{sect:trps_gradient}

\begin{figure*}[h]
    \centering
    \includegraphics[width=0.49\hsize]{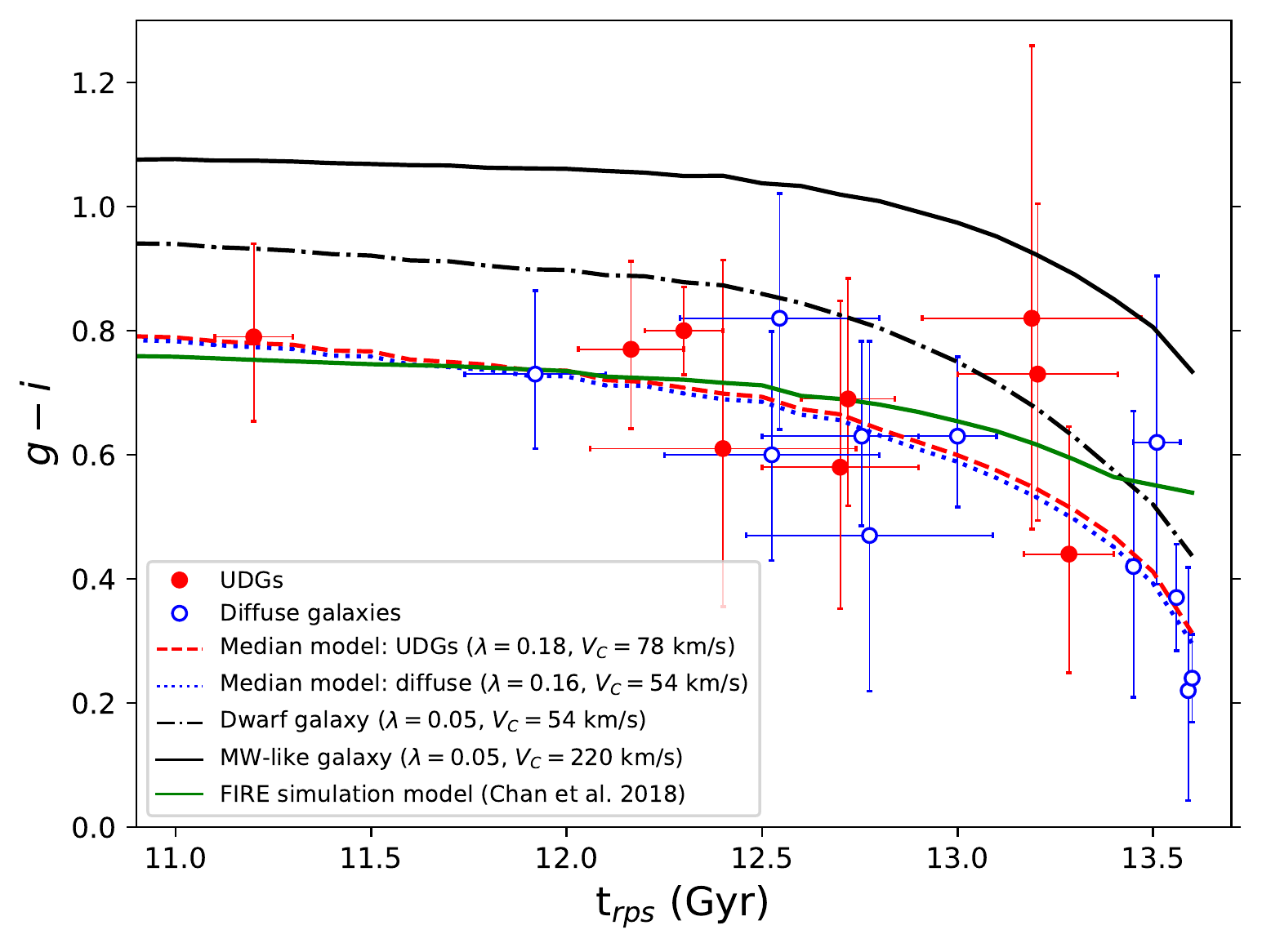}
    \includegraphics[width=0.49\hsize]{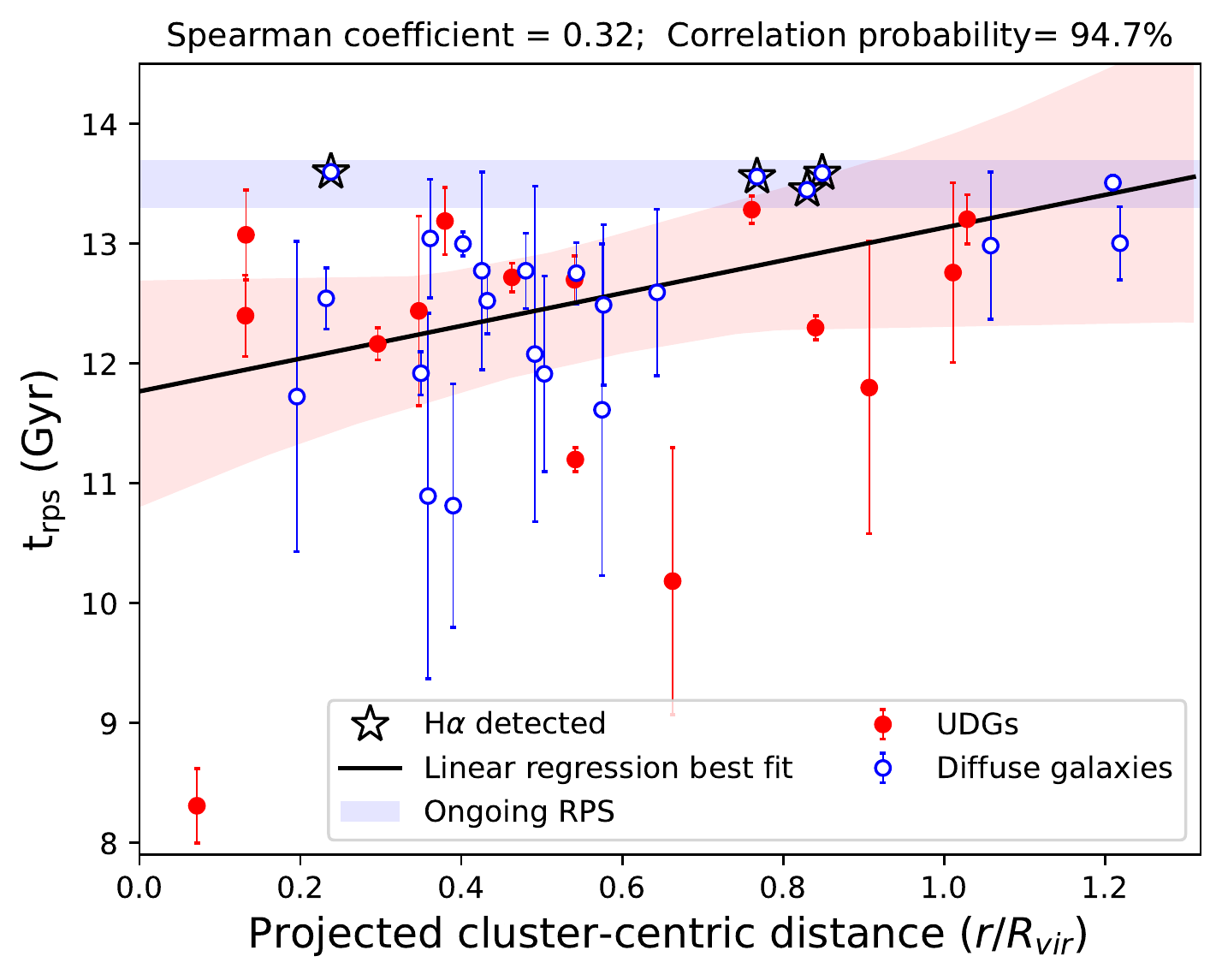}
    \caption{Comparision of the \trps{} values with the observed color and the cluster-centric distance. \textit{Left}: Observed $g-i$ color as a function of the best-fit \trps{} value from the models. The dashed red and dotted blue curves show the model integrated $g-i$ color for the median values of the UDGs and diffuse galaxies, respectively, as discussed in Sect. \ref{sec_models_distribparam}. The solid and dot-dashed black curves indicate the models for a representative dwarf and Milky Way-like galaxy, respectively \citep{boselli2014}. The solid green line is the quenching time model for a red UDG from \citet{chan2018} based on the FIRE simulation (see their Fig. 4, model \textit{m11b}). Sources with only upper limits in the color or with \trps{} uncertainty $>0.5$ Gyr are excluded from this plot. This plot also does not include one galaxy (ID 2365), which has \trps{} $\approx 8$ Gyr, a large error bar in the color, and a very crowded field upon visual inspection. \textit{Right}:  \trps{} values with respect to the projected distance from the cluster center, in units of the cluster virial radius ($R_{vir}=1.55$ Mpc; \citealt{ferrarese2012}). The blue-shaded area marks the region of ongoing RPS (\trps{} between 13.3 and 13.7 Gyr). The solid black line and the red-shaded region give the linear regression best fit and the $3\sigma$ scatter, respectively, as given in Table \ref{table:sample_gradientfits}. Sources with \trps{} uncertainty $>2.5$ Gyr are excluded from this plot.}
    \label{fig:trps_vs_g-i_distance}
\end{figure*}

The aging of the stellar population after a quenching episode results in the reddening of the colors, as indeed observed in Fig. \ref{fig:trps_vs_g-i_distance} (left panel).
In this figure, we also included the models corresponding to the median \Vc{} and \spin{} values of our subsamples, showing such a trend. Comparing them with a representative dwarf galaxy (small \spin{} and \Vc{}) and a Milky Way-like galaxy (small \spin{} and large \Vc{}) from \citet{boselli2014}, it is clear that the observed colors of our LSBs can be traced only with a model with an extended spin and low velocity. This trend is actually similar to the one found by \citet{chan2018} using the Feedback In Realistic Environments (FIRE) simulations for their galaxies with the largest spin ($\lambda \simeq 0.08$), with values comparable to those observed in our red UDGs. However, in our models we obtain a stronger variation of the $g-i$ color for the younger RPS events.
Although our median models go through the observed points, a single color cannot be used to pinpoint precisely the RPS time in individual galaxies, consistently with \citet{chan2018}. We also found that the $NUV-r$ color correlates better with \trps, but it is available only for a minority of galaxies, as NUV is not detected for many of them.

In Fig. \ref{fig:trps_vs_g-i_distance} (right panel), as expected, the \trps{} increases with decreasing cluster-centric distance, suggesting that LSBs located within the innermost regions have experienced an RPS event well before those now entering the cluster and located at its periphery. A linear fit and its corresponding correlation coefficients are given in Table \ref{table:sample_gradientfits}.

\begin{table*}[h]
\centering
    % \fontsize{7pt}{7pt}
    \caption{Correlation properties of the $u-i$ color and the \trps{} with the cluster-centric distance (in units of Mpc).}
    \begin{tabular}{|c|cc|cc|}
    \hline
    & \multicolumn{2}{c|}{\underline{\hspace{2cm} Linear fit \hspace{2cm}}} & \multicolumn{2}{c|}{\underline{\hspace{0.7cm} Spearman correlation \hspace{0.7cm}}} \\ 
    Relation & Slope & Intercept & $\rho$ &   Probability \\
    % & (mag Mpc$^{-1}$) & (mag) & & \\
    (1) & (2) & (3) & (4) & (5)\\
    \hline
    &&&&\\
    $u-i$ vs D$_{\rm M87}$ & $-0.21\pm0.18$ mag Mpc$^{-1}$ & $1.7
    \pm0.18$ mag & -0.41 & 97.7\%\\
    &&&&\\
    \trps{} vs D$_{\rm M87}$ & $0.87\pm0.37$ Gyr Mpc$^{-1}$ & $11.8\pm0.4$ Gyr & 0.32 & 94.7\%\\
    \hline
    \hline    
    \end{tabular}\label{table:sample_gradientfits}
    \tablefoot{(1) Correlating quantities; (2) Slope of the linear regression fit using {\tt linmix}; (3) Intercept of the linear regression fit; (4) Spearman correlation coefficient for the sample; (5) Probability that the two variables are correlated.}
\end{table*}

\section{Discussion}

\subsection{An evolutionary scenario for the formation of quiescent LSBs in clusters}

\begin{figure*}
    \centering
    \includegraphics[width=\hsize]{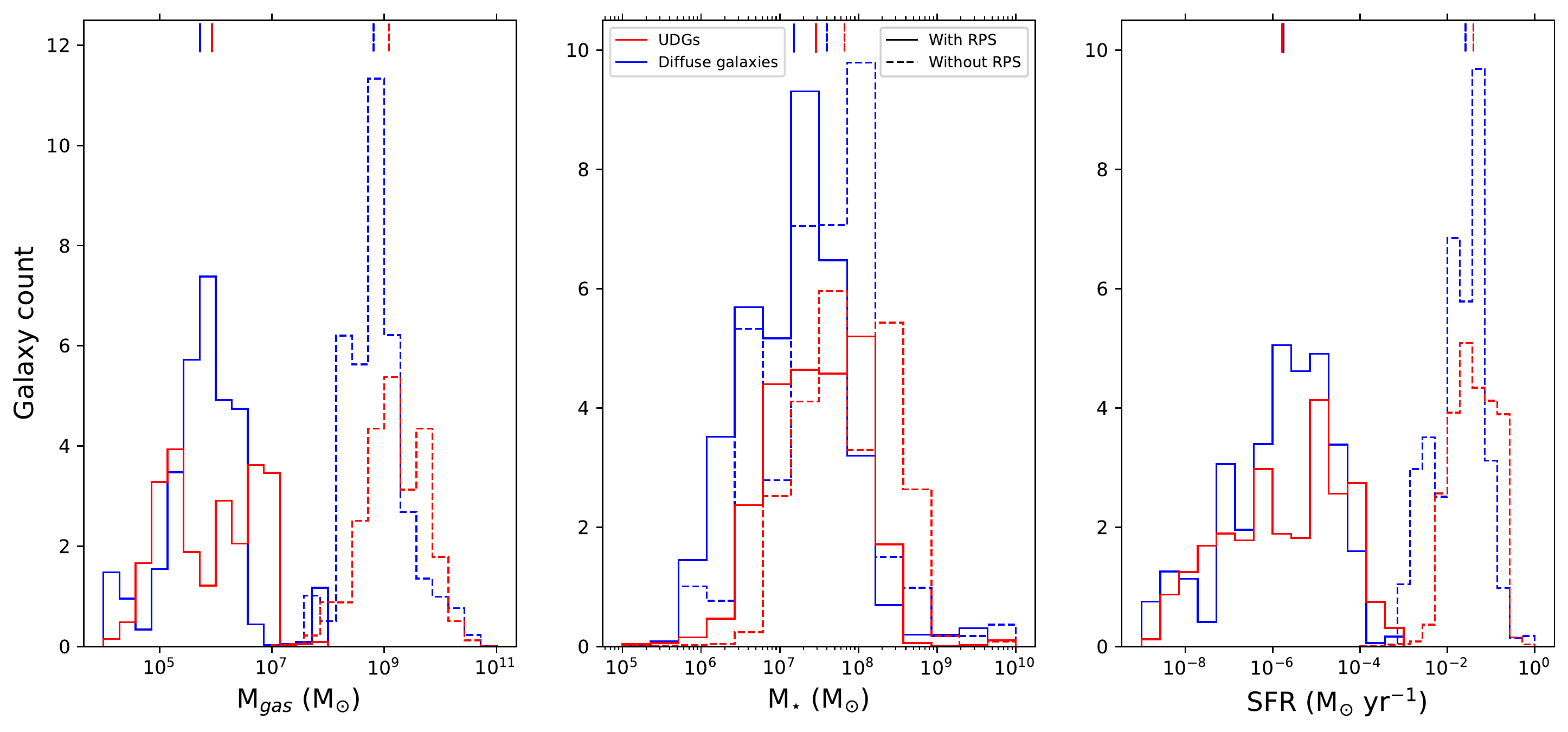}
    \caption{Distribution of the gas mass, stellar mass, and SFR of the sample obtained from the models, as given in Table \ref{table:model_params_full_sample}. The dashed and solid lines show the values before and after the RPS event, respectively.}
    \label{fig:histogram_mgas_mstar_sfr}
\end{figure*}

We have shown that models not including any effect of the environment poorly reproduce the properties of UDGs and diffuse galaxies. Under the assumption that an RPS event took place 
in the rich environment of Virgo, our multi-zone models are in good agreement with the observed profiles in \Ha{}, UV, and optical.
This naturally explains the trends observed in this paper if the LSB population inhabiting Virgo is dominated by objects that fell into the cluster as LSB gas-rich systems that lost their gas content after an RPS event (see Fig. \ref{fig:histogram_mgas_mstar_sfr}). Due to their shallow gravitational potential well, the stripping process has been rapid, transforming rotating gas rich systems into quiescent, smooth objects on very short timescales. This picture is consistent with the evolution of other dwarf systems in Virgo \citep[e.g.,][]{boselli2008a,Boselli2008b}, but also with that of more massive objects where the stripping process is however longer \citep[e.g.,][]{boselli2006,boselli2014,boselli2016,fossati2018}. As indicated in Fig. \ref{fig:histogram_mgas_mstar_sfr}, the RPS removed around three orders of magnitude of the gas (from gas mass of nearly $10^{9}$ \msun{} to $10^{6}$ \msun{}). This lack of gas drastically reduced the activity of star formation (the SFR decreased by around four orders of magnitude from $10^{-1}$ \msun{} yr$^{-1}$ to $10^{-5}$ \msun{} yr$^{-1}$), explaining the red colors of the UDGs (and diffuse galaxies) observed in Virgo. As extensively discussed in \citet{boselli_and_fossati2021}, the RPS process starts to be efficient at $\sim$ one virial radius, in particular in dwarf systems such as those analyzed in this work. The time necessary for galaxies to become red ($\lesssim 1$ Gyr), is short compared to the typical crossing time of the cluster ($\sim$ 1.7 Gyr, \citealt{boselli_and_gavazzi2006}), thus
explaining the predominance of red objects within the cluster (see \citealt{boselli2014}) and the outskirts preference for blue or HI bearing ones.

Although this agreement between RPS models and observations is striking, it is known that some UDGs are almost certainly formed through other processes, such as tidal forces and mergers, including a few objects in our sample (see, e.g., Section 4.3 of \citealt{lim2020}).

\subsection{Evidence from the analysis of individual objects}\label{sect:analysis_individual_galaxies}

This evolutionary picture is also consistent with the morphological properties of several representative objects. A clear example is the galaxy ID 1968 (VCC 1249; the most massive galaxy in our sample with $\log$\mstar{}$=8.6$) located in projection in the outer halo of the massive elliptical M49, the dominant galaxy of the Virgo cluster B substructure. Multifrequency observations consistently indicate that this LSB object has interacted with the massive elliptical \citep{sancisi1987,patterson1992,henning1993,battaia2012}. Although the interaction was probably dominated by a tidal perturbation, the dwarf galaxy lost its gas while crossing the hot X-ray emitting gas trapped within the halo of M49, suffering thus also an RPS event. \citet{battaia2012} estimated that the dwarf galaxy VCC 1249 abruptly reduced its star formation activity $\sim$200 Myr ago. This is consistent with our models, which suggest that VCC 1249 started to reduce its activity $\sim$240 Myr ago with a peak of quenching about 50 Myr ago (see Table \ref{table:model_params_full_sample}), with stripping still ongoing. The \Hi{} observations also, indeed, show that the \Hi{} gas is located in between the two objects, and forming stars as indicated by the presence of several compact \ion{H}{ii} regions \citep{battaia2012}. 

Other interesting objects are the galaxies ID 186, 261 and 1405 (see Fig. \ref{fig:alfalfa_detected_source}).
For ID 186, the color image of this galaxy combined with the distribution of \ion{H}{ii} regions revealed by the deep VESTIGE NB \Ha{} image, clearly indicates a perturbed morphology, with a banana-shaped structure in the eastern direction and a low-surface-brightness tail of very blue knots in the opposite direction. The relative position of this high density structure and of the low-surface-brightness tail, typical of galaxies suffering an RPS event \citep{boselli2021,boselli_and_fossati2021}, suggests that the galaxy is suffering RPS while moving toward the cluster center, located at 1.32 Mpc to the east (projected distance). The presence of atomic gas and of ionized gas in star forming regions indicate that the galaxy is at an early phase of its transformation already occurring at the periphery of the cluster. A similar perturbed morphology is also visible in the case of ID 261 and 1405, which are at a projected distance of 1.19 Mpc and 0.37 Mpc from the cluster center, respectively. These galaxies with a clear \Hi{} and \Ha{} detection without any obvious nearby companions could also be similar to the \Hi{} bearing diffuse galaxies that have been observed in several recent works \citep{leisman2017, prole2019, janowiecki2019}.

\section{Conclusions}\label{conclusion}

We extracted a sample of 64 LSBs from the NGVS catalog of the Virgo cluster \citep{ferrarese2020}. This sample of LSBs was selected following the procedure adopted in \citet{lim2020}, that is, selecting galaxies away from the cluster scaling relationships. The sample includes 26 UDGs already identified in \citet{lim2020} and 38 additional diffuse galaxies. We compared profiles obtained from the NGVS survey (in the optical), GUViCS (in the UV), and VESTIGE (in the \textit{r} band and \Ha{} NB) to multiwavelength galaxy evolution models.

Our main results are summarized as follows.

\begin{itemize}

    \item The spatial distribution of the sample within the cluster shows that UDGs are more concentrated in the cluster center compared to the diffuse galaxies, which are located more in the outskirts. 

     \item The optical colors of the sample indicates a predominantly red population, consistent with what is generally found in clusters. However, there is an indication of a color variation with the cluster-centric distance, where LSBs toward the edge of the cluster are bluer than the rest of the population.

    \item About 8\% of our sample (five galaxies) have \Hi{} counterparts in the ALFALFA survey (four of them with \Ha{} detection as well). Almost all of these sources are located toward the edge of the cluster. 
    
    \item The comparison with models successfully reproduces multiband color profiles and  suggests that the LSBs in our sample are predominantly dwarf galaxies (low velocities) that are extended (large spins) and experienced an RPS event on average 1.6 Gyr ago. A few sources are undergoing RPS events now as well.
    
    \item The RPS time also shows a variation with the cluster-centric distance, where galaxies closer to the center have older RPS events, while those in the cluster outskirts underwent RPS during much more recent epochs.
    
\end{itemize}

Previous studies have concluded that no single mechanism is responsible for the entire UDG class, with multiple processes likely having played a role. This work demonstrates the potential role played by ram pressure in producing red quiescent UDGs and other diffuse galaxies from progenitors that were gas-rich and blue (but already diffuse).
Our observations and empirical models can be tested in the future by considering data from future surveys that may reveal new LSBs in large optical surveys (e.g., LSST), but also with gas surveys (e.g., with the Square Kilometer Arrays), which could put limits on the gas mass in LSBs within clusters and outside.

\begin{acknowledgements}
We are grateful to the whole CFHT team who assisted us in the preparation and in the execution of the observations and in the calibration and data reduction: 
Todd Burdullis, Daniel Devost, Bill Mahoney, Nadine Manset, Andreea Petric, Simon Prunet, Kanoa Withington. We acknowledge financial support from "Programme National de Cosmologie and Galaxies" (PNCG) funded by CNRS/INSU-IN2P3-INP, CEA and CNES, France,
and from "Projet International de Coop\'eration Scientifique" (PICS) with Canada funded by the CNRS, France. Junais and Katarzyna Ma\l{}ek are also grateful for the support from the Polish National Science Centre via grant UMO-2018/30/E/ST9/00082. Alessia Longobardi is supported by Fondazione Cariplo, grant No 2018-2329. Studies on UDGs are gratefully supported within a French-Austrian cooperation project under grants no. 47549ZL (from the AMADEUS 2022 program by Campus France) and FR 07/2022 (by the Austrian Academic Exchange Service OeAD). Jin Koda acknowledges support from NSF through grants AST-1812847 and AST-2006600. Médéric Boquien gratefully acknowledges support by the ANID BASAL project FB210003 and from the FONDECYT regular grants 1170618 and 1211000.
\end{acknowledgements}

\bibliographystyle{aa}
\bibliography{bibliography}

\clearpage
\onecolumn

\begin{appendix}

\section{All the profiles and the model fits for the UDG and the diffuse galaxy samples}\label{appendix:profiles_and_bestfits}

\begin{figure*}[h]
    \centering
    %\ContinuedFloat
    \foreach \i in {1,...,2} {%
        \begin{subfigure}{0.46\textwidth}
            \includegraphics[width=\linewidth]{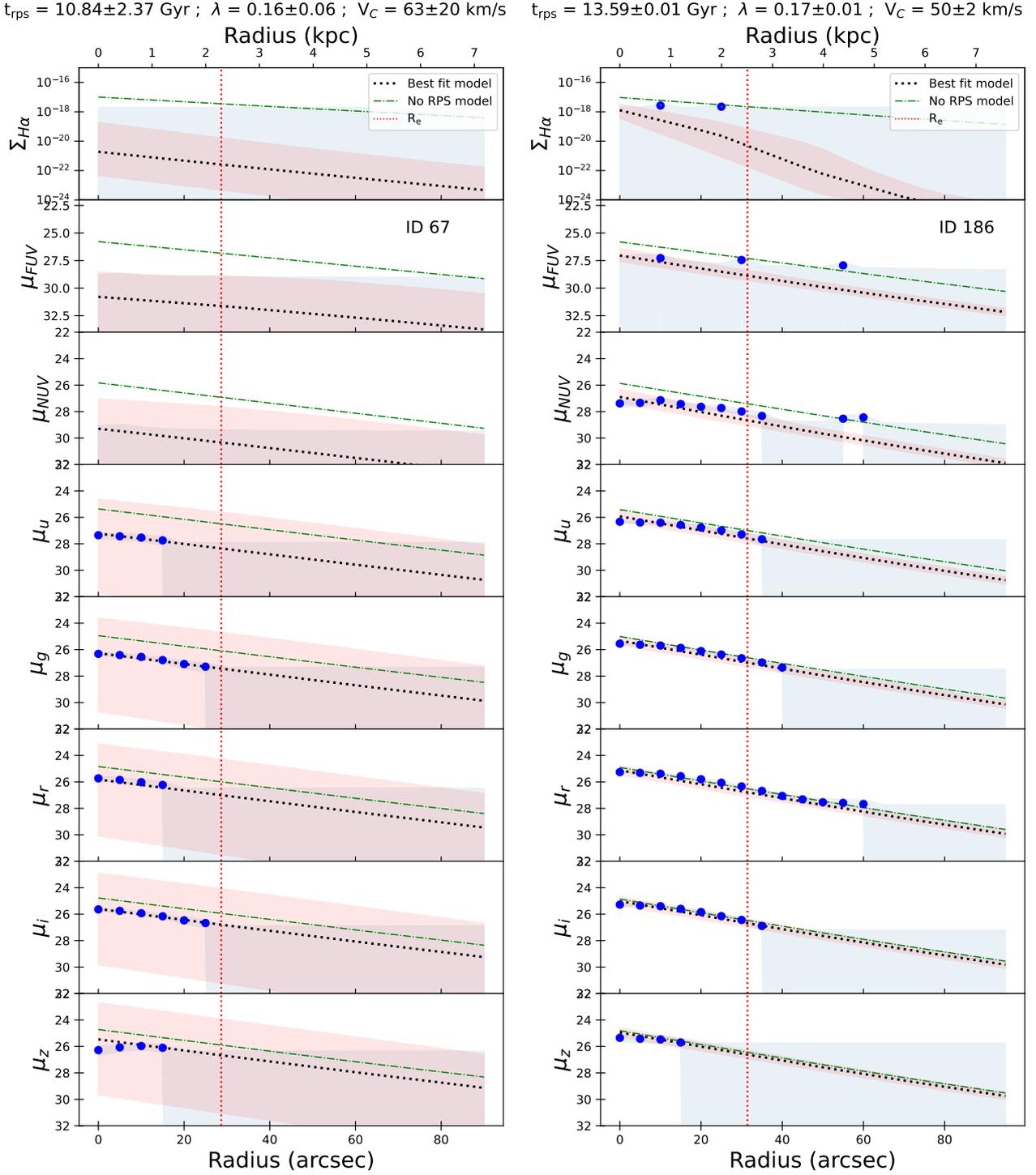}%
        \end{subfigure}\hspace{0em}}
    \caption{Radial surface-brightness profiles of all sources (blue filled dots) with their best-fit models (dotted black lines). The surface-brightness units are in \magperarcsec{} for all the bands except for \Ha{} (erg s$^{-1}$ cm$^{-2}$ arcsec$^{-2}$). The light-blue-shaded area marks the 1$\sigma$ error (for data points) and upper limits (3$\sigma$). The dotted black line indicates the best-fit model derived as described in Sect. \ref{sect:model_fitting_with_sample_profiles}, with its $3\sigma$ confidence level (red-shaded area). The dot-dashed green line shows the same model for an unperturbed system (without RPS). The vertical dotted red line gives the \textit{g}-band effective radius of the galaxy. The vertical gray-shaded regions shown for some galaxies are the regions excluded from the model fitting, where the disk is not dominant based on the decomposition discussed in Sect. \ref{sect:sb_decomposition}. The profiles shown here are corrected for foreground Galactic extinction and inclination.}
\label{fig:all_profiles}
\end{figure*}

This section includes a set of figures (Fig. \ref{fig:all_profiles}) of all the observed profiles and their best-fit models derived as described in Sect. \ref{sect:model_fitting_with_sample_profiles}.

\foreach \j in {1,...,31}{
        \pgfmathtruncatemacro\ystart{\ncounts[1,\j]}
        \pgfmathtruncatemacro\yend{\ncounts[1,\j]+1}
        
        \begin{figure*}[h]
            \centering
            %\ContinuedFloat
            \foreach \i in {\ystart,...,\yend}{%
                \begin{subfigure}{0.49\textwidth}
                    \includegraphics[width=\linewidth]{all_profiles_and_best_model_fits/galaxy\ID[1,\i]_bestmodel_profiles.pdf}%
                \end{subfigure}\hspace{0em}}
        \captionsetup{labelformat=empty}
        \caption{Fig. \ref{fig:all_profiles} continued.}% \label{fig:all_profiles}
        \end{figure*}
        }

\clearpage

\section{A subsample of Virgo cluster galaxies with high spin}\label{appendix:high_spin_sample}

Cosmological simulations indicate that galaxies are formed in halos having a wide range of spin parameters \citep{mo1998}, usually represented by a log-normal distribution peaking at 0.05 with a dispersion of 0.05. 
The peak around 0.05 correspond well to usual HSBs, while LSBs naturally correspond to the queue of large values ($\lambda$ $\geq$ 0.1). 
The methodology described in this work allows us to derive in a consistent way a spin parameter for each galaxy by fitting its surface-brightness profile in different bands. This technique could be in principle applied to all the Virgo cluster members cataloged in the NGVS to select a complete sample of galaxies with a high spin parameter. The large number of objects (3689), however, makes this approach prohibitive 
given the heavy data reduction procedure and the computational time necessary for the fits. 
Past works clearly link galaxies selected to be LSB to large spin parameters \citep{boissier2003,boissier2016}. For dwarf galaxies, however, the surface brightness itself does not necessarily indicates high spin parameters. To distinguish especially extended galaxies from regular dwarfs, different selection criteria have been proposed in the literature (such as the \citealt{lim2020} criteria adopted in the main body of the paper). \citet{vandokkum2015a} first proposed a criterion on size and surface brightness,  optimized to distinguish LSBs from regular dwarf galaxies in the Dragonfly Telescope images, characterized by a limited angular resolution. Such a selection has also been used in several other studies, and especially by \citet{koda2015}. \citet{boissierjunais2019} showed that the same models as the one presented here provided a good fit to size, surface brightness, and integrated colors of the sample of \citet{koda2015}, if large spins were indeed adopted. 

To check the robustness of our results, we thus applied this selection method to identify other objects with potential large spin parameters and we tried to see whether the trends observed in the UDG and diffuse galaxy samples are also shared by other objects characterized by a large $\lambda$. 
For this purpose, galaxies are selected from the NGVS catalog with a \textit{g}-band central surface brightness $\mu_{0} > 24$ \magperarcsec{} and  effective radius $R_{e,g} > 1.5$ kpc, leading to 114 
objects\footnote{We removed from this analysis the galaxy NGVSJ12:46:41.73+10:23:10.4, aka NGVS 3543 as studied in \citet{junais2021}, since it was recently identified as a foreground object by \citealt{jones2021} (see Appendix \ref{appendix:NGVS3543_and_AGC226178})}, out of which 45 are in common with the UDG and diffuse galaxy selection. We stress, however, that contrary to the UDG and diffuse galaxies selections adopted in Sect. \ref{sect:lsb_udg_sample_selection}, which are based on scaling relations drawn by a complete sample, this selection is not complete.

\begin{figure*}[h]
    \centering
    \includegraphics[width=0.9\hsize]{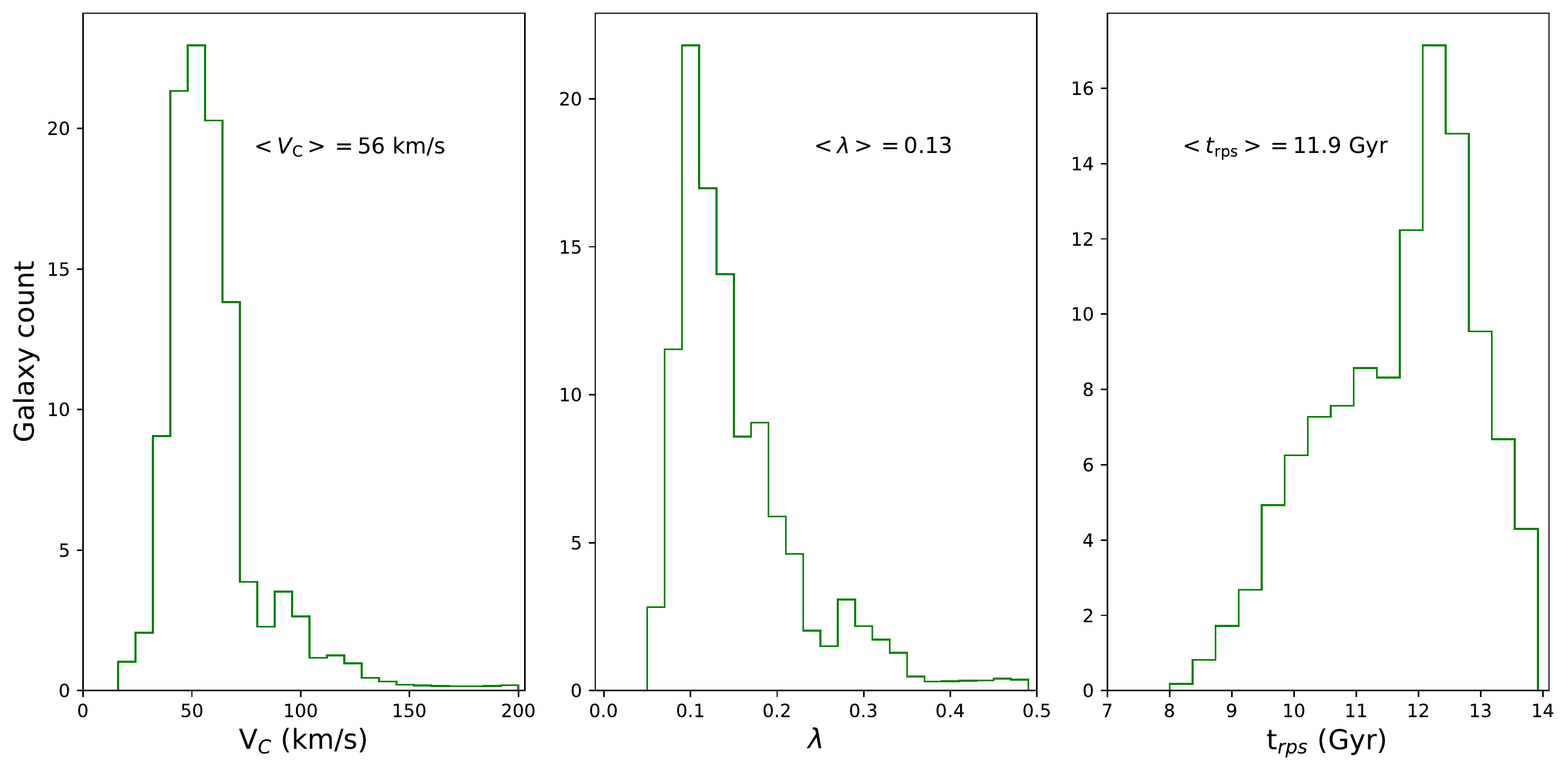}
    \caption{Distribution of the \Vc{}, \spin,{} and \trps{} parameters for the sample selected using a surface-brightness and size cut similar to that of \citet{koda2015}. The median values corresponding to each parameter are shown within each panel.}
    \label{fig:appendix_histogram_trps_vel_spin}
\end{figure*}

We applied the same analysis on this sample of 114 galaxies. As expected, this analysis confirms that the sample is composed of objects with large spin parameters (median $\lambda$ = 0.13) and shares the same trends observed for the UDGs and diffuse galaxy sample, with median parameters $V_C$ = 56 km s$^{-1}$ and $t_{rps}$ = 11.9 Gyr (see Fig. \ref{fig:appendix_histogram_trps_vel_spin}).
This sample also shows a similar color and \trps{} variation with cluster-centric distance as the one described in Sect. \ref{sect:distance_color_gradient} and Sect. \ref{sect:trps_gradient}
with redder galaxies and longer $t_{rps}$ in the center. The best-fit values (linear regression) and the Spearman correlation coefficients are given in Table \ref{table:appendix_large_spin_sample_gradientfits}. Eight galaxies are detected in HI by ALFALFA (one in common with the UDG and diffuse galaxy sample).

\begin{table*}[h]
\centering
    % \fontsize{7pt}{7pt}
    \caption{Correlation properties of the $u-i$ color and the \trps{} with the cluster-centric distance for the sample of high-spin galaxies.
    }
    \begin{tabular}{|c|cc|cc|}
    \hline
    & \multicolumn{2}{c|}{\underline{\hspace{2cm} Linear fit \hspace{2cm}}} & \multicolumn{2}{c|}{\underline{\hspace{0.7cm} Spearman correlation \hspace{0.7cm}}} \\ 
    Relation & Slope & Intercept & $\rho$ &   Probability \\
    % & (mag Mpc$^{-1}$) & (mag) & & \\
    (1) & (2) & (3) & (4) & (5)\\
    \hline
    &&&&\\
    $u-i$ vs D$_{\rm M87}$ & $-0.19\pm0.07$ mag Mpc$^{-1}$ & $1.82\pm0.07$ mag & -0.28 & 98.2\%\\
    &&&&\\
    \trps{} vs D$_{\rm M87}$ & $0.52\pm0.23$ Gyr Mpc$^{-1}$ & $11.82\pm0.23$ Gyr & 0.29 & 99.0\%\\
    \hline
    \hline    
    \end{tabular}\label{table:appendix_large_spin_sample_gradientfits}
    \tablefoot{(1) Correlating quantities; (2) Slope of the linear regression fit using {\tt linmix} ; (3) Intercept of the linear regression fit; (4) Spearman correlation coefficient for the sample; (5) Probability that the two variables are correlated.}
\end{table*}

%

%\newpage

\section{The system of NGVS 3543, \agc{}, and VCC 2034}\label{appendix:NGVS3543_and_AGC226178}

\begin{figure*}[h]
    \centering
    \includegraphics[width=0.49\hsize]{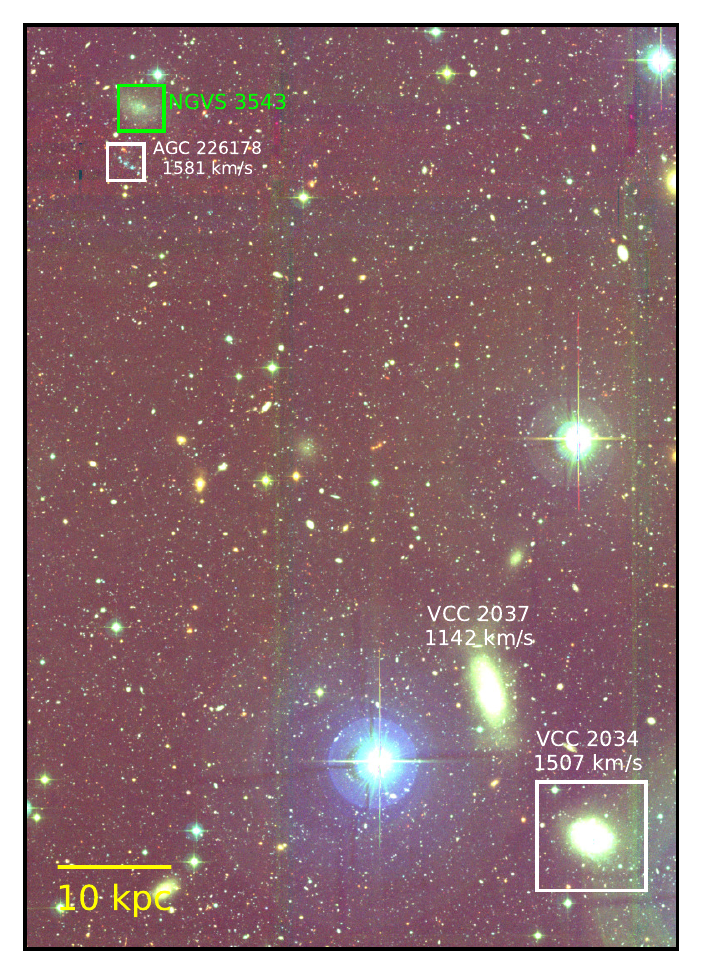}
    \includegraphics[width=0.49\hsize]{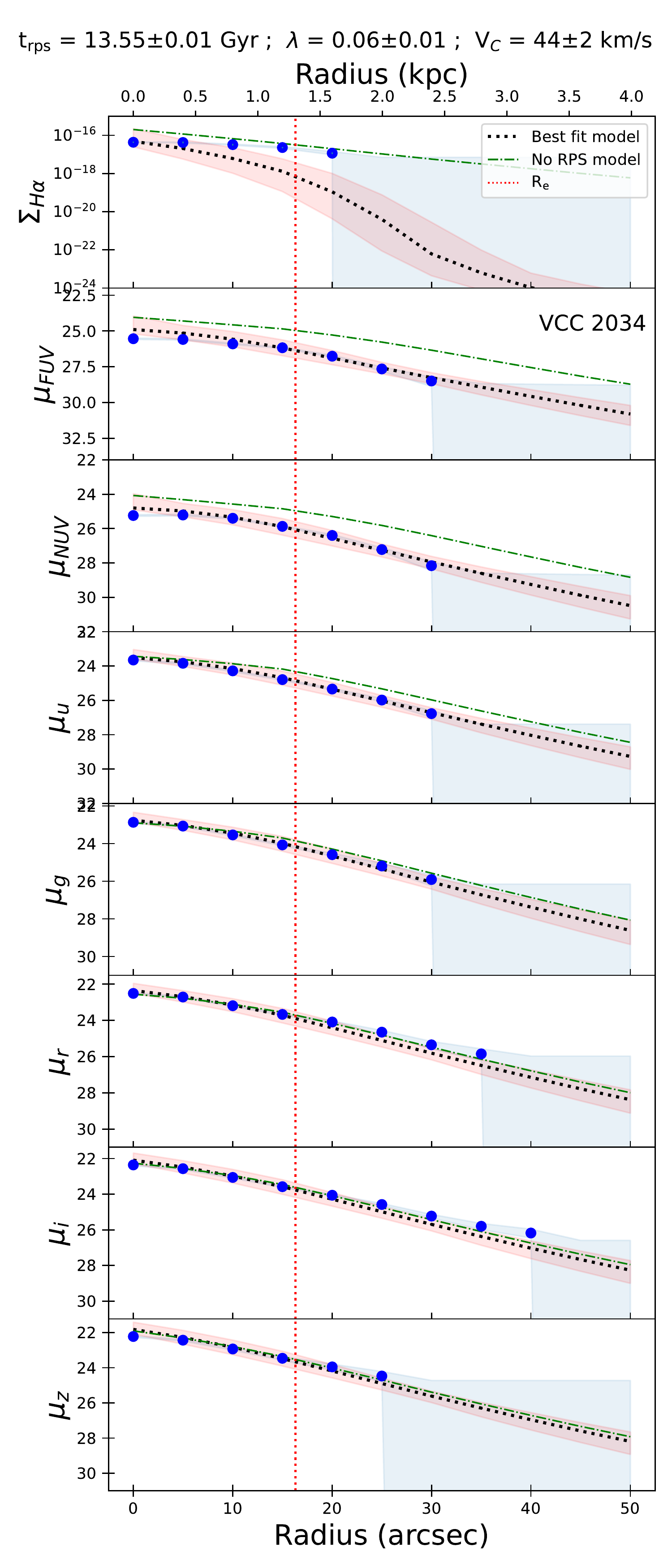}
    \caption{Optical image and surface brightness profiles of VCC 2034. \textit{Left:} NGVS \textit{u}, \textit{g}, \textit{i} color composite image of the system of VCC 2034 and \agc{} (marked inside the white boxes). The galaxy NGVS 3543, which is in projection to the Virgo cluster, is shown in the green box. The \Hi{} radial velocity measurements for a few sources from \citet{jones2021} are also marked, along with their names. \textit{Right:} Radial surface-brightness profiles of VCC 2034 with the best-fit models. The surface-brightness units are in \magperarcsec{} for all the bands except for \Ha{} (erg s$^{-1}$ cm$^{-2}$ arcsec$^{-2}$). The blue-shaded area marks the 1$\sigma$ error (for data points) and upper limits (3$\sigma$). The dotted black line indicates the best-fit model described in Sect. \ref{sect:model_fitting_with_sample_profiles} for a ram-pressure-stripped galaxy, with the red-shaded area the $3\sigma$ confidence level of the best-fit model. The dot-dashed green line shows the same model for an unperturbed system (without RPS). The vertical dotted red line gives the \textit{g}-band effective radius of the galaxy.}
    \label{fig:appendix_ngvs3543_vcc2034}
\end{figure*}

The system of NGVS 3543 and \agc{} was studied in \citet{junais2021}, who interpreted the peculiar distribution of the \Hi{} gas detected by ALFALFA and the star forming regions outside the stellar disk of NGVS 3543 as formed after an RPS event. This scenario was later questioned by \citet{jones2021} who showed, using recent \textit{Hubble} Space Telescope data, that the galaxy NGVS 3543 is located at only 10 Mpc and is thus not a member of the Virgo cluster. After a reanalysis of the ALFALFA and VLA data, these authors suggested that the star forming regions and the \Hi{} gas of \agc{} are rather associated with a gas stripping event from the dwarf irregular galaxy VCC 2034, which is about 70 kpc away (see Fig. \ref{fig:appendix_ngvs3543_vcc2034}).
Similarly, the MUSE data indicate that \agc{} is characterized by a very young stellar population ($\simeq$ 10-100 Myr) and hosts gas with a relatively high metallicity ($0.53\pm0.12$ $Z_{\odot}$), indicating a pre-enrichment from a galaxy with a stellar mass of $\sim$10$^8$ \msun{}.

\begin{table*}[h]
\centering
    \caption{Properties of the best RPS models, and the model with the same spin and velocity but without the RPS for the galaxy VCC 2034.}
    \begin{tabular}{|c|*{7}{c}|}
    \hline
    Galaxy & \trps{} & \spin{} & \Vc{} &  $\log$\mstar{} & $\log M_{gas}$ & $\log SFR$ & $Z_{gas}$ \\
    &  (Gyr) &  & (\kms{}) &  (\msun{}) & (\msun{}) & (\msun{} yr$^{-1}$) & ($Z_{\odot}$) \\
    (1) & (2) & (3) & (4) & (5) & (6) & (7) & (8)\\
    \hline
    &&&&&&&\\
    & \multicolumn{7}{c|}{\underline{\hspace{4.5cm} Best-fit model \hspace{4.5cm}}} \\
    \multirow{2}{*} {VCC 2034} & $13.55\pm0.01$ & $0.06\pm0.01$ & $44\pm2$ & $7.86\pm{0.04}$ & $7.09\pm{0.09}$ & $-2.61\pm{0.01}$ & $1.87\pm{0.38}$ \\
    \cline{2-8}
    &&&&&&&\\
    & \multicolumn{7}{c|}{\underline{\hspace{4.2cm} Model without RPS \hspace{4.2cm}}} \\
    & -- & $0.06\pm0.01$ & $44\pm2$ & $7.88\pm{0.03}$ & $8.49\pm{0.09}$ & $-1.36\pm{0.02}$ & $0.71\pm{0.14}$  \\
    \hline    
    \end{tabular}\label{table:appendix_vcc2034}
    \tablefoot{(1) Name of the galaxy; (2-4) Model parameters \trps{}, \spin{} and \Vc{}; (5-8) Stellar-mass, gas mass, SFR, and gas-phase metallicity predicted by the models. The uncertainties given in the best-fit models are from the confidence limits in \trps{}, \spin{} and \Vc{} parameters. For the non-RPS models, the uncertainties are from the error in \spin{} and \Vc{} alone.}
\end{table*}

We applied the RPS model to the dwarf galaxy VCC 2034 to see whether the 
scenario proposed by \cite{jones2021} for the origin of \agc{} is realistic.
For this purpose we derived the surface-brightness profiles in the NGVS, GUViCS, and VESTIGE bands and fitted them as described in Sect. \ref{sect:model_fitting_with_sample_profiles}. Table \ref{table:appendix_vcc2034} gives the resulting parameters of the modeling. The best-fit model (see Fig. \ref{fig:appendix_ngvs3543_vcc2034}) suggests that VCC 2034 is a low mass galaxy (but not an LSB because of its low spin) experiencing an ongoing RPS event that started $\sim$150 Myr ago. During this process, the galaxy lost about $3\times10^8$ \msun{} of gas, which could have seeded the formation of \agc{}, which has a gas mass of $\sim5\times10^7$ \msun{} (16\% of the gas lost by VCC 2034). Moreover, the metallicity of VCC 2034 before the RPS ($0.71\pm0.14$ $Z_{\odot}$) is also close to the one obtained for \agc{} by \citet{jones2021}. To conclude results from our modeling are consistent with the scenario proposed by \citet{jones2021} for the system of VCC 2034 and \agc{}.

% \FloatBarrier

\section{Additional tables}

\begin{landscape}
\setlength\LTcapwidth{\textheight} % default: 
\centering
% \fontsize{6pt}{6pt}
% \small
\begin{longtable}{|*{14}{c|}}
% \begin{table}
% \begin{tabular}{ccccccccccccc}
\caption{Properties of the selected sample of galaxies.}\label{Table:lsb_sample_ngvs_params}\\
% \hline
\hline
ID & NGVS name & VCC name & RA & DEC & D$_{\rm M87}$& g & \mstar{} & $R_{e,g}$ & $\mu_{0,g}$ & q & P.A & E (B-V) & Flag \\
 &  & & (deg) & (deg) & (Mpc) & (mag) & ($10^{7}$ \msun{}) & (kpc) & (\magperarcsec{}) &  & (deg) & (mag) &  \\
(1) & (2) & (3) & (4) & (5) & (6) & (7) & (8) & (9) & (10) & (11) & (12) & (13) & (14)\\
\hline
\endfirsthead
\caption*{Table \ref{Table:lsb_sample_ngvs_params} continued.}\\
\hline
ID & NGVS name & VCC name & RA & DEC & D$_{\rm M87}$& g & \mstar{} & $R_{e,g}$ & $\mu_{0,g}$ & q & P.A & E (B-V) & Flag \\
 &  & & (deg) & (deg) & (Mpc) & (mag) & ($10^{7}$ \msun{}) & (kpc) & (\magperarcsec{}) &  & (deg) & (mag) &  \\
(1) & (2) & (3) & (4) & (5) & (6) & (7) & (8) & (9) & (10) & (11) & (12) & (13) & (14)\\
\hline
\endhead
\hline
\endfoot
67 & NGVSJ12:12:21.23+13:44:16.6 & -- & 183.088 & 13.738 & 1.35 & 17.76 & 2.64 & 2.29 & 26.07 & 0.77 & -37.44 & 0.033 & 2 \\
186 & NGVSJ12:15:55.76+09:39:04.1 & VCC 169 & 183.982 & 9.651 & 1.32 & 16.97 & 2.29 & 2.51 & 24.71 & 0.6 & -79.74 & 0.018 & 2 \\
227 & NGVSJ12:16:32.59+13:09:51.3 & VCC 197 & 184.136 & 13.164 & 1.03 & 16.78 & 12.2 & 9.09 & 23.96 & 0.3 & 65.29 & 0.03 & 1 \\
261 & NGVSJ12:17:04.16+10:00:19.8 & VCC 217 & 184.267 & 10.006 & 1.19 & 15.84 & 7.88 & 2.91 & 23.5 & 0.52 & -83.93 & 0.021 & 2 \\
321 & NGVSJ12:18:05.98+07:38:02.6 & -- & 184.525 & 7.634 & 1.64 & 18.93 & 0.5 & 1.79 & 27.06 & 0.78 & 71.5 & 0.019 & 2 \\
421 & NGVSJ12:19:36.96+15:27:16.8 & VCC 360 & 184.904 & 15.455 & 1.18 & 16.61 & 5.31 & 3.01 & 25.56 & 0.81 & 81.84 & 0.023 & 1 \\
466 & NGVSJ12:20:11.38+11:53:56.5 & -- & 185.047 & 11.899 & 0.76 & 19.78 & 0.59 & 1.94 & 26.88 & 0.51 & 87.17 & 0.028 & 2 \\
590 & NGVSJ12:21:30.88+15:30:04.9 & VCC 481 & 185.379 & 15.501 & 1.11 & 17.56 & 5.5 & 2.2 & 25.58 & 0.7 & -39.03 & 0.023 & 2 \\
604 & NGVSJ12:21:38.40+06:16:59.7 & VCC 487 & 185.41 & 6.283 & 1.87 & 18.71 & 1.27 & 1.69 & 26.02 & 0.76 & -13.77 & 0.02 & 2 \\
646 & NGVSJ12:22:03.58+11:43:17.5 & -- & 185.515 & 11.722 & 0.65 & 18.34 & 2.79 & 1.79 & 26.68 & 1.0 & 88.31 & 0.029 & 2 \\
796 & NGVSJ12:23:04.59+12:00:53.4 & VCC 615 & 185.769 & 12.015 & 0.56 & 17.25 & 7.35 & 2.1 & 25.73 & 1.0 & 40.36 & 0.028 & 2 \\
892 & NGVSJ12:23:47.33+13:36:08.3 & -- & 185.947 & 13.602 & 0.6 & 18.61 & 2.68 & 2.04 & 26.7 & 0.72 & -4.76 & 0.047 & 2 \\
935 & NGVSJ12:24:01.84+13:51:56.4 & -- & 186.008 & 13.866 & 0.64 & 18.39 & 1.37 & 2.98 & 27.64 & 0.77 & -35.93 & 0.038 & 1 \\
964 & NGVSJ12:24:13.00+11:45:41.5 & VCC 707 & 186.054 & 11.762 & 0.5 & 17.55 & 7.16 & 1.79 & 25.15 & 0.95 & 14.74 & 0.038 & 2 \\
1008 & NGVSJ12:24:36.61+13:36:45.2 & -- & 186.153 & 13.613 & 0.56 & 19.65 & 0.77 & 1.6 & 27.27 & 0.62 & 69.75 & 0.04 & 2 \\
1017 & NGVSJ12:24:42.06+13:31:00.6 & -- & 186.175 & 13.517 & 0.54 & 17.6 & 5.84 & 2.98 & 26.03 & 0.76 & -1.61 & 0.042 & 1 \\
1160 & NGVSJ12:25:37.61+10:14:58.6 & VCC 811 & 186.407 & 10.25 & 0.72 & 16.96 & 7.28 & 2.77 & 26.05 & 0.95 & 21.24 & 0.028 & 1 \\
1164 & NGVSJ12:25:38.76+14:09:02.1 & VCC 824 & 186.411 & 14.151 & 0.62 & 16.01 & 15.3 & 2.24 & 23.33 & 0.84 & -36.15 & 0.047 & 2 \\
1346 & NGVSJ12:26:37.41+09:44:32.0 & -- & 186.656 & 9.742 & 0.82 & 20.7 & 0.25 & 1.21 & 28.22 & 0.9 & -28.4 & 0.024 & 2 \\
1352 & NGVSJ12:26:38.25+13:04:44.2 & VCC 927 & 186.659 & 13.079 & 0.35 & 19.45 & 1.63 & 0.97 & 22.71 & 0.81 & -70.36 & 0.029 & 1 \\
1397 & NGVSJ12:26:48.36+13:21:17.7 & -- & 186.702 & 13.355 & 0.4 & 19.1 & 1.73 & 2.06 & 27.25 & 0.78 & -77.18 & 0.026 & 1 \\
1405 & NGVSJ12:26:50.78+11:33:27.1 & VCC 950 & 186.712 & 11.558 & 0.37 & 15.01 & 12.3 & 2.39 & 22.79 & 0.33 & -48.18 & 0.031 & 2 \\
1424 & NGVSJ12:26:57.00+14:47:52.5 & -- & 186.738 & 14.798 & 0.74 & 19.1 & 0.81 & 2.06 & 26.83 & 0.49 & 51.69 & 0.034 & 2 \\
1476 & NGVSJ12:27:15.46+12:39:41.4 & VCC 987 & 186.814 & 12.662 & 0.26 & 17.77 & 7.88 & 2.48 & 26.51 & 0.68 & 0.05 & 0.03 & 2 \\
1479 & NGVSJ12:27:15.75+13:26:56.1 & -- & 186.816 & 13.449 & 0.39 & 19.42 & 1.16 & 1.23 & 26.43 & 0.83 & 11.46 & 0.025 & 1 \\
1529 & NGVSJ12:27:31.55+09:35:44.3 & VCC 1017 & 186.881 & 9.596 & 0.84 & 14.54 & 33.5 & 4.28 & 24.28 & 0.57 & 23.22 & 0.022 & 1 \\
1593 & NGVSJ12:27:55.22+12:22:09.5 & VCC 1052 & 186.98 & 12.369 & 0.2 & 16.16 & 20.8 & 3.93 & 25.77 & 0.78 & 41.59 & 0.027 & 1 \\
1633 & NGVSJ12:28:10.07+12:43:29.4 & -- & 187.042 & 12.725 & 0.21 & 18.36 & 2.17 & 2.93 & 27.45 & 0.73 & 88.93 & 0.024 & 1 \\
1687 & NGVSJ12:28:26.16+15:22:38.6 & -- & 187.109 & 15.377 & 0.88 & 20.03 & 0.37 & 1.19 & 27.47 & 1.0 & -36.97 & 0.026 & 2 \\
1719 & NGVSJ12:28:37.88+12:51:42.0 & -- & 187.158 & 12.862 & 0.21 & 19.17 & 0.6 & 2.49 & 27.84 & 0.59 & -63.21 & 0.02 & 1 \\
1846 & NGVSJ12:29:22.72+15:03:49.4 & VCC 1181 & 187.345 & 15.064 & 0.78 & 17.99 & 3.39 & 2.07 & 25.5 & 0.71 & -30.32 & 0.034 & 2 \\
1968 & NGVSJ12:30:00.61+07:55:45.8 & VCC 1249 & 187.503 & 7.929 & 1.29 & 14.46 & 40.6 & 3.06 & 22.3 & 0.67 & -1.25 & 0.022 & 2 \\
1993 & NGVSJ12:30:08.68+09:42:56.2 & -- & 187.536 & 9.716 & 0.77 & 17.09 & 28.6 & 3.73 & 26.42 & 0.58 & 88.39 & 0.021 & 1 \\
2001 & NGVSJ12:30:12.55+09:42:56.3 & -- & 187.552 & 9.716 & 0.77 & 18.7 & 4.47 & 1.49 & 26.39 & 1.0 & 79.08 & 0.021 & 2 \\
2046 & NGVSJ12:30:24.43+13:58:54.5 & VCC 1287 & 187.602 & 13.982 & 0.46 & 15.93 & 16.3 & 3.63 & 24.81 & 0.98 & 19.74 & 0.036 & 1 \\
2079 & NGVSJ12:30:37.30+10:20:53.0 & -- & 187.655 & 10.348 & 0.59 & 17.45 & 5.54 & 4.18 & 27.28 & 0.79 & -60.49 & 0.034 & 1 \\
2269 & NGVSJ12:31:48.01+12:21:33.1 & -- & 187.95 & 12.359 & 0.07 & 19.87 & 0.33 & 2.77 & 28.04 & 0.31 & 71.84 & 0.024 & 1 \\
2343 & NGVSJ12:32:12.92+09:18:56.7 & VCC 1421 & 188.054 & 9.316 & 0.89 & 17.33 & 3.52 & 2.18 & 25.54 & 0.9 & -14.17 & 0.021 & 2 \\
2351 & NGVSJ12:32:15.43+11:23:52.6 & -- & 188.064 & 11.398 & 0.3 & 19.21 & 0.89 & 1.61 & 25.44 & 0.64 & 13.93 & 0.057 & 2 \\
2365 & NGVSJ12:32:22.52+12:19:32.1 & -- & 188.094 & 12.326 & 0.11 & 18.16 & 4.37 & 5.15 & 28.08 & 0.5 & 89.1 & 0.028 & 1 \\
2458 & NGVSJ12:33:02.20+13:42:14.0 & -- & 188.259 & 13.704 & 0.41 & 20.03 & 0.46 & 2.54 & 28.38 & 0.49 & -65.27 & 0.036 & 1 \\
2531 & NGVSJ12:33:29.59+15:14:02.8 & -- & 188.373 & 15.234 & 0.84 & 17.51 & 6.23 & 2.29 & 25.95 & 0.9 & 35.97 & 0.033 & 1 \\
2572 & NGVSJ12:33:51.04+09:04:46.7 & -- & 188.463 & 9.08 & 0.98 & 20.23 & 0.33 & 1.28 & 27.46 & 0.79 & -74.74 & 0.02 & 2 \\
2621 & NGVSJ12:34:15.55+11:28:00.8 & VCC 1551 & 188.565 & 11.467 & 0.36 & 17.91 & 3.39 & 2.02 & 25.99 & 0.82 & -52.14 & 0.036 & 2 \\
2690 & NGVSJ12:34:49.20+05:54:13.4 & -- & 188.705 & 5.904 & 1.89 & 19.39 & 0.66 & 2.01 & 27.02 & 0.46 & 63.37 & 0.02 & 2 \\
2731 & NGVSJ12:35:12.86+07:03:22.4 & -- & 188.804 & 7.056 & 1.57 & 18.09 & 1.31 & 3.73 & 27.57 & 0.86 & 33.87 & 0.02 & 1 \\
2887 & NGVSJ12:36:37.40+11:09:13.1 & VCC 1681 & 189.156 & 11.154 & 0.54 & 16.4 & 16.0 & 2.52 & 23.89 & 0.84 & -67.59 & 0.031 & 2 \\
2999 & NGVSJ12:37:49.73+07:49:23.1 & -- & 189.457 & 7.823 & 1.41 & 19.5 & 0.73 & 1.49 & 27.22 & 1.0 & -37.31 & 0.023 & 1 \\
3032 & NGVSJ12:38:09.93+10:47:17.6 & -- & 189.541 & 10.788 & 0.69 & 20.66 & 0.11 & 1.51 & 27.7 & 0.69 & 24.37 & 0.021 & 1 \\
3088 & NGVSJ12:38:54.29+10:14:31.6 & VCC 1776 & 189.726 & 10.242 & 0.84 & 17.0 & 5.64 & 2.24 & 25.25 & 0.81 & -88.57 & 0.02 & 2 \\
3112 & NGVSJ12:39:20.34+11:26:20.0 & -- & 189.835 & 11.439 & 0.66 & 20.16 & 0.35 & 1.47 & 27.1 & 0.56 & -59.7 & 0.034 & 2 \\
3116 & NGVSJ12:39:21.99+12:05:34.6 & -- & 189.842 & 12.093 & 0.61 & 20.76 & 0.13 & 1.2 & 26.43 & 0.7 & 73.36 & 0.045 & 2 \\
3128 & NGVSJ12:39:31.78+11:27:15.3 & VCC 1798 & 189.882 & 11.454 & 0.67 & 17.94 & 2.41 & 2.04 & 25.97 & 0.76 & 75.42 & 0.034 & 2 \\
3146 & NGVSJ12:39:48.00+07:18:47.2 & -- & 189.95 & 7.313 & 1.59 & 18.12 & 2.39 & 2.54 & 26.48 & 0.64 & -61.81 & 0.024 & 1 \\
3190 & NGVSJ12:40:21.21+12:43:02.8 & VCC 1835 & 190.088 & 12.717 & 0.68 & 18.23 & 1.55 & 2.58 & 26.7 & 0.58 & 88.38 & 0.04 & 2 \\
3225 & NGVSJ12:40:56.41+14:15:16.3 & -- & 190.235 & 14.255 & 0.89 & 19.46 & 0.22 & 1.88 & 26.93 & 0.74 & -39.45 & 0.038 & 1 \\
3233 & NGVSJ12:40:58.86+14:15:57.6 & -- & 190.245 & 14.266 & 0.89 & 19.5 & 0.2 & 1.69 & 27.7 & 0.85 & 5.14 & 0.037 & 2 \\
3265 & NGVSJ12:41:30.86+11:40:55.8 & VCC 1882 & 190.379 & 11.682 & 0.78 & 18.58 & 1.76 & 2.17 & 24.43 & 0.48 & 71.21 & 0.033 & 2 \\
3272 & NGVSJ12:41:39.34+09:12:30.5 & VCC 1884 & 190.414 & 9.208 & 1.19 & 16.29 & 17.1 & 3.34 & 25.58 & 0.82 & 67.42 & 0.02 & 1 \\
3356 & NGVSJ12:42:56.59+13:20:49.4 & -- & 190.736 & 13.347 & 0.89 & 18.29 & 1.49 & 2.31 & 26.66 & 0.65 & 86.74 & 0.025 & 2 \\
3365 & NGVSJ12:43:07.09+16:28:42.0 & -- & 190.78 & 16.478 & 1.46 & 19.38 & 0.42 & 1.95 & 27.8 & 0.74 & -81.73 & 0.021 & 2 \\
3379 & NGVSJ12:43:20.88+14:02:02.9 & -- & 190.837 & 14.034 & 1.0 & 18.77 & 1.1 & 1.99 & 26.48 & 0.7 & 66.98 & 0.027 & 2 \\
3548 & NGVSJ12:46:55.48+10:10:56.7 & VCC 2045 & 191.731 & 10.182 & 1.3 & 15.64 & 40.4 & 1.6 & 21.73 & 0.33 & 75.54 & 0.032 & 1 \\
3633 & NGVSJ12:49:38.67+15:03:16.6 & -- & 192.411 & 15.055 & 1.52 & 19.16 & 1.19 & 1.76 & 27.09 & 0.72 & -82.5 & 0.029 & 2 \\

% \end{tabular}
% \end{table}
\end{longtable}
\tablefoot{(1) Source ID based on the position in the NGVS catalog. (2) Name of the source in NGVS. (3) VCC name. (4-5) Coordinates of the source in J2000. (6) Projected distance of the galaxy from M87. (7) \textit{g}-band magnitude. (8) Stellar mass of the galaxies (see Sect. \ref{sect:ngvs_survey}). (9) Effective radius of the source. (10) \textit{g}-band central surface brightness. (11) Axis ratio. (12) Position angle. (13) Foreground galactic extinction from \citet{schlegel1998}. (14) Flag for the galaxy type, where the flags 1 and 2 corresponds to UDGs from \citet{lim2020} and Diffuse galaxies, respectively. All the geometrical parameters given in this table are taken from the NGVS catalog.}
\end{landscape}

%-------------------------------------------------

% \begin{landscape}
\setlength\LTcapwidth{\textheight} % default: 
%\centering
% \small
% \fontsize{7pt}{7pt}
\begin{longtable}{|*{9}{c|}}
% \begin{table}
% \begin{tabular}{ccccccccccccc}
\caption{Photometric measurements of the sample.}\label{table:lsb_sample_magnitudes}\label{page:photometry_table}\\
\hline
ID  & $R_{\rm last}$ & \textit{u} & \textit{g} & \textit{r} & \textit{i} & \textit{z} & \textit{NUV} & \textit{FUV} \\
    & (kpc)  &    (mag)    &    (mag) &    (mag)  &   (mag)   &   (mag)   & (mag)   &   (mag)   \\
(1)  & (2) & (3) & (4)  & (5)    &   (6)  &   (7)  &   (8)  &   (9) \\
\hline
\endfirsthead
\caption*{Table \ref{table:lsb_sample_magnitudes} continued.}\\
\hline
ID  & $R_{\rm last}$ & \textit{u} & \textit{g} & \textit{r} & \textit{i} & \textit{z} & \textit{NUV} & \textit{FUV} \\
    & (kpc)  &    (mag)    &    (mag) &    (mag)  &   (mag)   &   (mag)   & (mag)   &   (mag)   \\
(1)  & (2) & (3) & (4)  & (5)    &   (6)  &   (7)  &   (8)  &   (9) \\
\hline
\endhead
\hline
\endfoot
67 & 2.0 & $19.58 \pm 0.3$ & $18.61 \pm 0.21$ & $18.03 \pm 0.29$ & $17.98 \pm 0.18$ & $18.02 \pm 0.34$ & $21.48 \pm 0.36$ & > 21.09 \\
186 & 3.2 & $17.79 \pm 0.19$ & $17.14 \pm 0.12$ & $16.82 \pm 0.08$ & $16.91 \pm 0.14$ & > 16.52 & $18.55 \pm 0.07$ & $18.68 \pm 0.13$ \\
227 & 4.0 & > 18.98 & $18.24 \pm 0.14$ & $17.68 \pm 0.13$ & $17.47 \pm 0.27$ & $17.43 \pm 0.27$ & > 21.87 & > 21.96 \\
261 & 7.2 & $16.39 \pm 0.08$ & $15.74 \pm 0.05$ & $15.36 \pm 0.05$ & $15.37 \pm 0.07$ & $15.37 \pm 0.14$ & $17.22 \pm 0.05$ & $17.27 \pm 0.14$ \\
321 & 1.6 & $20.53 \pm 0.34$ & $19.76 \pm 0.23$ & $19.16 \pm 0.2$ & $19.2 \pm 0.26$ & > 19.12 & > 21.22 & > 21.89 \\
421 & 5.2 & $17.6 \pm 0.18$ & $16.73 \pm 0.14$ & $16.25 \pm 0.15$ & $16.28 \pm 0.15$ & $16.32 \pm 0.16$ & > 19.42 & > 19.94 \\
466 & 2.8 & $20.92 \pm 0.38$ & $20.3 \pm 0.21$ & $19.82 \pm 0.37$ & > 19.08 & > 18.31 & > 20.93 & > 21.19 \\
590 & 2.0 & $19.26 \pm 0.3$ & $18.27 \pm 0.16$ & $17.7 \pm 0.08$ & $17.54 \pm 0.08$ & $17.44 \pm 0.06$ & > 21.72 & > 22.33 \\
604 & 2.0 & $20.1 \pm 0.18$ & $19.22 \pm 0.12$ & $18.7 \pm 0.17$ & $18.58 \pm 0.22$ & > 18.06 & > 21.54 & > 20.17 \\
646 & 2.0 & $20.24 \pm 0.28$ & $19.02 \pm 0.18$ & $18.58 \pm 0.22$ & > 18.1 & > 17.92 & > 21.8 & > 21.84 \\
796 & 2.4 & > 18.89 & $17.74 \pm 0.12$ & $17.14 \pm 0.1$ & $17.0 \pm 0.1$ & $16.92 \pm 0.11$ & > 20.72 & > 21.06 \\
892 & 2.4 & > 19.93 & $18.93 \pm 0.13$ & $18.41 \pm 0.19$ & $18.06 \pm 0.24$ & $17.99 \pm 0.16$ & -- & -- \\
935 & 1.6 & > 20.61 & $20.0 \pm 0.28$ & $19.52 \pm 0.3$ & > 19.32 & > 19.01 & -- & -- \\
964 & 3.2 & $18.9 \pm 0.3$ & $17.69 \pm 0.08$ & $17.08 \pm 0.1$ & $16.58 \pm 0.29$ & $16.87 \pm 0.3$ & > 20.73 & > 20.98 \\
1008 & 2.0 & > 20.89 & $20.13 \pm 0.14$ & $19.46 \pm 0.18$ & $19.0 \pm 0.3$ & > 19.27 & > 21.82 & > 21.64 \\
1017 & 0.8 & $21.08 \pm 0.29$ & $19.94 \pm 0.11$ & $19.51 \pm 0.12$ & $19.22 \pm 0.09$ & $19.1 \pm 0.16$ & > 23.34 & > 23.3 \\
1160 & 4.4 & $18.15 \pm 0.2$ & $16.98 \pm 0.11$ & $16.52 \pm 0.12$ & $16.28 \pm 0.15$ & $16.38 \pm 0.3$ & > 19.72 & > 19.65 \\
1164 & 3.6 & $17.01 \pm 0.24$ & $16.19 \pm 0.1$ & $15.7 \pm 0.08$ & $15.56 \pm 0.08$ & $15.48 \pm 0.11$ & > 19.33 & -- \\
1346 & 0.4 & $23.46 \pm 0.28$ & $22.8 \pm 0.18$ & $22.14 \pm 0.18$ & $21.85 \pm 0.26$ & > 21.51 & > 24.15 & > 24.15 \\
1352 & 1.6 & $20.23 \pm 0.16$ & $19.2 \pm 0.12$ & $18.58 \pm 0.14$ & $18.43 \pm 0.15$ & $18.22 \pm 0.18$ & > 21.84 & > 21.94 \\
1397 & 1.2 & > 21.21 & $20.78 \pm 0.24$ & $20.33 \pm 0.22$ & > 19.41 & > 17.31 & > 22.74 & > 23.33 \\
1405 & 7.2 & $15.86 \pm 0.05$ & $15.29 \pm 0.05$ & $15.07 \pm 0.05$ & $15.05 \pm 0.05$ & $14.98 \pm 0.05$ & $16.82 \pm 0.05$ & $17.08 \pm 0.05$ \\
1424 & 2.8 & > 20.66 & $19.41 \pm 0.13$ & $18.84 \pm 0.2$ & $18.92 \pm 0.25$ & > 18.73 & > 21.51 & > 21.8 \\
1476 & 2.4 & $19.67 \pm 0.37$ & $18.52 \pm 0.24$ & $17.94 \pm 0.22$ & $17.76 \pm 0.3$ & $17.45 \pm 0.35$ & > 21.7 & > 22.03 \\
1479 & 1.2 & $21.23 \pm 0.16$ & $20.18 \pm 0.16$ & $19.58 \pm 0.2$ & $19.52 \pm 0.34$ & $19.43 \pm 0.33$ & > 22.68 & > 22.53 \\
1529 & 14.8 & $15.2 \pm 0.18$ & $14.44 \pm 0.12$ & $14.05 \pm 0.14$ & $13.85 \pm 0.22$ & > 13.38 & > 17.47 & > 18.2 \\
1593 & 3.6 & > 17.94 & $17.0 \pm 0.2$ & $16.43 \pm 0.09$ & $16.39 \pm 0.2$ & $16.32 \pm 0.33$ & $20.66 \pm 0.3$ & > 20.82 \\
1633 & 2.8 & $20.11 \pm 0.3$ & $19.1 \pm 0.18$ & $18.52 \pm 0.2$ & > 18.37 & > 17.23 & > 21.32 & > 21.84 \\
1687 & 0.8 & > 21.89 & $21.6 \pm 0.26$ & $20.89 \pm 0.21$ & > 20.29 & > 20.73 & > 23.27 & > 23.79 \\
1719 & 2.8 & > 20.38 & $19.6 \pm 0.24$ & $19.1 \pm 0.36$ & > 18.92 & > 18.59 & > 21.6 & > 22.47 \\
1846 & 2.0 & > 18.88 & $18.68 \pm 0.16$ & $18.2 \pm 0.11$ & $17.96 \pm 0.08$ & $17.67 \pm 0.12$ & > 21.55 & > 22.49 \\
1968 & 3.6 & $15.82 \pm 0.06$ & $15.03 \pm 0.12$ & $14.66 \pm 0.17$ & $14.6 \pm 0.2$ & $14.47 \pm 0.26$ & $17.6 \pm 0.15$ & > 18.73 \\
1993 & 1.6 & > 19.2 & $19.26 \pm 0.28$ & > 19.36 & $18.9 \pm 0.15$ & $17.4 \pm 0.2$ & > 22.29 & > 21.96 \\
2001 & 4.8 & > 16.57 & > 17.23 & > 16.7 & > 16.43 & > 15.11 & > 19.43 & > 18.32 \\
2046 & 6.8 & $16.9 \pm 0.2$ & $15.74 \pm 0.08$ & $15.2 \pm 0.14$ & $14.96 \pm 0.11$ & $14.8 \pm 0.14$ & > 18.73 & > 19.72 \\
2079 & 3.6 & > 18.94 & $18.12 \pm 0.26$ & $17.52 \pm 0.13$ & $17.29 \pm 0.29$ & > 17.08 & > 20.42 & $19.91 \pm 0.26$ \\
2269 & 2.8 & > 21.67 & $20.58 \pm 0.2$ & $19.86 \pm 0.21$ & > 19.69 & > 19.41 & > 22.54 & > 22.82 \\
2343 & 2.0 & > 18.87 & $18.02 \pm 0.16$ & $17.57 \pm 0.14$ & $17.44 \pm 0.11$ & > 17.41 & > 21.7 & > 22.16 \\
2351 & 1.6 & $20.88 \pm 0.2$ & $19.82 \pm 0.11$ & $19.3 \pm 0.08$ & $19.19 \pm 0.12$ & $19.23 \pm 0.24$ & > 23.09 & > 22.64 \\
2365 & 0.4 & > 24.15 & $23.17 \pm 0.15$ & $21.9 \pm 0.06$ & $21.66 \pm 0.14$ & $21.3 \pm 0.2$ & > 25.85 & $26.11 \pm 0.25$ \\
2458 & 2.0 & > 22.07 & $21.16 \pm 0.22$ & $20.58 \pm 0.37$ & > 20.21 & > 18.76 & > 22.46 & > 22.52 \\
2531 & 4.0 & $18.61 \pm 0.18$ & $17.58 \pm 0.08$ & $16.97 \pm 0.07$ & $16.78 \pm 0.12$ & $16.65 \pm 0.14$ & > 19.87 & > 18.68 \\
2572 & 0.8 & > 22.15 & $21.63 \pm 0.3$ & > 21.24 & $20.94 \pm 0.2$ & > 20.28 & > 23.41 & > 24.13 \\
2621 & 3.2 & > 18.74 & $18.19 \pm 0.08$ & $17.41 \pm 0.1$ & $17.36 \pm 0.17$ & $17.39 \pm 0.29$ & > 20.54 & > 19.95 \\
2690 & 3.2 & $20.84 \pm 0.3$ & $19.58 \pm 0.11$ & $19.1 \pm 0.22$ & > 18.82 & > 18.64 & > 20.97 & > 20.39 \\
2731 & 4.4 & > 19.14 & $18.5 \pm 0.18$ & > 17.97 & > 17.62 & > 17.48 & > 20.23 & > 20.3 \\
2887 & 3.6 & $17.93 \pm 0.15$ & $16.76 \pm 0.08$ & $16.12 \pm 0.06$ & $16.03 \pm 0.1$ & $15.94 \pm 0.1$ & $20.48 \pm 0.34$ & > 20.59 \\
2999 & 1.2 & $21.36 \pm 0.29$ & $20.42 \pm 0.24$ & $19.94 \pm 0.08$ & $19.85 \pm 0.16$ & > 19.55 & > 22.02 & > 22.01 \\
3032 & 0.4 & > 23.51 & $23.46 \pm 0.33$ & $23.08 \pm 0.34$ & > 23.0 & > 21.97 & > 24.56 & > 25.16 \\
3088 & 4.0 & > 18.17 & $17.2 \pm 0.08$ & $16.82 \pm 0.12$ & $16.57 \pm 0.12$ & $16.34 \pm 0.18$ & > 19.9 & > 20.33 \\
3112 & 2.0 & $21.34 \pm 0.34$ & $20.44 \pm 0.14$ & $19.81 \pm 0.26$ & $19.86 \pm 0.28$ & > 19.53 & > 21.64 & > 21.88 \\
3116 & 0.4 & $24.24 \pm 0.34$ & $23.74 \pm 0.27$ & > 23.18 & > 22.15 & > 21.46 & $25.54 \pm 0.29$ & > 25.01 \\
3128 & 2.8 & $19.26 \pm 0.08$ & $18.19 \pm 0.12$ & $17.71 \pm 0.08$ & $17.59 \pm 0.14$ & $17.56 \pm 0.16$ & > 20.93 & > 20.74 \\
3146 & 4.0 & $18.82 \pm 0.38$ & $18.18 \pm 0.15$ & $17.42 \pm 0.18$ & $17.43 \pm 0.21$ & > 17.03 & > 20.29 & > 19.85 \\
3190 & 2.4 & > 19.61 & $19.04 \pm 0.22$ & $18.48 \pm 0.28$ & $18.41 \pm 0.26$ & > 17.16 & > 21.8 & > 20.84 \\
3225 & 1.2 & > 21.09 & $20.99 \pm 0.22$ & > 20.64 & > 20.12 & > 19.71 & > 23.08 & > 22.64 \\
3233 & 1.2 & $21.52 \pm 0.34$ & $20.78 \pm 0.22$ & > 20.21 & > 19.9 & > 19.74 & > 22.95 & > 22.17 \\
3265 & 3.2 & $19.77 \pm 0.09$ & $18.83 \pm 0.05$ & $18.21 \pm 0.08$ & $18.2 \pm 0.14$ & $18.1 \pm 0.28$ & > 21.64 & > 21.59 \\
3272 & 5.6 & $17.37 \pm 0.31$ & $16.37 \pm 0.1$ & $15.73 \pm 0.08$ & $15.47 \pm 0.13$ & $15.48 \pm 0.16$ & > 19.06 & > 18.69 \\
3356 & 2.8 & $19.64 \pm 0.36$ & $18.68 \pm 0.12$ & $18.21 \pm 0.12$ & $18.18 \pm 0.18$ & $17.86 \pm 0.28$ & > 20.85 & > 19.49 \\
3365 & 2.0 & > 20.15 & $20.11 \pm 0.26$ & > 19.68 & > 19.51 & > 18.11 & > 21.31 & > 22.16 \\
3379 & 2.8 & $19.9 \pm 0.2$ & $18.94 \pm 0.14$ & $18.24 \pm 0.24$ & $18.32 \pm 0.37$ & > 17.45 & > 21.27 & > 20.78 \\
3548 & 4.8 & $16.73 \pm 0.05$ & $15.57 \pm 0.05$ & $14.93 \pm 0.05$ & $14.77 \pm 0.05$ & $14.64 \pm 0.05$ & $19.63 \pm 0.22$ & > 20.63 \\
3633 & 1.6 & > 20.69 & $19.92 \pm 0.18$ & $19.46 \pm 0.21$ & $19.21 \pm 0.16$ & > 19.2 & > 21.94 & > 21.58 \\

\end{longtable}
\tablefoot{(1) ID of the source. (2) Last detected radius in the \textit{g} band above $3\sigma$ sky level. (3-7) \textit{u}, \textit{g}, \textit{r}, \textit{i}, and \textit{z} band magnitudes. (8-9) GALEX \textit{NUV} and \textit{FUV} magnitudes. The upper limits ($3\sigma$) in the broadband magnitudes are denoted with > symbol. The sources with no GALEX UV data are marked with the "-" symbol. All the magnitudes were measured within an aperture of the last \textit{g}-band observed radius (column 2) and corrected for Galactic extinction.}
% \end{landscape}

%---------------------------------------------------------------

\begin{landscape}
\setlength\LTcapwidth{\textheight} % default: 4in 
\centering
\fontsize{9pt}{9pt}
\begin{longtable}{|c|*{8}{c}|*{4}{c}|}

\caption{Properties of the best RPS models, and the model with the same spin and velocity but without the RPS.}\label{table:model_params_full_sample}\label{page:models_table}\\
\hline

& \multicolumn{8}{c|}{\underline{\hspace{4.5cm} Best-fit model \hspace{4.5cm}}} & \multicolumn{4}{c|}{\underline{\hspace{2cm} Same model without RPS \hspace{2cm}}}\\ 

ID & \trps{} & \spin{} & \Vc{} &  $\chi^{2}_{\nu}$ & $\log$\mstar{} & $\log M_{gas}$ & $\log SFR$ & $Z_{gas}$ & $\log$\mstar{} & $\log M_{gas}$ & $\log SFR$ & $Z_{gas}$  \\
 &  (Gyr) &  & (\kms{}) &  &  (\msun{}) & (\msun{}) & (\msun{} yr$^{-1}$) & ($Z_{\odot}$) & (\msun{}) & (\msun{}) & (\msun{} yr$^{-1}$) & ($Z_{\odot}$)\\
(1) & (2) & (3) & (4) & (5) & (6) & (7) & (8) & (9) & (10) & (11)  &  (12) & (13)\\
\hline
\endfirsthead
\caption*{Table \ref{table:model_params_full_sample} continued.}\\
\hline
& \multicolumn{8}{c|}{\underline{\hspace{4.5cm} Best-fit model \hspace{4.5cm}}} & \multicolumn{4}{c|}{\underline{\hspace{2cm} Same model without RPS \hspace{2cm}}}\\ 

ID & \trps{} & \spin{} & \Vc{} &  $\chi^{2}_{\nu}$ & $\log$\mstar{} & $\log M_{gas}$ & $\log SFR$ & $Z_{gas}$ & $\log$\mstar{} & $\log M_{gas}$ & $\log SFR$ & $Z_{gas}$  \\
 &  (Gyr) &  & (\kms{}) &  &  (\msun{}) & (\msun{}) & (\msun{} yr$^{-1}$) & ($Z_{\odot}$) & (\msun{}) & (\msun{}) & (\msun{} yr$^{-1}$) & ($Z_{\odot}$)\\
(1) & (2) & (3) & (4) & (5) & (6) & (7) & (8) & (9) & (10) & (11)  &  (12) & (13)\\
\hline
\endhead
\hline
\endfoot
67 & $10.84\pm2.37$ & $0.16\pm0.06$ & $63\pm20$ & 0.27 & $7.17\pm{0.70}$ & $5.70\pm{0.12}$ & $-5.97\pm{0.27}$ & $0.32\pm{0.04}$ & $7.70\pm{0.26}$ & $9.01\pm{0.44}$ & $-1.50\pm{0.22}$ & $0.19\pm{0.07}$ \\
186 & $13.59\pm0.01$ & $0.17\pm0.01$ & $50\pm2$ & 5.70 & $7.27\pm{0.01}$ & $6.35\pm{0.47}$ & $-4.20\pm{0.52}$ & $0.36\pm{0.03}$ & $7.29\pm{0.01}$ & $8.75\pm{0.05}$ & $-1.87\pm{0.02}$ & $0.12\pm{0.01}$ \\
227 & $10.18\pm1.11$ & $0.12\pm0.02$ & $59\pm6$ & 0.36 & $7.27\pm{0.23}$ & $6.24\pm{0.12}$ & $-5.01\pm{0.12}$ & $0.47\pm{0.03}$ & $7.81\pm{0.11}$ & $8.97\pm{0.14}$ & $-1.40\pm{0.10}$ & $0.28\pm{0.06}$ \\
261 & $13.56\pm0.01$ & $0.15\pm0.01$ & $62\pm2$ & 20.56 & $7.73\pm{0.00}$ & $6.06\pm{0.24}$ & $-4.56\pm{0.26}$ & $0.57\pm{0.07}$ & $7.75\pm{0.00}$ & $9.03\pm{0.05}$ & $-1.44\pm{0.00}$ & $0.18\pm{0.01}$ \\
321 & $12.98\pm0.61$ & $0.30\pm0.07$ & $79\pm38$ & 0.82 & $7.29\pm{0.83}$ & $5.33\pm{1.29}$ & $-6.61\pm{2.65}$ & $0.22\pm{0.10}$ & $7.45\pm{0.70}$ & $9.21\pm{0.67}$ & $-1.77\pm{0.60}$ & $0.07\pm{0.01}$ \\
421 & $13.29\pm0.12$ & $0.18\pm0.01$ & $64\pm2$ & 1.59 & $7.58\pm{0.02}$ & $5.04\pm{0.48}$ & $-7.09\pm{0.80}$ & $0.31\pm{0.03}$ & $7.66\pm{0.00}$ & $9.08\pm{0.04}$ & $-1.54\pm{0.00}$ & $0.13\pm{0.01}$ \\
466 & $12.08\pm1.40$ & $0.20\pm0.04$ & $34\pm9$ & 0.77 & $6.21\pm{0.53}$ & $4.15\pm{0.50}$ & $-8.40\pm{1.24}$ & $0.20\pm{0.08}$ & $6.56\pm{0.22}$ & $8.24\pm{0.35}$ & $-2.56\pm{0.21}$ & $0.08\pm{0.03}$ \\
590 & $11.26\pm1.84$ & $0.13\pm0.02$ & $61\pm12$ & 0.23 & $7.41\pm{0.54}$ & $6.02\pm{0.09}$ & $-5.32\pm{0.08}$ & $0.43\pm{0.02}$ & $7.84\pm{0.20}$ & $8.99\pm{0.26}$ & $-1.36\pm{0.18}$ & $0.24\pm{0.05}$ \\
604 & $13.51\pm0.06$ & $0.19\pm0.01$ & $34\pm2$ & 5.42 & $6.51\pm{0.07}$ & $3.34\pm{0.08}$ & $-9.74\pm{0.08}$ & $0.13\pm{0.00}$ & $6.57\pm{0.05}$ & $8.25\pm{0.07}$ & $-2.55\pm{0.04}$ & $0.08\pm{0.00}$ \\
646 & $10.58\pm2.58$ & $0.15\pm0.05$ & $69\pm29$ & 0.12 & $7.25\pm{0.98}$ & $5.94\pm{0.52}$ & $-5.28\pm{0.54}$ & $0.40\pm{0.02}$ & $7.83\pm{0.47}$ & $9.04\pm{0.61}$ & $-1.38\pm{0.42}$ & $0.22\pm{0.06}$ \\
796 & $10.89\pm1.52$ & $0.11\pm0.01$ & $65\pm10$ & 1.08 & $7.59\pm{0.41}$ & $6.43\pm{0.20}$ & $-4.65\pm{0.12}$ & $0.55\pm{0.02}$ & $8.05\pm{0.15}$ & $9.07\pm{0.20}$ & $-1.17\pm{0.14}$ & $0.32\pm{0.04}$ \\
892 & $10.81\pm1.01$ & $0.15\pm0.03$ & $72\pm13$ & 0.71 & $7.53\pm{0.33}$ & $6.43\pm{0.20}$ & $-4.93\pm{0.07}$ & $0.41\pm{0.04}$ & $8.03\pm{0.15}$ & $9.22\pm{0.24}$ & $-1.21\pm{0.12}$ & $0.23\pm{0.05}$ \\
935 & $10.80\pm2.80$ & $0.48\pm0.33$ & $316\pm283$ & 0.04 & $7.95\pm{2.35}$ & $5.77\pm{1.04}$ & $-6.39\pm{0.74}$ & $0.22\pm{0.02}$ & $8.55\pm{1.78}$ & $10.02\pm{1.77}$ & $-0.91\pm{1.45}$ & $0.11\pm{0.04}$ \\
964 & $10.16\pm1.77$ & $0.10\pm0.01$ & $62\pm11$ & 1.39 & $7.44\pm{0.50}$ & $6.37\pm{0.33}$ & $-4.61\pm{0.28}$ & $0.57\pm{0.03}$ & $8.05\pm{0.18}$ & $9.01\pm{0.25}$ & $-1.18\pm{0.16}$ & $0.36\pm{0.05}$ \\
1008 & $13.05\pm0.49$ & $0.23\pm0.02$ & $37\pm8$ & 5.36 & $6.39\pm{0.39}$ & $3.91\pm{0.68}$ & $-8.82\pm{1.23}$ & $0.15\pm{0.07}$ & $6.55\pm{0.27}$ & $8.37\pm{0.29}$ & $-2.57\pm{0.26}$ & $0.05\pm{0.00}$ \\
1017 & $12.44\pm0.79$ & $0.08\pm0.03$ & $30\pm9$ & 2.67 & $6.80\pm{0.38}$ & $5.19\pm{0.21}$ & $-5.62\pm{0.61}$ & $0.80\pm{0.32}$ & $6.96\pm{0.32}$ & $8.02\pm{0.43}$ & $-2.17\pm{0.31}$ & $0.44\pm{0.20}$ \\
1160 & $12.72\pm0.12$ & $0.16\pm0.01$ & $64\pm2$ & 2.00 & $7.57\pm{0.02}$ & $6.08\pm{0.04}$ & $-5.46\pm{0.12}$ & $0.38\pm{0.03}$ & $7.75\pm{0.00}$ & $9.08\pm{0.04}$ & $-1.44\pm{0.00}$ & $0.16\pm{0.01}$ \\
1164 & $13.00\pm0.10$ & $0.09\pm0.01$ & $54\pm2$ & 0.71 & $7.81\pm{0.01}$ & $6.11\pm{0.12}$ & $-4.84\pm{0.27}$ & $0.80\pm{0.11}$ & $7.93\pm{0.02}$ & $8.81\pm{0.06}$ & $-1.29\pm{0.02}$ & $0.43\pm{0.07}$ \\
1346 & $10.80\pm2.80$ & $0.17\pm0.10$ & $48\pm20$ & 0.33 & $6.70\pm{0.73}$ & $5.68\pm{0.57}$ & $-5.30\pm{0.75}$ & $0.38\pm{0.17}$ & $7.18\pm{0.27}$ & $8.56\pm{0.62}$ & $-1.98\pm{0.24}$ & $0.25\pm{0.18}$ \\
1352 & $10.42\pm2.42$ & $0.09\pm0.03$ & $44\pm15$ & 0.94 & $7.02\pm{0.77}$ & $5.82\pm{0.44}$ & $-5.04\pm{0.24}$ & $0.69\pm{0.08}$ & $7.61\pm{0.33}$ & $8.48\pm{0.48}$ & $-1.58\pm{0.31}$ & $0.44\pm{0.12}$ \\
1397 & $10.80\pm2.80$ & $0.24\pm0.15$ & $72\pm47$ & 0.04 & $6.79\pm{1.29}$ & $5.26\pm{0.62}$ & $-5.75\pm{1.02}$ & $0.26\pm{0.04}$ & $7.38\pm{0.70}$ & $8.88\pm{1.01}$ & $-1.82\pm{0.61}$ & $0.14\pm{0.08}$ \\
1405 & $13.60\pm0.01$ & $0.16\pm0.01$ & $76\pm2$ & 2.42 & $8.04\pm{0.00}$ & $7.87\pm{0.10}$ & $-2.38\pm{0.10}$ & $0.52\pm{0.05}$ & $8.04\pm{0.00}$ & $9.30\pm{0.04}$ & $-1.19\pm{0.01}$ & $0.20\pm{0.02}$ \\
1424 & $12.78\pm0.32$ & $0.21\pm0.02$ & $42\pm5$ & 1.57 & $6.65\pm{0.18}$ & $5.07\pm{0.10}$ & $-7.08\pm{0.22}$ & $0.23\pm{0.03}$ & $6.86\pm{0.11}$ & $8.55\pm{0.15}$ & $-2.28\pm{0.10}$ & $0.07\pm{0.01}$ \\
1476 & $10.54\pm2.54$ & $0.21\pm0.10$ & $111\pm53$ & 0.09 & $7.82\pm{1.01}$ & $6.51\pm{0.57}$ & $-4.83\pm{0.51}$ & $0.34\pm{0.02}$ & $8.36\pm{0.54}$ & $9.64\pm{0.71}$ & $-0.96\pm{0.42}$ & $0.20\pm{0.07}$ \\
1479 & $10.71\pm2.71$ & $0.13\pm0.06$ & $48\pm24$ & 0.12 & $6.76\pm{1.05}$ & $5.11\pm{0.49}$ & $-6.89\pm{1.00}$ & $0.40\pm{0.11}$ & $7.34\pm{0.51}$ & $8.51\pm{0.77}$ & $-1.83\pm{0.48}$ & $0.27\pm{0.13}$ \\
1529 & $12.70\pm0.20$ & $0.13\pm0.01$ & $97\pm3$ & 4.96 & $8.49\pm{0.03}$ & $6.95\pm{0.09}$ & $-4.16\pm{0.20}$ & $0.68\pm{0.06}$ & $8.62\pm{0.01}$ & $9.60\pm{0.05}$ & $-0.69\pm{0.01}$ & $0.36\pm{0.04}$ \\
1593 & $12.40\pm0.34$ & $0.18\pm0.01$ & $90\pm4$ & 0.57 & $8.03\pm{0.08}$ & $6.60\pm{0.06}$ & $-4.90\pm{0.16}$ & $0.40\pm{0.02}$ & $8.24\pm{0.04}$ & $9.54\pm{0.08}$ & $-1.03\pm{0.03}$ & $0.18\pm{0.02}$ \\
1633 & $10.64\pm2.64$ & $0.19\pm0.07$ & $69\pm29$ & 0.41 & $7.08\pm{0.96}$ & $5.56\pm{0.33}$ & $-6.37\pm{0.45}$ & $0.28\pm{0.01}$ & $7.64\pm{0.44}$ & $9.05\pm{0.60}$ & $-1.55\pm{0.37}$ & $0.15\pm{0.05}$ \\
1687 & $10.80\pm2.80$ & $0.44\pm0.36$ & $310\pm290$ & 0.22 & $7.78\pm{2.52}$ & $5.61\pm{1.20}$ & $-6.27\pm{0.55}$ & $0.29\pm{0.08}$ & $8.38\pm{1.94}$ & $9.66\pm{2.14}$ & $-1.06\pm{1.60}$ & $0.20\pm{0.08}$ \\
1719 & $13.08\pm0.38$ & $0.48\pm0.27$ & $190\pm134$ & 0.28 & $8.15\pm{1.08}$ & $6.38\pm{0.82}$ & $-6.20\pm{0.37}$ & $0.22\pm{0.03}$ & $8.27\pm{1.01}$ & $9.94\pm{1.03}$ & $-1.14\pm{0.75}$ & $0.07\pm{0.01}$ \\
1846 & $10.56\pm2.17$ & $0.12\pm0.03$ & $55\pm11$ & 1.96 & $7.19\pm{0.55}$ & $6.06\pm{0.32}$ & $-4.92\pm{0.35}$ & $0.51\pm{0.05}$ & $7.84\pm{0.24}$ & $8.84\pm{0.28}$ & $-1.37\pm{0.21}$ & $0.30\pm{0.09}$ \\
1968 & $13.45\pm0.01$ & $0.08\pm0.01$ & $58\pm2$ & 5.31 & $8.08\pm{0.04}$ & $5.61\pm{0.27}$ & $-5.09\pm{0.53}$ & $1.47\pm{0.30}$ & $8.11\pm{0.03}$ & $8.89\pm{0.06}$ & $-1.12\pm{0.02}$ & $0.55\pm{0.10}$ \\
1993 & $9.70\pm1.80$ & $0.23\pm0.04$ & $119\pm13$ & 10.21 & $7.99\pm{0.35}$ & $6.93\pm{0.18}$ & $-4.59\pm{0.18}$ & $0.31\pm{0.02}$ & $8.63\pm{0.14}$ & $9.89\pm{0.15}$ & $-0.70\pm{0.12}$ & $0.17\pm{0.03}$ \\
2001 & $13.48\pm0.01$ & $0.15\pm0.01$ & $20\pm2$ & -0.00 & $5.91\pm{0.05}$ & $2.69\pm{0.06}$ & $-10.38\pm{0.05}$ & $3.81\pm{3.65}$ & $5.96\pm{0.04}$ & $7.61\pm{0.06}$ & $-3.11\pm{0.05}$ & $0.09\pm{0.01}$ \\
2046 & $12.16\pm0.14$ & $0.14\pm0.01$ & $88\pm2$ & 1.19 & $8.18\pm{0.00}$ & $6.86\pm{0.04}$ & $-4.31\pm{0.13}$ & $0.55\pm{0.04}$ & $8.41\pm{0.01}$ & $9.48\pm{0.04}$ & $-0.87\pm{0.02}$ & $0.29\pm{0.03}$ \\
2079 & $13.19\pm0.28$ & $0.34\pm0.04$ & $99\pm9$ & 4.19 & $7.79\pm{0.11}$ & $5.40\pm{0.60}$ & $-7.45\pm{1.21}$ & $0.17\pm{0.06}$ & $7.90\pm{0.08}$ & $9.65\pm{0.11}$ & $-1.36\pm{0.06}$ & $0.07\pm{0.01}$ \\
2269 & $10.21\pm1.61$ & $0.30\pm0.06$ & $97\pm21$ & 1.47 & $7.34\pm{0.52}$ & $6.25\pm{0.34}$ & $-5.88\pm{0.24}$ & $0.22\pm{0.02}$ & $7.93\pm{0.22}$ & $9.60\pm{0.28}$ & $-1.34\pm{0.17}$ & $0.08\pm{0.01}$ \\
2343 & $11.61\pm1.39$ & $0.11\pm0.03$ & $47\pm9$ & 0.52 & $7.19\pm{0.38}$ & $5.82\pm{0.04}$ & $-5.32\pm{0.30}$ & $0.54\pm{0.12}$ & $7.57\pm{0.13}$ & $8.64\pm{0.30}$ & $-1.61\pm{0.11}$ & $0.30\pm{0.10}$ \\
2351 & $11.72\pm1.29$ & $0.13\pm0.02$ & $37\pm7$ & 0.93 & $6.59\pm{0.45}$ & $5.24\pm{0.10}$ & $-6.31\pm{0.11}$ & $0.35\pm{0.02}$ & $7.00\pm{0.18}$ & $8.33\pm{0.25}$ & $-2.13\pm{0.17}$ & $0.17\pm{0.04}$ \\
2365 & $8.31\pm0.31$ & $0.24\pm0.04$ & $116\pm19$ & 22.81 & $7.68\pm{0.27}$ & $6.77\pm{0.23}$ & $-4.97\pm{0.15}$ & $0.24\pm{0.01}$ & $8.46\pm{0.16}$ & $9.86\pm{0.22}$ & $-0.85\pm{0.11}$ & $0.14\pm{0.02}$ \\
2458 & $10.79\pm2.79$ & $0.28\pm0.17$ & $88\pm66$ & 0.11 & $6.72\pm{1.60}$ & $5.02\pm{0.76}$ & $-7.23\pm{1.11}$ & $0.20\pm{0.05}$ & $7.35\pm{1.01}$ & $8.87\pm{1.21}$ & $-1.88\pm{0.88}$ & $0.12\pm{0.04}$ \\
2531 & $11.20\pm0.10$ & $0.11\pm0.01$ & $58\pm2$ & 2.32 & $7.46\pm{0.00}$ & $6.35\pm{0.01}$ & $-4.68\pm{0.11}$ & $0.57\pm{0.06}$ & $7.89\pm{0.01}$ & $8.92\pm{0.05}$ & $-1.32\pm{0.01}$ & $0.32\pm{0.04}$ \\
2572 & $10.80\pm2.80$ & $0.35\pm0.24$ & $232\pm210$ & 0.35 & $7.61\pm{2.50}$ & $5.39\pm{1.14}$ & $-7.15\pm{0.26}$ & $0.23\pm{0.02}$ & $8.23\pm{1.89}$ & $9.58\pm{1.91}$ & $-1.20\pm{1.57}$ & $0.12\pm{0.06}$ \\
2621 & $12.54\pm0.26$ & $0.14\pm0.01$ & $48\pm2$ & 4.40 & $7.16\pm{0.05}$ & $5.73\pm{0.06}$ & $-5.73\pm{0.16}$ & $0.39\pm{0.03}$ & $7.39\pm{0.01}$ & $8.69\pm{0.06}$ & $-1.78\pm{0.01}$ & $0.17\pm{0.01}$ \\
2690 & $13.01\pm0.30$ & $0.22\pm0.01$ & $35\pm5$ & 6.36 & $6.30\pm{0.26}$ & $4.47\pm{0.40}$ & $-7.95\pm{0.63}$ & $0.20\pm{0.02}$ & $6.47\pm{0.19}$ & $8.27\pm{0.19}$ & $-2.64\pm{0.18}$ & $0.06\pm{0.00}$ \\
2731 & $12.76\pm0.75$ & $0.29\pm0.07$ & $68\pm22$ & 0.84 & $7.10\pm{0.52}$ & $5.00\pm{0.43}$ & $-7.80\pm{1.12}$ & $0.18\pm{0.07}$ & $7.29\pm{0.35}$ & $9.09\pm{0.43}$ & $-1.89\pm{0.31}$ & $0.06\pm{0.01}$ \\
2887 & $11.92\pm0.18$ & $0.09\pm0.01$ & $58\pm2$ & 2.10 & $7.77\pm{0.03}$ & $6.54\pm{0.02}$ & $-4.22\pm{0.14}$ & $0.81\pm{0.10}$ & $8.06\pm{0.02}$ & $8.92\pm{0.08}$ & $-1.17\pm{0.01}$ & $0.46\pm{0.07}$ \\
2999 & $11.80\pm1.22$ & $0.16\pm0.03$ & $42\pm7$ & 1.73 & $6.69\pm{0.35}$ & $5.25\pm{0.07}$ & $-6.40\pm{0.24}$ & $0.30\pm{0.04}$ & $7.02\pm{0.18}$ & $8.54\pm{0.22}$ & $-2.12\pm{0.16}$ & $0.13\pm{0.03}$ \\
3032 & $10.80\pm2.80$ & $0.46\pm0.35$ & $458\pm60$ & 0.03 & $10.27\pm{0.61}$ & $8.10\pm{1.83}$ & $-3.78\pm{3.65}$ & $1.61\pm{1.45}$ & $10.37\pm{0.71}$ & $11.37\pm{0.23}$ & $0.51\pm{0.57}$ & $1.10\pm{1.00}$ \\
3088 & $12.75\pm0.26$ & $0.13\pm0.01$ & $54\pm2$ & 3.28 & $7.46\pm{0.05}$ & $5.92\pm{0.12}$ & $-5.44\pm{0.26}$ & $0.45\pm{0.04}$ & $7.64\pm{0.00}$ & $8.84\pm{0.05}$ & $-1.55\pm{0.01}$ & $0.22\pm{0.03}$ \\
3112 & $12.78\pm0.82$ & $0.23\pm0.04$ & $38\pm10$ & 1.51 & $6.30\pm{0.50}$ & $3.96\pm{0.64}$ & $-8.48\pm{1.56}$ & $0.18\pm{0.06}$ & $6.52\pm{0.31}$ & $8.35\pm{0.36}$ & $-2.60\pm{0.29}$ & $0.06\pm{0.00}$ \\
3116 & $10.80\pm2.80$ & $0.11\pm0.10$ & $298\pm220$ & -0.00 & $9.18\pm{2.03}$ & $7.92\pm{1.69}$ & $-3.06\pm{2.40}$ & $4.04\pm{3.77}$ & $9.61\pm{1.62}$ & $10.04\pm{1.69}$ & $0.04\pm{1.28}$ & $1.73\pm{1.58}$ \\
3128 & $12.52\pm0.28$ & $0.14\pm0.01$ & $49\pm3$ & 2.76 & $7.19\pm{0.11}$ & $5.72\pm{0.08}$ & $-5.74\pm{0.18}$ & $0.38\pm{0.04}$ & $7.40\pm{0.08}$ & $8.74\pm{0.11}$ & $-1.76\pm{0.06}$ & $0.17\pm{0.02}$ \\
3146 & $13.21\pm0.21$ & $0.21\pm0.01$ & $54\pm4$ & 1.00 & $7.13\pm{0.13}$ & $4.73\pm{0.59}$ & $-7.75\pm{1.02}$ & $0.22\pm{0.03}$ & $7.23\pm{0.09}$ & $8.86\pm{0.10}$ & $-1.93\pm{0.07}$ & $0.08\pm{0.00}$ \\
3190 & $10.80\pm2.80$ & $0.28\pm0.15$ & $103\pm54$ & 0.19 & $7.38\pm{0.94}$ & $5.60\pm{0.48}$ & $-7.11\pm{1.23}$ & $0.21\pm{0.07}$ & $7.94\pm{0.42}$ & $9.41\pm{0.67}$ & $-1.33\pm{0.33}$ & $0.14\pm{0.07}$ \\
3225 & $10.80\pm2.80$ & $0.28\pm0.15$ & $107\pm87$ & 0.09 & $6.84\pm{1.97}$ & $4.73\pm{0.73}$ & $-8.21\pm{0.44}$ & $0.20\pm{0.04}$ & $7.48\pm{1.36}$ & $8.99\pm{1.44}$ & $-1.79\pm{1.17}$ & $0.09\pm{0.04}$ \\
3233 & $10.80\pm2.80$ & $0.46\pm0.34$ & $237\pm183$ & 0.09 & $8.20\pm{1.53}$ & $5.93\pm{0.41}$ & $-6.44\pm{0.92}$ & $0.26\pm{0.09}$ & $8.49\pm{1.26}$ & $10.08\pm{1.24}$ & $-0.98\pm{0.98}$ & $0.16\pm{0.09}$ \\
3265 & $11.91\pm0.81$ & $0.12\pm0.01$ & $37\pm4$ & 1.47 & $6.79\pm{0.21}$ & $5.57\pm{0.07}$ & $-5.72\pm{0.17}$ & $0.43\pm{0.04}$ & $7.17\pm{0.06}$ & $8.37\pm{0.14}$ & $-1.98\pm{0.06}$ & $0.22\pm{0.04}$ \\
3272 & $10.04\pm1.46$ & $0.14\pm0.01$ & $96\pm12$ & 2.36 & $8.04\pm{0.37}$ & $6.98\pm{0.22}$ & $-4.11\pm{0.19}$ & $0.53\pm{0.04}$ & $8.54\pm{0.18}$ & $9.59\pm{0.17}$ & $-0.77\pm{0.15}$ & $0.32\pm{0.04}$ \\
3356 & $12.49\pm0.67$ & $0.18\pm0.02$ & $49\pm4$ & 0.98 & $7.00\pm{0.18}$ & $5.37\pm{0.26}$ & $-6.48\pm{0.52}$ & $0.29\pm{0.04}$ & $7.24\pm{0.04}$ & $8.75\pm{0.11}$ & $-1.92\pm{0.04}$ & $0.11\pm{0.02}$ \\
3365 & -- & $0.34\pm0.04$ & $47\pm11$ & 10.04 & $6.58\pm{0.32}$ & $8.65\pm{0.33}$ & $-2.56\pm{0.29}$ & $0.03\pm{0.00}$ & -- & -- & -- & -- \\
3379 & $12.59\pm0.70$ & $0.20\pm0.01$ & $51\pm10$ & 1.24 & $6.99\pm{0.41}$ & $5.23\pm{0.29}$ & $-6.87\pm{0.48}$ & $0.27\pm{0.02}$ & $7.22\pm{0.26}$ & $8.79\pm{0.25}$ & $-1.94\pm{0.24}$ & $0.10\pm{0.01}$ \\
3548 & $12.30\pm0.10$ & $0.08\pm0.01$ & $64\pm2$ & 18.13 & $8.06\pm{0.02}$ & $6.76\pm{0.07}$ & $-3.85\pm{0.23}$ & $1.06\pm{0.19}$ & $8.29\pm{0.03}$ & $9.02\pm{0.06}$ & $-0.98\pm{0.02}$ & $0.58\pm{0.10}$ \\
3633 & $10.45\pm2.46$ & $0.18\pm0.05$ & $73\pm27$ & 0.20 & $7.20\pm{0.92}$ & $5.92\pm{0.51}$ & $-5.75\pm{0.34}$ & $0.30\pm{0.02}$ & $7.78\pm{0.45}$ & $9.15\pm{0.52}$ & $-1.44\pm{0.39}$ & $0.15\pm{0.04}$ \\

% \end{tabular}
% \end{table}
\end{longtable}
\tablefoot{(1) Name of the source; (2-4) Best-fit model \trps{}, \spin{} and \Vc{} values; (5) Reduced \chisq{} value of the best-fit model (ratio of the total \chisq{} by the number of data points used for the \chisq{} computation); (6-9) Stellar-mass, gas mass, SFR, and gas-phase metallicity from the best-fit model; (10-13) Variation of these quantities in the absence of an RPS, for the same best-fit model (this should not be confused with the independent non-RPS fits that we discarded as poor fits in Sect. \ref{sect:model_fitting_with_sample_profiles}). The uncertainties given in the RPS models are from the confidence limits in \trps{}, \spin{} and \Vc{} parameters. For the non-RPS models, the uncertainties are from the error in \spin{} and \Vc{} alone. For ID 3365, the best-fit model is a model without RPS and is indicated by the $-$ symbol.}
\end{landscape}

\end{appendix}
\end{document}